%% file: hepph.tex
\documentclass[a4paper,11pt]{article}
\usepackage[english]{babel}
\usepackage{jheppub}
\usepackage{amsmath,amssymb,graphicx,textcomp,enumerate,alltt,xspace,multirow}

\input texmacros.tex

\newcommand{\subtopic}[1]{\vspace{4mm}}

\preprint{
\begin{flushright}
 ZU-TH 22/16   
\end{flushright}
}

\title{An NLO+PS generator for ${\boldsymbol{t \bar t}}$ and ${\boldsymbol{W t}}$ production and decay including non-resonant and interference effects} 

\author[a]{Tom\'a\v{s} Je\v{z}o,} 
\author[c]{Jonas M. Lindert,}
\author[b]{Paolo Nason,}
\author[a]{Carlo Oleari}
\author[c]{and Stefano Pozzorini}
\emailAdd{tomas.jezo@mib.infn.it}
\emailAdd{lindert@physik.uzh.ch}
\emailAdd{paolo.nason@mib.infn.it}
\emailAdd{carlo.oleari@mib.infn.it}
\emailAdd{pozzorin@physik.uzh.ch}

\affiliation[a]{
  Universit\`a di Milano-Bicocca and INFN, Sezione di Milano-Bicocca,
  Piazza della Scienza 3,\\ 20126 Milano, Italy
}

\affiliation[b]{
  INFN, Sezione di Milano-Bicocca, Piazza della Scienza 3, 20126 Milano, Italy
}

\affiliation[c]{
  Physics Institute, Universit\"at Z\"urich, Z\"urich, Switzerland
}

\abstract{ We present a Monte Carlo generator that implements
  significant theoretical improvements in the simulation of top-quark
  pair production and decay at the LHC.  Spin correlations and
  off-shell effects in top-decay chains are described in terms of
  exact matrix elements for
  $pp\to \fourl \,\bbbar$ at NLO QCD, where 
the leptons $\ell$ and $l$ belong to 
different families, and $b$ quarks are massive.
Thus, the
  contributions from $t\bar t$ and $Wt$ single-top production as well as
  their quantum interference are fully included.
  Matrix elements are matched to the \PythiaEight~parton shower using a
  recently proposed method that allows for a consistent treatment of
  resonances in the \POWHEG{} framework.
  These theoretical improvements are
  especially important for the interpretation of precision
  measurements of the top-quark mass, for single-top analyses in the
  $Wt$ channel, and for $t\bar t$ and $Wt$ backgrounds in the presence
  of jet vetoes or cuts that enhance off-shell effects.  The new
  generator is based on a process-independent interface of the
  \OpenLoops amplitude generator with the \POWHEGBOX{} framework.  }


\begin{document}
\maketitle

\flushbottom

\section{Introduction}

The production of top-quark pairs plays a key role in the physics program of
the LHC. On the one hand, this process can be exploited
for detailed studies of top-quark properties and interactions, for
precision tests of the Standard Model~(SM), and for measurements of
fundamental parameters such as the top-quark mass.
On the other hand, it represents a challenging background in many SM
studies and searches of physics beyond the Standard Model~(BSM).
The sensitivity of such analyses can depend in a critical way 
on the precision of theoretical simulations,
and given that any experimental measurement is
performed at the level of top-decay products,
precise theoretical predictions are needed for the full process of
$\ttbar$ production and decay, including, if possible, also irreducible
backgrounds and interference effects.
This is especially
important in the context of precision measurements of the top-quark
mass.

After the discovery of the Higgs boson and the measurement of its mass, the
allowed values of the $W$-boson and top-quark masses are strongly correlated,
and a precise determination of both parameters would lead to a SM test of
unprecedented precision~\cite{Agashe:2014kda}.  At present there is some
tension, at the $1.6\,\sigma$ level, between the indirect top-mass
determination from electroweak precision data~($177\pm 2.1$~GeV) and the
combination of direct measurements at the Tevatron and the LHC~($173.24\pm
0.95$~GeV).
The precise value of the top-quark mass is particularly crucial to the issue
of vacuum stability in the Standard Model~\cite{Degrassi:2012ry}.  At high
scales, the Higgs quartic coupling $\lambda$ evolves to increasingly small
values as $m_t$ grows, and it is remarkable that above about $m_t=171$~GeV,
i.e.~very close to the present world average, $\lambda$ becomes negative at
the Planck scale, rendering the electroweak vacuum meta-stable, while for $m_t>176$~GeV
the electroweak vacuum becomes unstable.

The most precise top-mass measurements are based upon fits of $m_t$-dependent
Monte Carlo predictions to certain kinematic distributions.
For a precise $m_t$ determination, it is crucial to rely on Monte Carlo
generators that describe $\ttbar$ production and decay, including the shape
of top resonances, on the basis of higher-order scattering amplitudes. These
are given in terms of a theoretically well-defined top-mass parameter in an
unambiguous way, and can provide more reliable estimates of perturbative
theoretical uncertainties.

Perturbative predictions for inclusive $\ttbar$ production are available up
to next-to-next-to leading order~(NNLO) in QCD~\cite{Czakon:2013goa,
  Czakon:2015owf}, and the next-to-leading order~(NLO) electroweak
corrections are also known~\cite{Beenakker:1993yr, Bernreuther:2006vg,
  Bernreuther:2008md, Kuhn:2006vh, Hollik:2011ps, Kuhn:2013zoa,
  Pagani:2016caq}.  Calculations at NLO QCD exist also for $\ttbar$
production in association with one~\cite{Dittmaier:2007wz} or
two~\cite{Bredenstein:2009aj, Bredenstein:2010rs, Bevilacqua:2009zn,
  Bevilacqua:2010ve, Bevilacqua:2011aa} extra jets.  The present state-of-the
art accuracy of $\ttbar$ generators is NLO QCD, and inclusive generators
matching NLO QCD matrix elements to parton showers~(NLO+PS, from now on) have
been available for quite some time: in~\citere{Frixione:2003ei}, based
upon the \MCatNLO~\cite{Frixione:2002ik} method, and
in~\citere{Frixione:2007nw}, based upon the \POWHEG{}
method~\cite{Nason:2004rx, Frixione:2007vw}.  In the following we will refer
to the latter as the \hvq{} generator.\footnote{ \hvq{} is the name of the
  corresponding directory in the \POWHEGBOX{} package. The \hvq{} code is also
  available under the \VTWO{} package.}  More recent generators can provide NLO
QCD precision also for $\ttbar$ production in association with up to one or
two additional jets~\cite{Kardos:2011qa, Alioli:2011as, Frederix:2012ps,
  Kardos:2013vxa, Cascioli:2013era, Hoeche:2013mua, Hoeche:2014qda}.
Top-quark decays are known at NNLO QCD~\cite{Brucherseifer:2013iv}, but so
far they have always been implemented at lower precision in complete
calculations of top-pair production and decay.  The vast majority of such
calculations rely on the narrow-width approximation~(NWA), where matrix
elements for $\ttbar$ production and decay factorise. 
Various generators based on the NWA
approximation~\cite{Frixione:2003ei, Frixione:2007nw, Kardos:2011qa,
  Alioli:2011as, Frederix:2012ps, Kardos:2013vxa, Cascioli:2013era,
  Hoeche:2013mua, Hoeche:2014qda} apply NLO QCD corrections only to $\ttbar$
production and include finite-width effects and spin correlations
in an approximate way using the method of~\citere{Frixione:2007zp}.
The best
available NWA fixed-order calculations implement NLO QCD corrections to the
production and decay parts with exact spin
correlations~\cite{Bernreuther:2004jv, Melnikov:2009dn, Campbell:2012uf}.
The \ttNLOdec{}\footnote{The name \ttNLOdec{} refers to the corresponding
  directory in the \VTWO{} package.}  generator of~\citere{Campbell:2014kua}
implements the results of~\citere{Campbell:2012uf} using
the \POWHEG{} method~\cite{Nason:2004rx, Frixione:2007vw}. Finite width and
interference effects are implemented in an approximate way, using LO
 $pp\to \WWbb$ matrix elements. Thus, in
the resonance region it provides NLO corrections to both production and decay,
including NLO corrections to $W$ hadronic decays, and implements full spin
correlations.  In addition, it can be operated both in the five-flavour number
scheme~(5FNS) and in the four-flavour number scheme~(4FNS).

A complete description of $\ttbar$ production and decay beyond the NWA
requires the calculation of the full set of Feynman diagrams that
contribute to the production of $\WWbb$ final states, including also
leptonic or hadronic $W$-boson decays.  The existing predictions at
NLO QCD~\cite{Bevilacqua:2010qb, Denner:2010jp, Denner:2012yc,
  Heinrich:2013qaa, Frederix:2013gra, Cascioli:2013wga} deal with the
different-flavour dilepton channel, $pp\to\enmn\,\bbbar$.
Besides an exact NLO treatment of spin correlations and off-shell effects
associated with the top-quark and $W$-boson resonances, such calculations
account for non-factorisable NLO effects~\cite{Beenakker:1999ya,
  Melnikov:1995fx, Falgari:2013gwa} and provide an exact NLO description
of the top resonance, including quantum corrections to the top propagator.
Moreover, in addition to doubly-resonant topologies of $\ttbar$ type, also
genuine non-resonant effects stemming from topologies with less than two top
or $W$-propagators are included, as well as quantum interferences between
different topologies.

The first NLO calculations of the $pp\to\enmn\,\bbbar$
process~\cite{Bevilacqua:2010qb, Denner:2010jp, Denner:2012yc,
  Heinrich:2013qaa} have been performed in the 5FNS, where $b$ quarks are
treated as massless particles. 
In the meanwhile, NLO QCD predictions in the 5FNS are
available also for  $\enmn\,\bbbar$ production in association with 
one extra jet~\cite{Bevilacqua:2015qha}.
Due to the presence of collinear $g\to \bbbar$ singularities, the
applicability of these calculations in the 5FNS is limited to observables
that involve at least two hard $b$ jets.  This restriction can be
circumvented through NLO calculations\footnote{For a discussion at LO
  see~\citere{Kauer:2001sp}.} in the 4FNS, where $b$ quarks are treated as
massive partons~\cite{Frederix:2013gra, Cascioli:2013wga}.
In addition to a more reliable description of $b$-quark kinematics, these
calculations give access to the full $\enmn\,\bbbar$ phase space, including
regions where one or both $b$ quarks become unresolved.  This is crucial in
order to describe top backgrounds in presence of jet vetoes.
Moreover, inclusive $\enmn\,\bbbar$ calculations in the 4FNS guarantee a
consistent theoretical treatment of single-top $\Wt$ production at
NLO.

In the 5FNS, $\Wt$ and $\ttbar$ production and decay involve partonic
channels of type $gb\to \WWb$ and $gg\to \WWbb$, respectively.  The $gg\to
\WWbb$ channel at LO is part of the NLO radiative corrections to the $gb\to
\WWb$ one,
thus yielding a NLO correction that, being $t\bar{t}$ mediated, is much
larger than the Born term. This led to the proposal of various
methods~\cite{Zhu:2001hw, Campbell:2005bb, Frixione:2008yi, White:2009yt} to
define single top cross sections not including the resonant $t\bar{t}$
contribution. 
However, the separation of $tW$ and $t\bar{t}$  production
breaks gauge invariance and 
does not allow for a consistent treatment of interference effects.
On the other hand, in the 4FNS the $pp\to \enmn\,\bbbar$
calculations provide a unified NLO description of $\ttbar$ and $\Wt$
production, with a fully consistent treatment of their quantum
interference~\cite{Cascioli:2013wga}.  Single-top production in the 4FNS is
described by topologies with a single top propagator and a collinear $g\to
\bbbar$ splitting in the initial state.
The fact that $g\to \bbbar$ splittings are accounted for by the matrix
elements guarantees a more precise modelling of the spectator $b$ quark,
while the simultaneous presence of $\Wt$ and $\ttbar$ channels, starting from
LO, ensures a perturbatively stable description of both contributions, as
well as a NLO accurate prediction for their interference.

A generator based on the \POWHEG{} method and $pp\to \enmn\,\bbbar$ matrix
elements at NLO in the 5FNS has been presented in~\citere{Garzelli:2014dka}.
However, the matching of parton showers to matrix elements that involve
top-quark resonances poses nontrivial technical and theoretical
problems~\cite{Jezo:2015aia} that have not been addressed
in~\citere{Garzelli:2014dka} and which cannot be solved within the original
formulations of the \POWHEG{} or \MCatNLO{} methods.
The problem is twofold. On the one hand, when interfacing a generator to a
shower, if we do not specify which groups of final-state particles arise from
the decay of the same resonance, the recoil resulting from shower emissions
leads to arbitrary shifts of the resonance invariant masses, whose magnitude
can largely exceed the top-quark width, resulting in unphysical distortions
of the top line shape~\cite{Jezo:2015aia}.
On the other hand, in the context of the infrared-subtraction and matching
procedures, the standard mappings that connect the Born and real-emission
phase spaces affect the top resonances in a way that drastically deteriorates
the efficiency of infrared~(IR) cancellations and jeopardises the consistency
of the matching method~\cite{Jezo:2015aia}.

A general NLO+PS matching technique that allows for a consistent treatment of
resonances has been introduced, and applied to $t$-channel single-top
production, in~\citere{Jezo:2015aia}.  This approach will be referred to as
resonance-aware matching. It is based on the \POWHEG\footnote{A related
  approach within the \MCatNLO{} framework has been presented and also
  applied to $t$-channel single-top production in~\citere{Frederix:2016rdc}.}
method and is implemented in the \RES{} framework, which represents an
extension of the \POWHEGBOX~\cite{Alioli:2010xd}.  In this framework each
component of the cross section~(i.e.~Born, virtual and real) is separated
into the sum of contributions that are 
dominated by well-defined resonance histories, 
such that in the narrow-width limit 
each parton can be uniquely attributed either to the 
decay products of a certain resonance or to the production subprocess.
Within each contribution the subtraction procedure is organized in
such a way that the off-shellness of resonant $s$-channel propagators is preserved,
and resonance information on the final-state particles can be communicated to the shower program that
handles further radiation and hadronization.  This avoids uncontrolled
resonance distortions, ensuring a NLO accurate description of the top line
shape. The resonance-aware approach 
also improves the efficiency of infrared subtraction and phase-space integration
in a dramatic way.

In this paper we present a NLO+PS generator, that we dub \bbfourl{} in the
following, based on NLO matrix elements for $pp\to\enmn\,\bbbar$ in the 4FNS
matched to \PythiaEight{}~\cite{Sjostrand:2007gs, Sjostrand:2014zea} using
the resonance-aware \POWHEG{} method.  This new generator combines, for the
first time, the following physics features:
\begin{itemize}
\item[-] consistent NLO+PS treatment of top resonances, including quantum
  corrections to top propagators and off-shell top-decay chains;
\item[-] exact spin correlations at NLO, interference between NLO radiation
  from top production and decays, full NLO accuracy in $\ttbar$ production
  and decays;
\item[-] unified treatment of $\ttbar$ and $\Wt$ production with interference
  at NLO;
\item[-] improved modelling of $b$-quark kinematics thanks to $b$-quark mass
  effects;
\item[-] access to phase-space regions with unresolved $b$ quarks and/or jet
  vetoes.
\end{itemize}
These improvements are of particular interest for precision top-mass
measurements, for $Wt$ analyses, and for top backgrounds in the presence of
jet vetoes or in the off-shell regime.
Technically, the \bbfourl{} generator is based on
\OpenLoops{}~\cite{OLhepforge} matrix elements.  To this end we have
developed a general and fully-flexible \POWHEGBOX{}+\OpenLoops{} interface,
which allows one to set up NLO+PS generators for any desired process.

The paper is organized as follows.
In \refse{sec:reminder} we briefly review the 
resonance-aware matching method.
In \refse{sec:BOXRES} we discuss new developments in the
\RES{} framework that have been relevant for the present work.
In \refse{sec:description} we discuss various aspects of the \bbfourl{}
generator, including scope, usage, interface to \PythiaEight{}, and
consistency checks.
In \refse{sec:setup} we detail the setup employed for the
phenomenological studies presented in the subsequent sections.  There we
compare the \bbfourl{} generator to the previously available \POWHEG{}
generators, the \hvq{} and \dec{} ones, and we present technical studies that
show the impact of the resonance-aware matching and of other improvements
implemented in \bbfourl{}.
Specifically, in \refse{sec:ttbarPhenomenology} we consider observables
that are directly sensitive to top-quark resonances and top-decay products,
while in \refse{sec:WtPhenomenology} we investigate the $\enmn\,\bbbar$
cross section in the presence of jet vetoes that enhance its single-top
content.
Our conclusions are presented in \refse{sec:conc}.

The \RES{} framework together with the \bbfourl{} generator can be downloaded at \url{http://powhegbox.mib.infn.it}.

\section{Resonance-aware subtraction and matching}
\label{sec:reminder}

In the following we recapitulate the problems that arise in
processes where intermediate narrow resonances can radiate as they decay,
and summarize the ideas and methodology behind the 
resonance-aware algorithm of~\citere{Jezo:2015aia}. 
We refer the reader to the original publication
for the description of the method in full detail.

Commonly used IR subtraction methods for the calculation of NLO 
corrections~\cite{Frixione:1995ms,Catani:1996vz,Catani:2002hc}
are based upon some procedure of momentum reshuffling for the construction of collinear and
infrared counterterms. More specifically, given the kinematics of the
real-emission process, and having specified a particular collinear
region~(i.e.~a pair of partons that are becoming collinear), there is a
well-defined mapping that constructs a Born-like kinematic
configuration~(called the ``underlying Born'' configuration) as a function of
the real one. The mapping is such that, in the strict collinear limit, the Born
configuration is obtained from the real one by appropriately merging the
collinear partons.  In the traditional methods, these mappings do not
necessarily preserve the virtuality of possible intermediate $s$-channel
resonances.  If we consider the collinear region of two partons arising from
the decay of the same $s$-channel resonance, the typical difference in the
resonance virtuality between the real kinematics and the underlying-Born one
is of order $m^2/E$, where $m$ is the mass of the two-parton system, and $E$
is its energy.  Because of this, the cancellation between the real
contribution and the subtraction term becomes effective only if $m^2/E <
\Gamma$, where $\Gamma$ is the width of the resonance.
As long as $\Gamma$ is above zero, the traditional NLO calculations do
eventually converge, thanks to the fact that in the strict collinear limit
the cancellation takes place. However, convergence becomes more problematic
as the width of the resonance decreases.

The presence of radiation in resonance decays causes even more severe
problems in NLO+PS frameworks.  In \POWHEG{}, radiation is generated
according to the formula
\begin{eqnarray}
  \mathd \sigma & = & \bar{B} (\Phi_{\mathrm{B}}) \,\mathd \Phi_{\mathrm{B}}  \left[
  \Delta (q_{\tmop{cut}}) + \sum_{\alpha} \Delta (k^{\alpha}_{\sss T}) 
  \frac{R_{\alpha} (\Phi_{\alpha} (\Phi_{\mathrm{B}}, \Phi_{\tmop{rad}}))}{B
  (\Phi_{\mathrm{B}})} \,\mathd \Phi_{\tmop{rad}} \right].  \label{eq:powheg}
\end{eqnarray}
The first term in the square bracket corresponds to the probability that no
radiation is generated with hardness above an infrared cutoff
$q_{\tmop{cut}}$, and its kinematics corresponds to the Born one. Each
$\alpha$ in the sum labels a collinear singular region of the real cross
section. The full real matrix element is
decomposed into a sum of terms
\begin{equation}
  R = \sum_{\alpha} R_{\alpha}\,,
\end{equation}
where each $R_{\alpha}$ is singular only in the region labelled by $\alpha$.
The real phase space $\Phi_{\alpha} (\Phi_{\mathrm{B}}, \Phi_{\tmop{rad}})$
depends upon the singular region $\alpha$ and is given as a function of the
Born kinematics $\Phi_{\mathrm{B}}$ and three radiation variables
$\Phi_{\tmop{rad}}$.  The inverse of $\Phi_{\alpha}$
implements the previously mentioned mapping of the real kinematics into an
underlying Born one.  Thus, 
for a given $\Phi_{\mathrm{B}}$ and $\Phi_{\tmop{rad}}$, 
each term in the sum inside the square bracket in~\refeq{eq:powheg}
is associated with a
different real phase-space point. For each $\alpha$, $k^{\alpha}_{\sss T}$ is
defined as the hardness of the collinear splitting characterized by the
kinematics $\Phi_{\alpha} (\Phi_{\mathrm{B}}, \Phi_{\tmop{rad}})$. It 
usually corresponds to the relative transverse momentum of the two collinear
partons.

The Sudakov form factor, $\Delta$, is such that the square bracket in~\refeq{eq:powheg}, after performing the integrals in $\mathd
\Phi_{\tmop{rad}}$, becomes exactly
equal to one~(a property sometimes called {\it unitarity}
of the real radiation).
 In general we have
\begin{eqnarray}
  \Delta (q) & = & \prod_{\alpha} \Delta_{\alpha} (q)\,, 
\end{eqnarray}
with
\begin{eqnarray}
  \Delta_{\alpha} (q) & = & \exp \left[ - \int_{k^{\alpha}_{\sss T} > q}
  \frac{R_{\alpha} (\Phi_{\alpha} (\Phi_{\mathrm{B}}, \Phi_{\tmop{rad}}))}{B
  (\Phi_{\mathrm{B}})}\, \mathd \Phi_{\tmop{rad}} \right] . \label{eq:Sudakov}
\end{eqnarray}
In order to achieve NLO accuracy, the $\bar{B} (\Phi_{\mathrm{B}})$ factor must
equal the NLO inclusive cross section at given underlying Born kinematics,
\begin{eqnarray}
  \bar{B} (\Phi_{\mathrm{B}}) & = & B (\Phi_{\mathrm{B}}) + V (\Phi_{\mathrm{B}}) +
  \sum_{\alpha} \int R_{\alpha} (\Phi_{\alpha} (\Phi_{\mathrm{B}}\,,
  \Phi_{\tmop{rad}})) \,\mathd \Phi_{\tmop{rad}}\,, 
\end{eqnarray}
where both the second and third term on the right hand side are infrared
divergent, but the sum, being an inclusive cross section, is finite. The
cancellation of singularities is achieved with the usual subtraction
techniques.

We are now in a position to discuss the problems that arise in processes with
radiation in decays of resonances. In order to do this, we focus on the $W^- W^+ b
\bar{b}$ production process. As an example of the problem, we consider a 
real emission contribution where a gluon $g$ is
radiated, such that the mass of the $W^+ b g$ and $W^-\bar{b}$ systems are very
close to the top nominal mass. 
We call $\alpha_b$ the singular region
corresponding to $b$ and $g$, and $\alpha_{\bar{b}}$ the region corresponding
to the $\bar{b}$ and $g$ becoming collinear, respectively.
If we consider the
case when the $b$ and $\bar b$ partons are relatively close in direction, as
$g$ becomes collinear to the $b$ or the $\bar{b}$ parton, two components will
dominate the real cross section, $R_{\alpha_b}$ and $R_{\alpha_{\bar{b}}}$,
in a proportion that is determined by how close the gluon is to the $b$ or to
the $\bar{b}$ partons. If the gluon is not much closer to the $b$ region with
respect to the $\bar{b}$ one, the $R_{\alpha_{\bar{b}}}$ contribution will be
comparable or larger than the $R_{\alpha_b}$ one.  We now observe that, for
the same real kinematic configuration, we have two singular regions and two
corresponding underlying-Born configurations.  In the $\alpha_b$ singular
region, the underlying Born is obtained by merging the $b g$ system into a
single $b$, while in the $\alpha_{\bar b}$ region it is the ${\bar b} g$
system that is merged into a single $\bar{b}$. It is therefore clear that, in
the $\alpha_b$ merging, the resonance virtualities are nearly preserved in
the underlying Born, while in the $\alpha_{\bar b}$ one the resonances will
be far off-shell.  The $R_{\alpha_{\bar{b}}} / B$ terms appearing both in~\refeq{eq:powheg} 
and~(\ref{eq:Sudakov}) will become very large, the top
resonances being on-shell in the numerator and off-shell in the
denominator. However, in the \POWHEG{} framework, these ratios
should be either small~(of order $\alpha_s$) or should approach the
Altarelli-Parisi splitting functions for the method to work.

It is thus clear that, if resonances are present, the traditional
decomposition into singular regions must be revised.
In particular, each $\alpha$ should become associated
to a specific resonance structure of the event, such that
 collinear partons originate from the same resonance. Furthermore,
the phase space mapping $\Phi_{\alpha} (\Phi_{\mathrm{B}}, \Phi_{\tmop{rad}})$
should preserve the virtuality of the intermediate resonances.
This is, in brief, what was done in~\citere{Jezo:2015aia}.

The resonance-aware formalism also offers the opportunity to modify and
further improve the \POWHEG{} radiation formula. We make, for the moment, the
assumption that each decaying resonance has only one singular region, and the
radiation not originating from a resonance decay also has only one singular
region. This is the case, for example, for the resonance structure of the
process $g g \rightarrow (t \rightarrow W^+ b) (\bar{t} \rightarrow W^-
\bar{b})$, since in \POWHEG{} the initial-state-radiation~(ISR) regions are
combined into a single one. We consider the formula
\begin{eqnarray}
  \mathd \sigma & = & \bar{B} (\Phi_{\mathrm{B}}) \,\mathd \Phi_{\mathrm{B}} 
  \prod_{\alpha = \alpha_b, \alpha_{\bar{b}}, \alpha_{\rm\tiny ISR}} \left[
  \Delta_{\alpha} (q_{\tmop{cut}}) + \Delta_{\alpha} (k^{\alpha}_{\sss T}) 
  \frac{R_{\alpha} (\Phi_{\alpha} (\Phi_{\mathrm{B}},
  \Phi^{\alpha}_{\tmop{rad}}))}{B (\Phi_{\mathrm{B}})} \,\mathd
  \Phi^{\alpha}_{\tmop{rad}} \right],  \label{eq:allrad}
\end{eqnarray}
where, by writing $\Phi^{\alpha}_{\tmop{rad}}$, we imply that the radiation
variables are now independent for each singular region. By expanding the
product, we see that we get a term with no emissions at all, as in
\refeq{eq:powheg}, plus terms with multiple~(up to three) emissions. It can be shown
that, as far as the hardest radiation is concerned, formula~(\ref{eq:allrad}) is
equivalent to formula~(\ref{eq:powheg}). To this end, one begins by
rewriting~\refeq{eq:allrad} as a sum of three terms, with appropriate
$\theta$ functions such that each term represents the case where the hardest
radiation comes from one of the three regions. It is easy then to integrate
in each term all radiations but the hardest, thus recovering the full Sudakov
form factor appearing in the second term in the square bracket of~\refeq{eq:powheg}.

The \bbfourl{} generator can generate radiation using the improved
multiple-radiation scheme of formula~(\ref{eq:allrad}) or the conventional
single-radiation approach of~\refeq{eq:powheg}.  In events generated with
multiple emissions included, the hardest radiation from all sources
(i.e.~production, $t$ and $\bar{t}$ decays) may be present.  The
\POWHEG{} generated event is then completed by a partonic shower Monte
Carlo program that attaches further radiation to the event.
The interface to the shower must be such that the shower does not generate
radiation in production, in $t$ decay and in $\bar{t}$ decay that is harder
than the one generated by \POWHEG{} in production, $t$ and $\bar{t}$ decay,
respectively.\footnote{
We note that this method guarantees full  NLO accuracy, including exact spin correlations,
only at the level of each individual emission, while 
correlation effects between multiple QCD emissions are handled in approximate form. 
Nevertheless it should be clear that~\refeq{eq:allrad}
represents a significant improvement with respect 
to pure parton showering after the first emission.}

\section{The \RES{} framework}
\label{sec:BOXRES}
In this section we illustrate features that have been added to the \RES{}
package since the publication of~\citere{Jezo:2015aia}, and discuss some
issues that were not fully described there.

\subsection*{Automatic generation of resonance histories}
In the \RES{} implementation of~\citere{Jezo:2015aia}, the initial
subprocesses and the associated resonance structures were set up by hand. We
have now added an algorithm for the automatic generation of all relevant
resonance histories for a given process at a specified perturbative order.
Thanks to this feature, the user only needs to provide a list of
subprocesses, as was the case in the \VTWO{} package. This is a considerable
simplification, in view of the fact that, when electroweak processes are
considered, the number of resonance histories can increase substantially. 
Details of this feature are given in~\refapp{app:resonance_histories}.

\subsection*{Colour assignment}
Events that are passed to a shower generator for subsequent
showering must include colour-flow information in the limit of large
number of colours. In the \VTWO{} framework, colours are assigned with a
probability proportional to the corresponding component of the colour flow
decomposition of the amplitude. 
The extension of this approach to the \RES{} framework requires some care
due to possible inconsistencies between the  
colour assignment and the partitioning 
into resonance histories.
This issue and its systematic solution are discussed in detail in~\refapp{app:colourassignment}.

\subsection*{\POWHEG{}+\OpenLoops interface}
All tree and one-loop amplitudes implemented in the \bbfourl{} generator are based on the
\OpenLoops program ~\cite{OLhepforge} in combinations with \Collier~\cite{Denner:2016kdg}
or \CutTools~\cite{Ossola:2007ax} and \OneLOop \cite{vanHameren:2010cp}.
In the framework of the present work a new
general process-independent interface between the \POWHEGBOX{} and \OpenLoops
has been developed. It allows for a straightforward implementation of 
a multitude of  NLO multi-leg processes matched to parton showers including QCD
and, in the future, also NLO electroweak corrections~\cite{Kallweit:2014xda,
  Kallweit:2015dum}. Technical details and a brief documentation of this new
interface can be found in~\refapp{sec:openloops_interface}.

\section{Description of the generator}
\label{sec:description}
The implementation of combined off-shell \ttbar and \Wt production in the
\RES{} framework presented in this paper is based on all possible Feynman
diagrams contributing to the process \ppllllbbX at NLO accuracy in QCD,
i.e.~up to order $\as^3 \aem^4$.
All  bottom-mass effects have been fully taken into
account and for the consistent treatment of top-, $W$-, and $Z$-resonances at
NLO we rely on the automated implementation of the complex-mass
scheme~\cite{Denner:1999gp, Denner:2005fg} within \OpenLoops.
\subsection{Resonance histories}\label{sec:reshists}
\begin{figure}[tb]
\begin{center}
\includegraphics[width=0.7\textwidth]{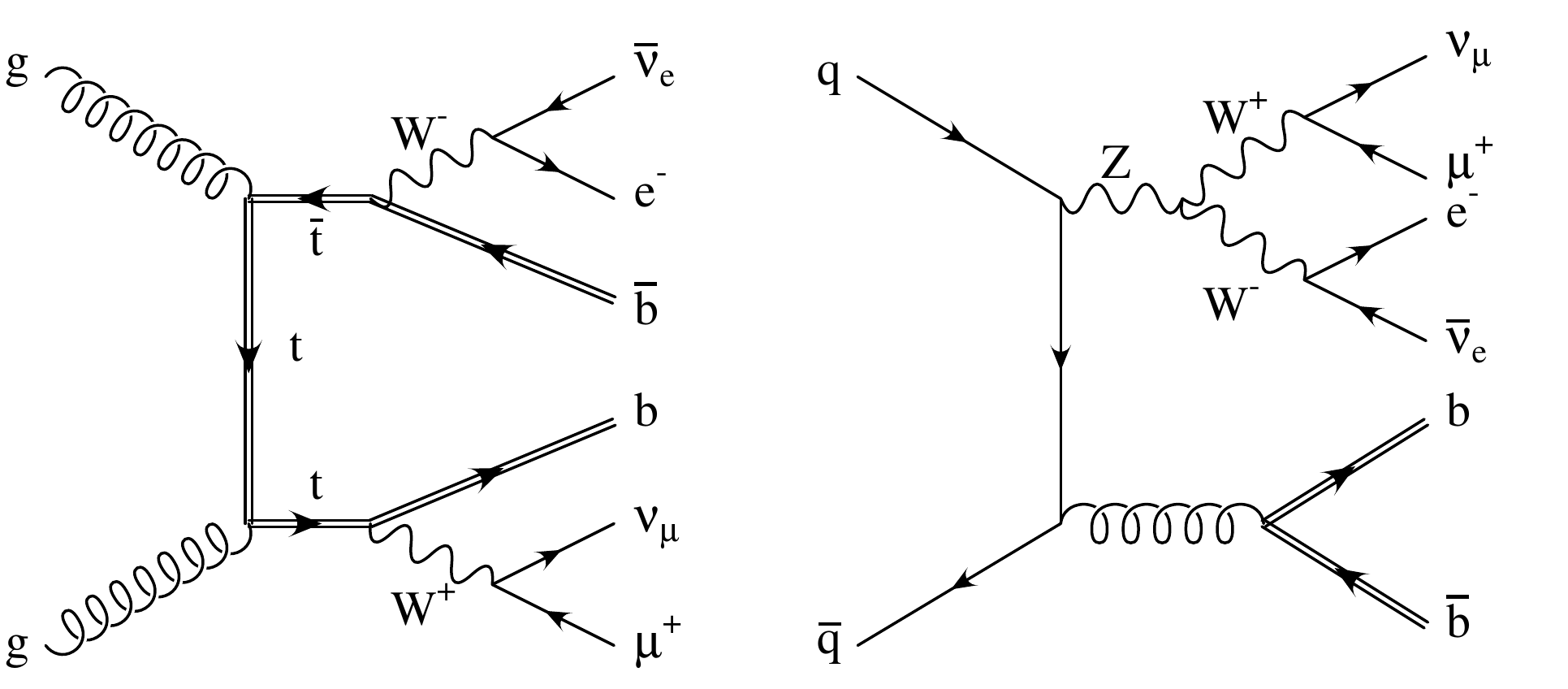}
\end{center}
\caption{Sample Feynman graphs corresponding to the two resonance histories
  relevant for $pp \to \mu^+ \nu_\mu e^- \bar{\nu}_e \,\bbbar$ production.}
\label{fig:born_res_hist} 
\end{figure}

The Born level resonance structure for \ppllllbb{} at $O(\as^2 \aem^4)$ 
is actually very simple. Indeed, it is sufficient to consider
two kinds of resonance histories. In~\reffi{fig:born_res_hist}
we show two corresponding Feynman diagrams for the process $pp \to \mu^+
\nu_\mu e^- \bar{\nu}_e \,\bbbar$ .

Internally, according to the \RES{} conventions~\cite{Jezo:2015aia},
the resonance histories are described by the arrays
\begin{verbatim}
  flav_1    = [i,  j,  6, -6,  24, -24, -13,  14,  11, -12,  5, -5],
  flavres_1 = [0,  0,  0,  0,   3,   4,   5,   5,   6,   6,  3,  4],
  flav_2    = [i,  j, 23, 24, -24, -13,  14,  11, -12,   5, -5],
  flavres_2 = [0,  0,  0,  3,   3,   4,   4,   5,   5,   0,  0],
\end{verbatim}
for all 
relevant
choices of initial parton flavours \verb!i,j!. In
\verb!flav! we store the identities of the initial- and final-state
particles, with intermediate resonances, if they exist, labelled according to
the Monte Carlo numbering scheme~(gluons are labeled by zero in the
\POWHEGBOX). In \verb!flavres!, for each particle, we give the position of
the resonance from which it originates. For partons associated with the production subprocess
\verb!flavres! is set to zero.

\begin{figure}[tb]
\begin{center}
\includegraphics[width=0.35\textwidth]{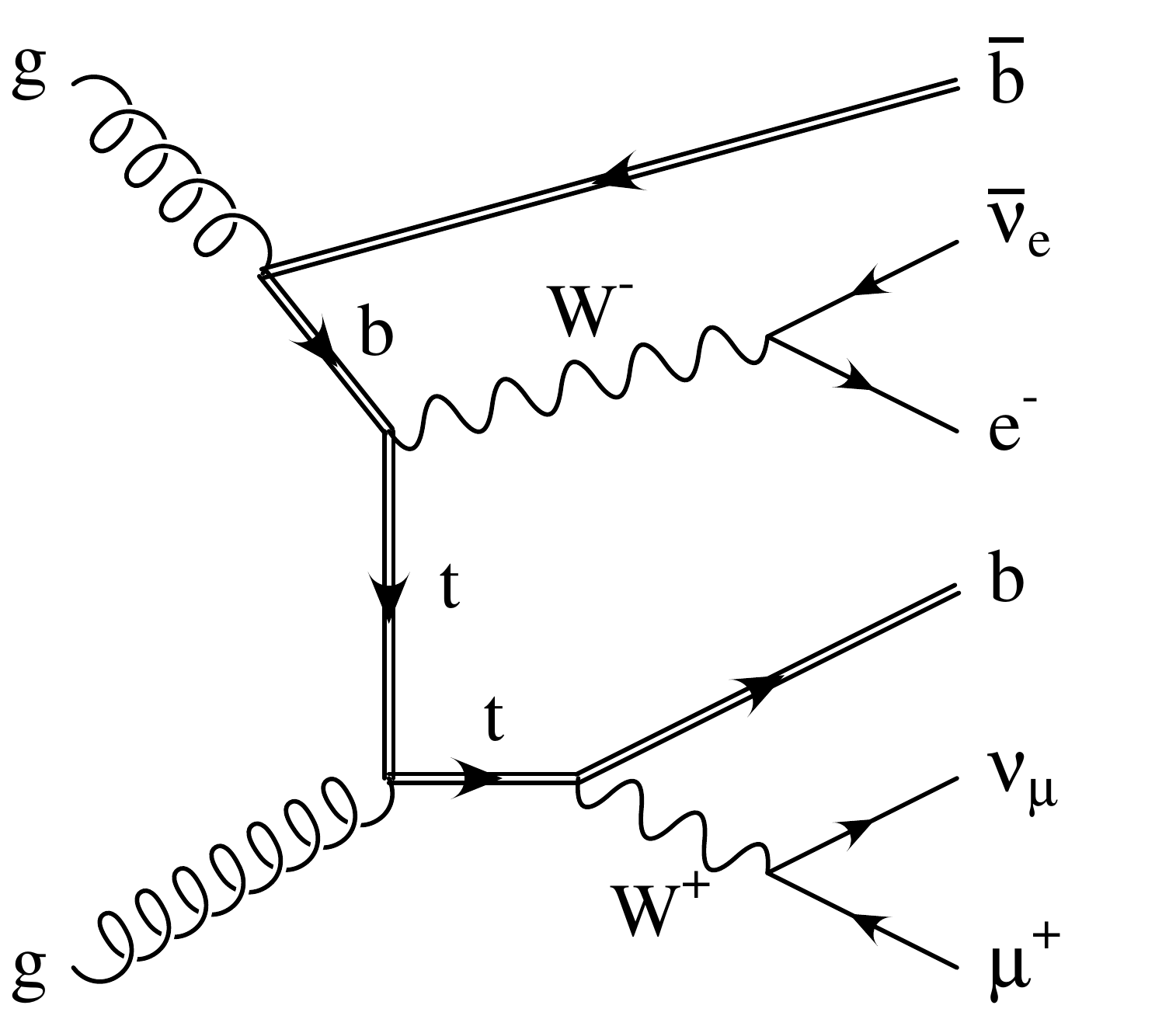}
\end{center}
\caption{Representative Born diagram for $Wt$ production.}
\label{fig:born_Wt} 
\end{figure}

The resonance structures that differ only by the external parton flavours are
collected into resonance groups, so that, in the present case, we have only
two resonance groups.
We remark that there is no need of a unique correspondence  between resonance structures and possible
combinations of resonant propagators in individual Feynman diagrams.
What is required is that all resonances present in any given Feynman graph are also
present in an associated resonance structure, but not vice versa.
For example, in the present implementation of the \bbfourl{} generator 
the consistent treatment of single-top topologies like the one in~\reffi{fig:born_Wt} 
is guaranteed through resonance histories of $\ttbar$ type
(\verb!flav_1,flavres_1!),
which involve an additional $\bar{t}\to {\bar b} W^-$ resonance.
This does not lead to any problems, since the corresponding
subtraction kinematics, which preserves the mass of the ${\bar b} W^-$ system,
is perfectly adequate also for single-top topologies.

The \RES{} code automatically recognizes resonance histories that can be
collected into the same resonance group. It also includes a subroutine for
the automatic generation of an 
adequate phase-space sampling for each resonance group.
In this context, rather than relying upon standard Breit-Wigner
sampling,  care is taken that also the off-shell regions are 
adequately populated.
This is essential in resonance histories of the kind
shown in the right graph of~\reffi{fig:born_res_hist}, where the
generation of the $W$ virtualities according to their Breit-Wigner shape
would well probe the region where an off-shell $Z$ decays into two on-shell
$W$'s, but not the regions where an on-shell $Z$ decays into an on-shell $W$
and an off-shell one.  It also guarantees that cases like the diagram in~\reffi{fig:born_Wt} 
are properly sampled.
The interested reader can find more technical details by inspecting the code
itself.

\subsection{The complex-mass scheme}
\label{sec:cms}
In our calculation all intermediate massive particles are consistently
treated in the complex-mass scheme~\cite{Denner:1999gp, Denner:2005fg},
where the widths of unstable particles are absorbed into 
the imaginary part of the corresponding mass parameters,
\begin{equation}
\label{eq:complexmasses1}
\mu^2_i=M_i^2-\ri\Gamma_iM_i \qquad\mbox{for}\ \ i=\PW,\PZ,\Pt,\PH.
\end{equation}
This choice implies a complex-valued weak mixing angle,
\begin{equation}
\label{eq:defsintheta1}
\sin\theta_W^2=1-\cos\theta_W^2=1-\frac{\mu_\PW^2}{\mu_\PZ^2}\,,
\end{equation}
and guarantees gauge invariance at NLO~\cite{Denner:2005fg}. 

\subsection{The decoupling and \MSB{} schemes}
\label{sect:decoupling}
When performing a fixed-order calculation with massive quarks, one can define
two consistent renormalization schemes that describe the same physics: the
usual \MSB{} scheme, where all flavours are treated on equal footing, and a
mixed scheme~\cite{Collins:1978wz}, that we call decoupling scheme, in
which the $\nlf$ light flavours are subtracted in the \MSB{} scheme, while
heavy-flavour loops are subtracted at zero momentum. In this scheme, heavy
flavours decouple at low energies.

In the calculation of the $\fourl\,\bbbar$ 
hard scattering cross section we treat the bottom quark
as massive and, correspondingly, $\nlf$ is equal to four.
The renormalization
of the virtual contributions is performed in the decoupling scheme with a
four-flavour running $\as$. 
For consistency, the evolution of parton distribution
functions~(PDFs) should 
be performed with four active flavours, so
that, in particular, no bottom-quark density is present and no bottom-quark
initiated processes have to be considered.
However, given that the process at hand is characterised by typical
scales far above the $b$-quark threshold, it is more convenient 
to convert our results to the $\MSB$ scheme
in such a way that they can be expressed in terms of the $\MSB$
strong coupling constant, running with five active flavours, and also with five-flavour PDFs.

The procedure for such a switch of schemes is well known, and was discussed
in~\citere{Cacciari:1998it}. 
For $\fourl \,\bbbar$ production, we need to transform 
the $q\bar q$ and $gg$ squared Born amplitudes ${\cal B}_{qq}$ and ${\cal B}_{gg}$,
computed in the decoupling scheme, in the following way
\begin{eqnarray}
{\cal B}_{qq}&\rightarrow& \left[1
- \frac{4}{3}\,\TF \,\frac{\as}{2\pi}\, \log \(\frac{\mur^2}{\mb^2}\)\right] \, {\cal B}_{qq}\,,\\
{\cal B}_{gg}&\rightarrow& 
\left[1+\frac{4}{3}\,\TF \,\frac{\as}{2\pi}\, \log \(\frac{\muf^2}{\mur^2}\)\right] \, {\cal
  B}_{gg}\,.
\end{eqnarray}
where 
$\mur$ and $\muf$ are the renormalization
and factorization scales, respectively, and $\mb$ is the bottom-quark
mass. The contribution of the $b$ parton densities, that are present in the
five-flavour scheme, should not be included in this context.

\subsection{The virtual corrections}

The virtual contributions have been generated using the new interface of the
\POWHEGBOX{} with the \OpenLoops amplitude generator, as described in~\refapp{sec:openloops_interface}. 
While \OpenLoops guarantees a very fast evaluation of
one-loop matrix elements, the overall efficiency of the 
generator can be significantly improved by minimising the
number of phase space points that require the calculation
of virtual contributions.
As detailed in \refapp{app:virtuals}, 
this is is achieved by evaluating the virtual and real-emission contributions with 
independent statistical accuracies optimised according to the respective relative weights.
Moreover, when generating events, a reweighting method can be used in order
to restrict virtual evaluations to the small fraction of phase space points that 
survive the unweighting procedure.

\subsection{Interface to the shower}
The generator presented in this work shares many common features with the one
of~\citere{Campbell:2014kua}. In particular, in both generators, Les
Houches events include resonance information, and an option for a multiple
radiation scheme is implemented, denoted as \verb!allrad! scheme, according to the
corresponding {\tt powheg.input} flag.  As
explained in~\citere{Campbell:2014kua} and reviewed in~\refse{sec:reminder}, 
when this scheme is activated, the mechanism of
radiation generation is modified. Rather than keeping only the hardest
radiation arising from all singular regions, the program stores several
``hardest radiations'': one that takes place at the production stage, and one
for the decay of each resonance that can radiate. All these radiations are
assembled into a single Les Houches event.  Thus, for example, in events with
the $t$ and $\bar{t}$ resonances, one can have up to three radiated partons:
one coming from the initial-state particles, one arising from the $b$ in the 
$t$-decay, and one from the $\bar{b}$ in the $\bar{t}$-decay.

When generating fully showered events, the hardness\footnote{Here and in the
  following by hardness we mean the relative transverse momentum of two
  partons arising from a splitting process, either in initial- or in
  final-state radiation.} of the shower must be limited in a way that depends
upon the origin of the radiating parton. If the radiating parton is not son
of a resonance, the hardness of the shower arising from it must be limited by
the hardness of the Les Houches radiation that arises in
production.\footnote{%
  By radiation in production we mean any radiation that
  does not arise from a decaying resonance. This can be initial-state
  radiation, but also radiation from final-state partons, as in the right
  diagram in~\reffi{fig:born_res_hist} and the one in~\reffi{fig:born_Wt},
 where the $b$'s do not arise from a decaying
  resonance.}
Radiation arising from partons originating from a resonance must have their
hardness limited by the hardness of the parton radiated from the resonance in
the Les Houches event.  This requires a shower interface that goes beyond the
Les Houches approach. In~\citere{Campbell:2014kua} a suitable procedure
has been conceived and implemented in \PythiaEight{}~\cite{Sjostrand:2007gs,
  Sjostrand:2014zea}. The interested reader can find all details in the
Appendix~A of ~\citere{Campbell:2014kua}. In essence, the procedure was to
examine the showered event, compute the transverse momentum of \PythiaEight{}
radiation in top decays, and veto it if higher than the corresponding
\POWHEG{} one.  Vetoing is performed by rejecting the showered event, and
generating a new \PythiaEight{} shower, initiated by the same Les Houches
event. This procedure was iterated until the showered event passes the veto.
In the present work, we have adopted this procedure in order to 
make a more meaningful comparison with the results of~\citere{Campbell:2014kua}.
However, we have
also verified that, by using \PythiaEight{} internal mechanism for vetoing
radiation from resonance decay, we get results that are fully compatible with
our default approach.\footnote{An interface to
  \Herwigseven{}~\cite{Bellm:2015jjp} is now under development.} This 
aspect and the comparison among the two methods are shown in~\refapp{sec:veto}.

\subsection{Traditional NLO+PS matching}
It is possible to run our new generator in a way that is fully equivalent to a
standard \POWHEG{} matching algorithm~(as implemented in the \VTWO{})
ignoring the resonance structure of the processes.  This is
achieved by including the line \verb!nores 1! in the \verb!powheg.input!
  file.\footnote{In this mode, our generator becomes similar to the
    implementation~\citere{Garzelli:2014dka}, except for our use of the four
    flavours scheme.}  Such an option is implemented only for the purpose of
  testing the new formalism with respect to the old one.

It turns out that, in the \verb!nores 1! mode, the program has much worse
convergence properties, most likely because of the less effective cancellation
of infrared singularities mentioned in~\refse{sec:reminder}.  We
find, for example, that in runs with equal statistics~(with about 15 million
calls) the absolute error in the \verb!nores 1! case is roughly seven times
larger than in the \verb!nores 0!~(default) case.  The generation of
events also slows down by a similar factor.

We stress again that, in the limit of small widths, 
the NLO+PS results
obtained in the
\verb!nores 1! 
mode are bound to become inconsistent, as
discussed in~\refse{sec:reminder} and, more extensively, in~\citere{Jezo:2015aia}.

\subsection{Consistency checks}\label{sec:Checks}
At the level of fixed-order NLO calculations, the traditional machinery of the
\POWHEGBOX{} is well tested and we trust corresponding results to be correct.
On the other hand, the NLO subtraction procedure implemented in the \RES{}
code is substantially different and still relatively new. As was done 
in~\citere{Jezo:2015aia} for $t$-channel single-top production, also for the
$\fourl\,\bbbar$ production presented here, we systematically validated the
fixed-order NLO results obtained with the \RES{} implementation by switching
on and off the generation of resonance structures. We found perfect agreement
between the two calculations.

Additionally, we performed a detailed comparison against the fixed-order NLO
results of~\citere{Cascioli:2013wga} and found agreement at the permil
level. Furthermore, via a numerical scan in the limit of the top width going
to zero, $\Gamma_t \to 0$, we verified that any
$\as\log\left(\Gamma_t\right)$ enhanced terms in the soft-gluon limit
successfully cancel between real and virtual contributions.  This last test
was performed for various light- and $b$-jet exclusive distributions which
are subject to sizable non-resonant/off-shell corrections.

\section{Phenomenological setup}
\label{sec:setup}

In this section we document the input parameters, acceptance cuts and 
generator settings that have been adopted for 
the numerical studies presented in~\refse{sec:ttbarPhenomenology}.
Moreover we introduce a systematic labelling scheme
for the various NLO+PS approximations that are going to be compared.

\subsection{Input parameters}
Masses and widths are assigned the following values
\begin{align}
  m_{Z} &= 91.188 \;\GeV\,,  &  \Gamma_{Z} &=  2.441 \;\GeV\,,\\
  m_{W} &= 80.419 \;\GeV\,,  &  \Gamma_{W} &=  2.048 \;\GeV\,, \\
  m_{H} &= 125 \;\GeV\,,  &  \Gamma_{H} &=  4.03\times10^{-3}\;\GeV\,, \\
  m_{t} &= 172.5 \;\GeV\,,   &  \Gamma_{t} &=  1.329 \;\GeV\,, \\
    m_{b} &= 4.75 \;\GeV\,.  &  
\end{align}
The electroweak couplings are derived from the gauge-boson masses and the
Fermi constant, $\GF=1.16585\times10^{-5}~\GeV^{-2}$, in the
$G_{\mu}$-scheme, via
\begin{equation}
\aem=\sqrt{2}\, \frac{G_\mu}{\pi} \left|\mu_{\sss W}^2\(1-\frac{\mu_{\sss
    W}^2}{\mu_{\sss Z}^2}\)\right|=\frac{1}{132.50698}\,, 
\end{equation}
where $\mu_{\sss W}$ and $\mu_{\sss Z}$ are complex masses given by~\refeq{eq:complexmasses1}.

The value of the
top-quark width we use is consistently calculated at NLO from all other input
parameters by computing the three-body decay widths $\Gamma(t\to f \bar f' b)$
into any pair of light fermions $f$ and $\bar f'$ and a massive $b$ quark. To this
end, we employ a numerical routine of the \MCFM implementation of~\citere{Campbell:2012uf}.

As parton distributions
we have adopted the five-flavour MSTW2008NLO PDFs~\cite{Martin:2009iq}, as
implemented in the~\citere{Buckley:2014ana},
with the corresponding five-flavour strong
coupling constant, and for their consistent combination with
four-flavour scheme parton-level cross sections 
the scheme transformation of \refse{sect:decoupling} was applied.
In the evaluation of the matrix elements, only the bottom
and the top quarks are massive. All the other quarks are treated as
massless. In addition, the Cabibbo-Kobayashi-Maskawa matrix is assumed to be
diagonal.

When generating events we adopt the following scale choice:
\begin{itemize}
  \item For resonance histories with a top pair we use
\begin{equation}
\label{eq:ttscale}
  \mur=\muf=\lq \(m_t^2+p_{{T},t}^2\)\(m_{\bar t}^2+p_{{T},{\bar t}}^2\)\rq ^{\frac{1}{4}}\;,
\end{equation}
where the (anti)top masses and transverse momenta are defined in the underlying Born phase space 
in terms of final state (off-shell) decay products.

  \item For resonance histories with an intermediate $Z$ we use
\begin{equation}
\label{eq:zscale}
  \mur=\muf=\frac{\sqrt{p_{Z}^2}}{2}\;,
\end{equation}
where $p_Z=p_{\ell^+}+p_{\nu_\ell}+p_{l^-}+p_{\bar\nu_l}$.
\end{itemize}
In addition, we set the value of the \POWHEGBOX{} parameter \verb!hdamp! to
the mass of the top quark. This setting yields a transverse-momentum
distribution of the top pair that is more sensitive to scale variations and
more consistent with data at large transverse momenta. It only affects
initial-state radiation. For a detailed description of this parameter, we
refer the reader to~\citere{Alioli:2008tz}.

\subsection{\PythiaEight{} settings}
We interface our \POWHEG{} generator to 
\PythiaEightPone{},\footnote{An interface to \PythiaEightPtwo{} is also
  available, but was not used for the present work.}
as illustrated in
Appendix~A of~\citere{Campbell:2014kua}, and so we perform the following
\PythiaEight{} calls:
\begin{verbatim}
    pythia.readString("SpaceShower:pTmaxMatch = 1");
    pythia.readString("TimeShower:pTmaxMatch = 1");
    pythia.readString("PartonLevel:MPI = off");
    pythia.readString("SpaceShower:QEDshowerByQ = off");       
    pythia.readString("SpaceShower:QEDshowerByL = off");     
    pythia.readString("TimeShower:QEDshowerByQ = off");        
    pythia.readString("TimeShower:QEDshowerByL = off");
\end{verbatim}
The first two calls are required when interfacing \PythiaEight{} to NLO+PS
generators.  The third call switches off multi-parton interactions and it is
only invoked for performance reasons: in fact, the shower of the events is
faster when multi-parton interactions are not simulated.
The remaining calls switch off the electromagnetic radiation in
\PythiaEight{}. This makes it easier to reconstruct the $W$ boson momentum,
since we do not need to dress the charged lepton, from vector boson decay,
with electromagnetic radiation. These settings are appropriate in the
present context since we do not make any comparison with data.

\PythiaEight{} provides by default matrix-element corrections~\cite{Norrbin:2000uu}.
In our case, they are relevant for radiation in the top decays, which are corrected
using $t\to Wb g$ tree level 
matrix elements. These corrections are also applied in subsequent
emissions in order to better model radiation from heavy flavours in general.
If not explicitly stated otherwise, we include the following setup calls
\begin{verbatim}
    pythia.readString("TimeShower:MEcorrections = on");
    pythia.readString("TimeShower:MEafterFirst = on");
\end{verbatim}
These corrections never modify the Les Houches event weight. They only affect
the radiation generated by the shower. Thus, leaving these flags on does not
lead to over-counting. If the second flag is off, matrix-element corrections
are applied only to the first shower emission. If it is on, they are also
applied to subsequent radiation.  In fact, even if these corrections cannot
fully account for the structure of the matrix elements, they at least better
account for mass effects arising in radiation from the off-shell top quarks
and from the massive final-state $b$'s.

In our analysis, we keep $B$ hadrons stable, performing the corresponding
\PythiaEight{} setup calls. Aside from these, all remaining settings are left
to the defaults of \PythiaEightPone{}.

\subsection{Generators and labels}
In~\refse{sec:ttbarPhenomenology} we compare three different generators that
implement an increasingly precise treatment of $\ttbar$
production and decay:
\begin{itemize}
\item the \hvq{} generator of~\citere{Frixione:2007nw};
\item the \DEC{} generator of~\citere{Campbell:2014kua};
\item the new \bbfourl{} generator, which we consider as our best prediction.
\end{itemize}
\begin{table}%
\begin{center}
\begin{tabular}{l|ccc}
label		 &   \TTBAR                  & \TTBARDEC               & \BBFLRES      \\ \hline\hline
generator	 &   \hvq~\cite{Frixione:2007nw}                  & \dec~\cite{Campbell:2014kua}	             & \bbfourl{} \\
framework	 & \POWHEGBOX{}            & \VTWO{}                 & \RES{}     \\
NLO matrix elements   & $ t\bar t$  & $ t(\to \ell^+\nu_{\sss \ell} b)\,\bar t(\to l^-\bar{\nu}_{\sss l} \bar b)$  & $\fourl\,\bbbar$ \\
decay accuracy 	 & LO+PS                   & NLO+PS                  & NLO+PS     \\
NLO radiation     & single            & multiple    & multiple \\ 
spin correlations & approx.                & exact                   & exact      \\
off-shell $\ttbar$  effects   & BW  smearing & LO $ b\bar b 4\ell$ reweighting   & exact  \\ 
\Wt \& non-resonant effects & no          & LO $ b\bar b 4\ell$ reweighting   & exact  \\ 
$ b$-quark massive          &   yes         &  yes                & yes   \\
  \hline 
  \end{tabular} 
\end{center}
\caption{Labels and characteristic features of the three generators considered in this paper.}
\label{tab:generators}
\end{table}
The main physics features of the various generators 
and the labels that will be used to identify the corresponding predictions 
are listed in~\refta{tab:generators}.
All generators are run with their default settings and 
are interfaced to
\PythiaEightPone{}.
The \bbfourl{} generator implements
the scale choice 
of~\refeqs{eq:ttscale}{eq:zscale}, while in \DEC{} and \hvq{} a scale 
corresponding to~\refeq{eq:ttscale} is used.

In order to quantify the impact of various aspects of the 
resonance-aware approach, in~\refse{sec:ttbarPhenomenology} we will 
compare various settings of the \bbfourl{} generator where
some resonance-aware improvements are turned on and off or are replaced by 
certain approximations. Specifically, the following settings will be considered:
\begin{itemize}
\item[(a)] the resonance-aware formalism is switched on
with default settings;
\item[(b)] the resonance-aware formalism
is switched off,
which corresponds to using the traditional \POWHEG{} approach;
\item[(c)] the resonance-aware formalism is switched off, but a resonance
  assignment is guessed based on the kinematic structure of the events, 
  according to the method described in~\refapp{app:kinematic_guess};
\item[(d)] the resonance-aware formalism is switched on, but, 
instead of applying the multiple-radiation scheme of~\refeq{eq:allrad},
only a single radiation is generated with \POWHEG{} according to~\refeq{eq:powheg};
 \item[(e)] same as (d), but the resonance 
 information is stripped off in the \POWHEG{} Les Houches event file 
 before passing it to the showering program.  
\end{itemize}
The various \bbfourl{} settings and corresponding labels are summarised in~\refta{tab:bb4loptions}.
\begin{table}%
\begin{center}
  \begin{tabular}{lc||cccc}
&\BBFLRES{}  setting & resonance-aware        & radiation in &  flags in the \\     
  \phantom{xx}&  label  &matching  & production and decay    & {\tt powheg.input}\\ \hline

(a)&{\bbfourlPYdefault}  &   yes                    & multiple  & 1, 0, 0, 0 \\
(b)&{\noresPY}          &  no                       & single     & 0, 0, 0, 1 \\
(c)&{\noresiPY}        &  no (kinematic guess)          & single      &  0, 0, 1, 1 \\
(d)&{\bbfourlPYnoallrad}       &  yes               & single     &  0, 0, 0, 0 \\
(e)&{\stripresPY}        &  yes (stripped off)           & single     &  0, 1, 0, 0\\
\hline
\end{tabular} 
\end{center}
\caption{Labels for the various \BBFLRES{} predictions that are considered
  and compared in this paper.  In the last column we list the values of the
  \POWHEGBOX{} flags {\tt allrad}, {\tt stripres}, {\tt guessres}, {\tt
    nores}, to be specified in the {\tt powheg.input} file.}
\label{tab:bb4loptions}
\end{table}

\subsection{Physics objects}
In the subsequent sections we study various observables defined in terms of the following physics objects.
\begin{enumerate}[(a)]
\item \label{item:Bhad} We denote as $B$ and $\bar{B}$ hadron the hardest
  $b$-flavoured and $\bar b$-flavoured hadron in the event.
\item Final-state hadrons are recombined into jets using the 
{\tt FastJet} implementation~\cite{Cacciari:2011ma} of the anti-$\kt$ jet
 algorithm~\cite{Cacciari:2008gp}
with $R=0.5$.
\item We denote as $b$-jet~($\bj$) and
  anti-$b$-jet~($\bbj$) the jet that contains the hardest $B$ and $\bar{B}$
  hadron, respectively. 
  When examining results
  obtained with the hadronization switched off, jets are $b$-tagged based on
  $b$ quarks rather than $B$ hadrons.

\item Leptons, neutrinos and missing transverse energy are identical to their
  corresponding objects at matrix-element level, since we switched off QED
  radiation and hadron decays in \PythiaEight{}.

\item Reconstructed $W^+$ and $W^-$ bosons are identified with the
  corresponding off-shell lepton-neutrino pairs in the hard matrix
  elements.\footnote{Similarly as for top resonances, also $W$ resonances are
    identified with their off-shell decay products according to the resonance
    history of the event at hand. This information is written in the shower
    record and propagated through the shower evolution. In this way, possible
    QED radiation off charged leptons is included into the $W$-boson momentum
    at Monte Carlo truth level.  However, since electromagnetic radiation
    from \PythiaEight{} is turned off in our analysis, each $W$ boson
    coincides with a bare lepton--neutrino system.}

\item Reconstructed top and anti-top quarks are defined as off-shell $W^+\bj$
  and $W^-\bbj$ pairs, respectively, i.e.~$b$-jets and $W$-bosons are matched
  based on charge and $b$-flavour information at Monte-Carlo truth level. The
  same approach is used for $\ell^+\bj$ and $\l^-\bbj$ pairs.
\end{enumerate}
Unless stated otherwise, in kinematic distributions we always
perform an average over the $t$ and $\bar{t}$ case~(thus also on
lepton--antilepton, $b$--anti-$b$, etc.).

The top-pair observables in \refses{sec:ttNLOdec}{sec:RES_HVQ}
are computed by requiring the presence of a $b$ and a $\bar b$ jet with
\begin{eqnarray}
\label{eq:jet_cuts}
 \pt^{j} > 30~\GeV, \qquad \qquad |\eta^{j}|<2.5\,,
\end{eqnarray}
and applying the following leptonic cuts,
\begin{eqnarray}
\label{eq:leptonic_cuts}
\pt^{l} > 20~\GeV, \qquad |\eta^{l}|<2.5\,,  \qquad \pt^{\sss \mathrm{miss}} > 20~\GeV,
\end{eqnarray}
where $l=\ell^+, l^-$ and $\pt^{\sss\mathrm{miss}}$ is obtained from
the vector sum of the transverse momentum of the neutrinos in the final
state.

\section{Top-pair dominated observables}\label{sec:ttbarPhenomenology}

Here we present numerical predictions for 
$pp\to e^+\nu_e\mu^-\bar\nu_\mu\bbbar+X$ at $\sqrt{s}=$8\,TeV.
In particular, we study various observables that are sensitive
to the shape of top resonances.

\subsection{Comparison with traditional NLO+PS matching}
\label{sec:nores}

In the following, we compare nominal \bbfourl{} predictions, 
generated with default settings, with results
obtained by switching off the resonance-aware formalism~(i.e.~setting the
flag \verb!nores! to 1). In this way we get results that are fully equivalent
to a \VTWO{}~(or ``traditional'') implementation.
For this comparison we do not impose any cuts,
i.e.~we perform a fully inclusive analysis that involves, besides $t\bar t$ production, also
significant contributions from $\Wt$ single-top production.

Events generated with the traditional implementation do not contain any
information whatsoever about their resonance structures. We label the curves
obtained by showering these events as \nores{}. Because the resonance
information is not available, the shower generator will not preserve the
virtualities of the resonances.  
In order to further explore the usability of
the \nores{} results,  we also consider the possibility of reconstructing the
resonance information of the Les Houches event on the basis of its kinematic
proximity to one of the possible resonant configurations.
Specifically, we perform an educated guess of the resonance structure of the event, assigning
it to a $t{\bar t}$ or to a $Z$ resonance configuration~(see~\refse{sec:reshists}), 
and assigning the radiation either to the initial
state or to the outgoing $b$'s.
The curves obtained this way are labelled as \noresi{} and the procedure for
reconstructing the resonance information from the event kinematics is
detailed in~\refapp{app:kinematic_guess}.

\begin{figure}[htb]
\begin{center}
  \includegraphics[width=0.49\textwidth]{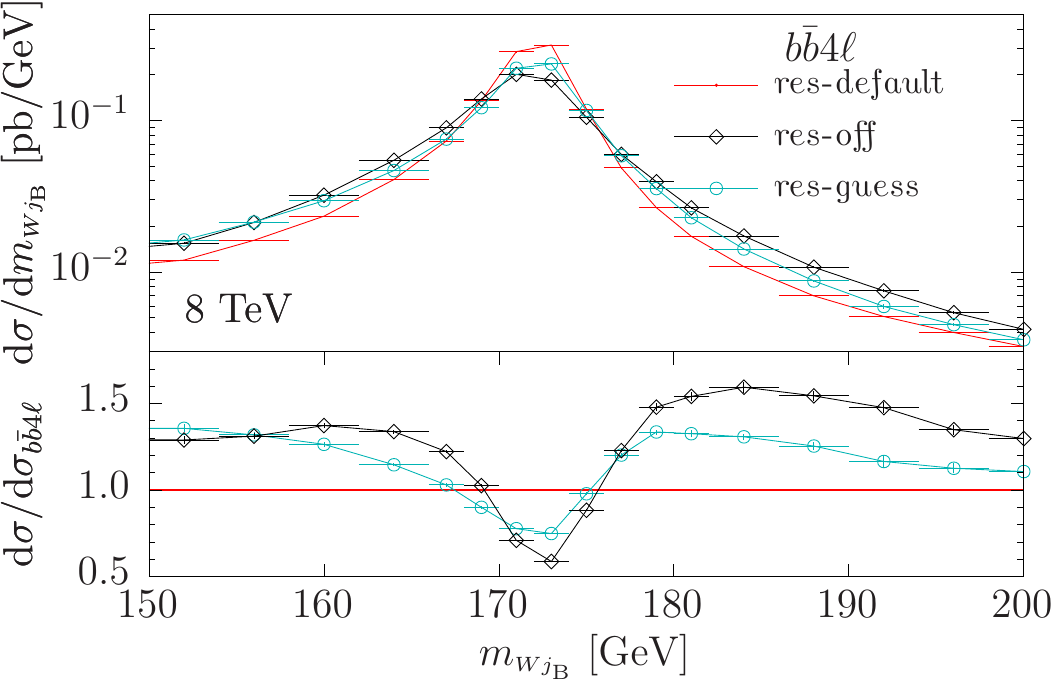}
  \includegraphics[width=0.49\textwidth]{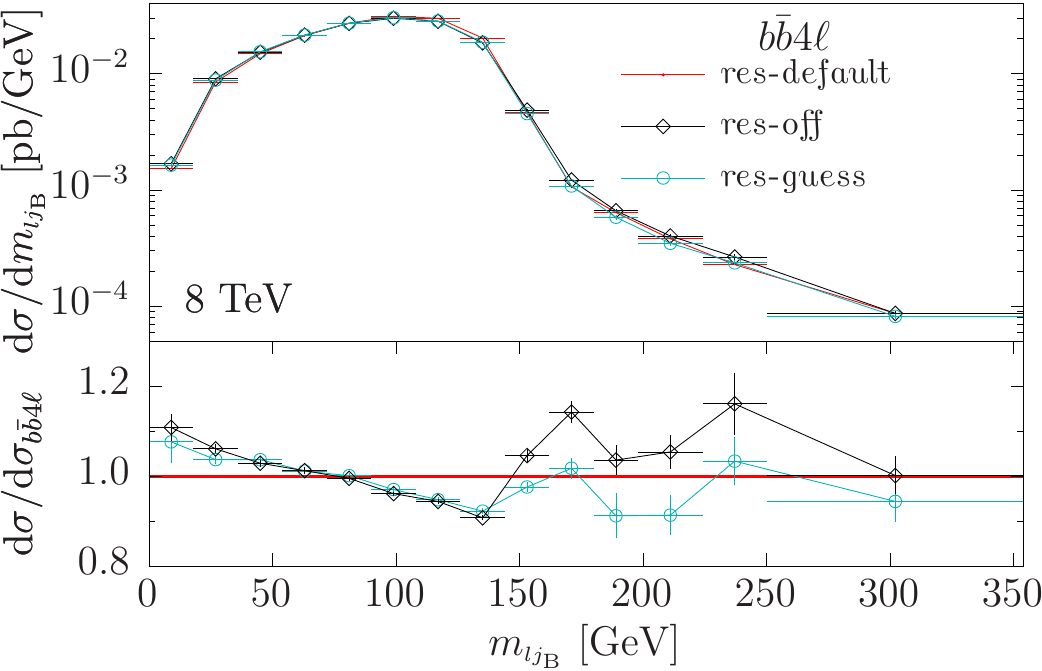}
\end{center}
\caption{NLO+PS predictions for the invariant mass of the $W\bj$~(left) and of the $l\bj$~(right)
  systems obtained with the new \bbfourl generator. We compare our default resonance-aware predictions
  (\resdefault) against the ``traditional'', i.e. resonance-unaware, implementation (\nores{}) and
  a prediction where the event-by-event resonance information is obtained from a guess based on kinematics.
  In the ratio plot we illustrate relative deviations with respect to \resdefault.}
\label{fig:m_w_jbot-res-nores} 
\end{figure}

We first consider, in~\reffi{fig:m_w_jbot-res-nores}, 
the invariant mass of the $W\bj$ and of the $l\bj$ systems.
In the \nores{} case, we observe that
the reconstructed mass peak has a wider shape. This is 
expected, since neither the \POWHEGBOX{} nor the shower program preserve the
virtuality of the top resonances. In the \noresi{} case 
the width of the peak is diminished, although not quite at the level of 
the resonance-aware prediction, labelled as \resdefault{}. We also 
observe a mild shift in the peak in the \noresi{} case, 
which improves the agreement with the \default{} result.
The distribution in the mass of the lepton-\bj{} system
also shows marked differences in shape in the region that is most
relevant for a top-mass determination, with more pronounced differences in
the \nores{} case.

\begin{figure}[htb]
\begin{center}
  \includegraphics[width=0.49\textwidth]{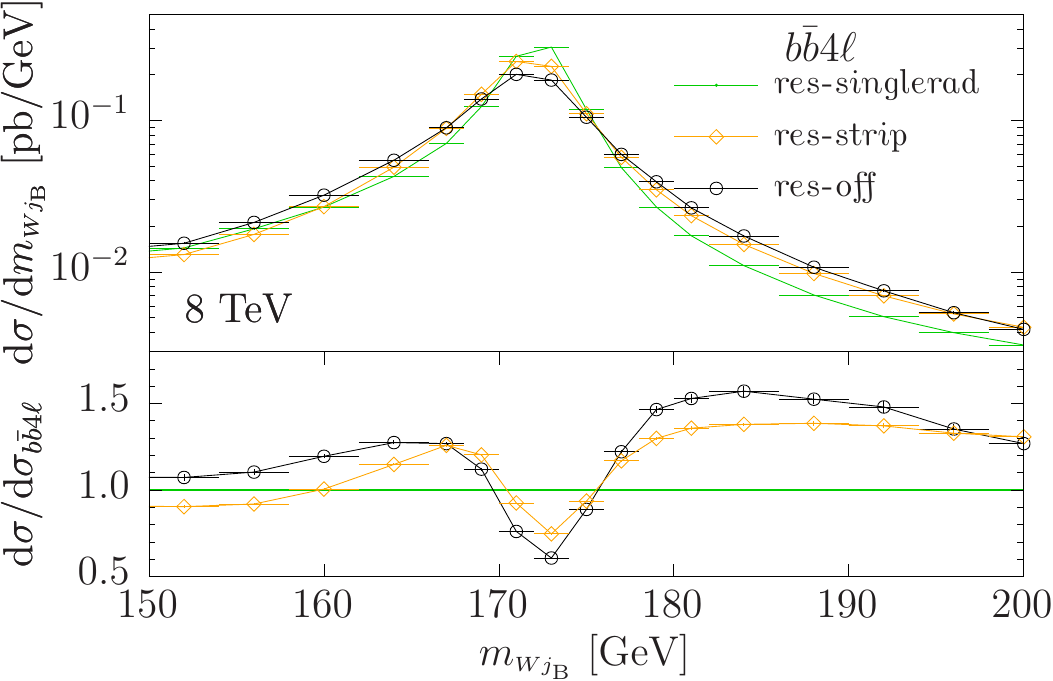}
\end{center}
\caption{Invariant mass of the $W\bj$ system obtained with the \bbfourl generator.
 We compare our resonance-aware predictions without employing the multiple radiation scheme (\bbfourlNoallrad{})
 against the ``traditional'', i.e. resonance-unaware, implementation (\nores{}) and
  a prediction where any resonance information is stripped off the Les Houches event file (\stripres{}).
  In the ratio plot we illustrate relative deviations with respect to \bbfourlNoallrad{}.}
\label{fig:m_w_jbot-res-nores-more} 
\end{figure}

The above findings suggest that the width of the peak is determined both by the shower
generator being aware of the resonances in the Les Houches event, and by the
hardest radiation generation being performed in a way that is consistent with
the resonance structure. 
In order to assess the effects 
that originate solely from resonance-aware matching and showering
in a more accurate way, in~\reffi{fig:m_w_jbot-res-nores-more}
we disable the multiple radiation scheme 
of~\refeq{eq:allrad} (by setting {\tt allrad 0})
and compare the resulting resonance-aware predictions (\bbfourlNoallrad{})
against the cases where resonance information
is removed from the Les Houches event before
showering~(\stripres{}) or the case where the resonance-aware system is 
completely switched off (\nores{}).
We find that the \stripres{} result lies between the \bbfourlNoallrad{} 
and the \nores{} ones, somewhat
closer to the latter, and the differences between 
the various predictions are considerable.
Therefore, we conclude that the observed widening of the 
peak in~\reffis{fig:m_w_jbot-res-nores}{fig:m_w_jbot-res-nores-more} can be 
attributed to both
shortcomings of a resonance unaware parton
shower matching: the parton shower reshuffling not
preserving the resonance masses, and the 
uncontrolled effects of resonances at the level of 
the first emission in the traditional \POWHEG{} approach.

\begin{figure}[tb]
\begin{center}
  \includegraphics[width=0.49\textwidth]{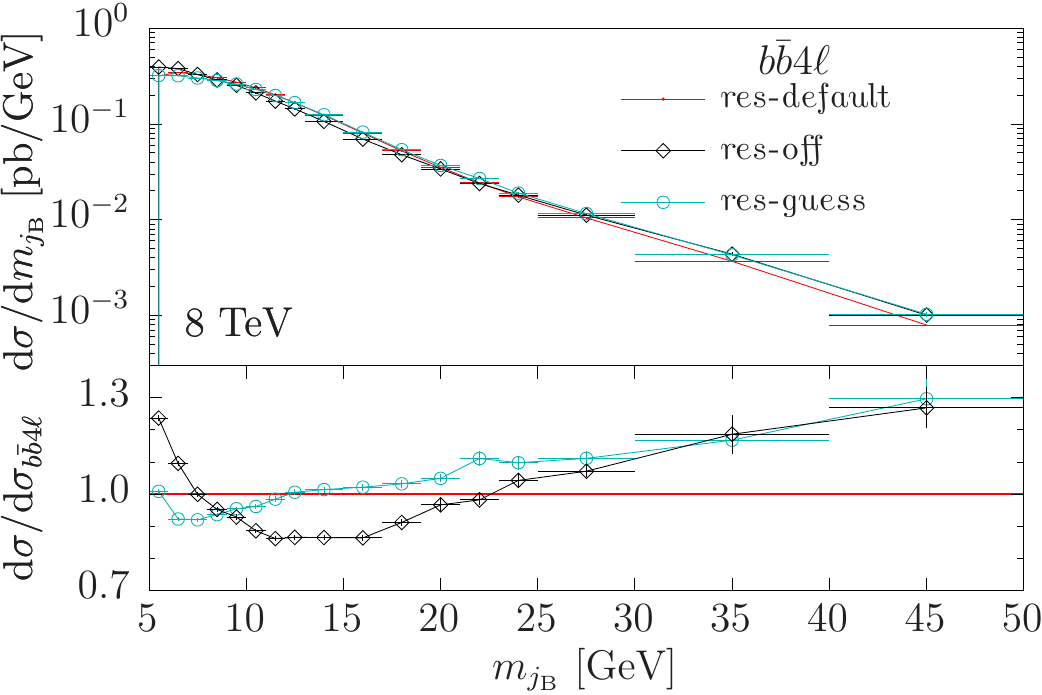}
  \includegraphics[width=0.49\textwidth]{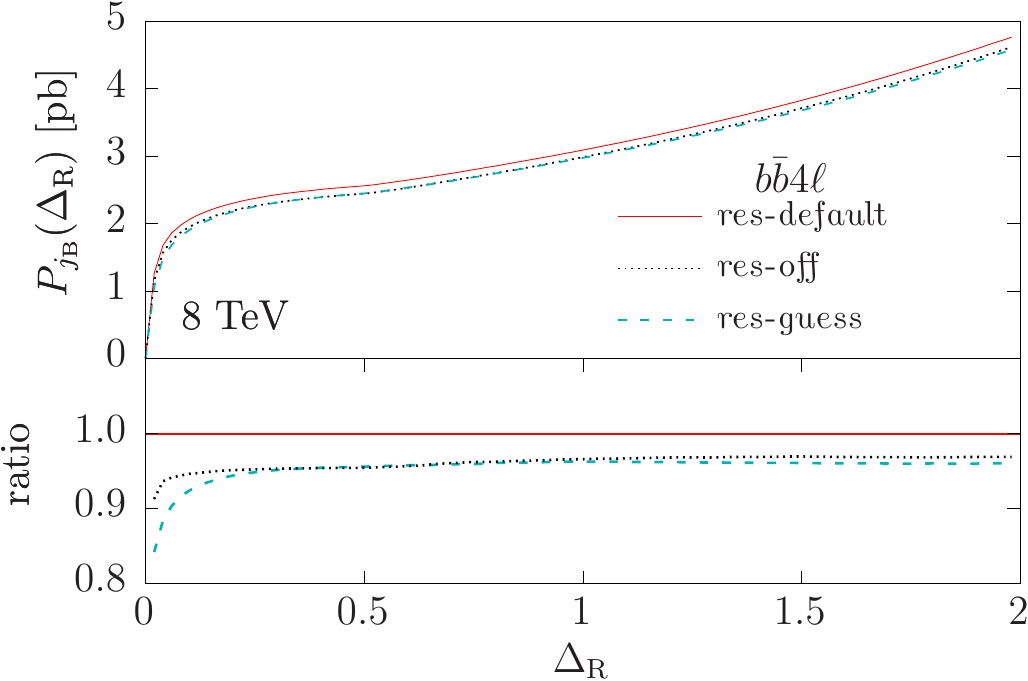}
\end{center}
\caption{Mass~(left) and profile~(right) of the $b$-jet $\bj$. Absolute predictions and ratios as 
in~\reffi{fig:m_w_jbot-res-nores}.}
\label{fig:bjet-allrad-nores} 
\end{figure}

In~\reffi{fig:bjet-allrad-nores} we display
the $\bj$ mass and profile,
defined as
\begin{equation}\label{eq:profile}
  \Etbj(\Delta_{\sss\mathrm R})= \int {\mathrm d}\sigma\;\frac{\sum_j
    \pt^{\sss j}\,
    \theta\!\(\Delta_{\sss \mathrm R}-\Delta_{\sss\mathrm R}^{(j,\bj)}\)}{\pt^\bj}\,.
\end{equation}
This observable corresponds to the cross section weighted by the fraction of the total
hadronic transverse momentum of the particles contained in a given cone
around the jet axis, with respect to the transverse momentum of the $b$-jet.
Again we observe marked differences among the \default{} and the \nores{}
results, and, to a lesser extent, between the \default{} and \noresi{}
ones. Both plots suggest that in the \nores{} case there is less activity
around the $B$ hadron, leading to smaller jet masses and to a slightly
steeper jet profile. The particularly pronounced shape distortion of the $\bj$
mass plot near 10~GeV in the \noresi{} case can be tentatively attributed to
the transition from the region where radiation~(generated with the
traditional method) does not change the mass of the resonance by an amount
comparable or larger than its width, to the region where it does, so that we
see the difference between the \noresi{} and \default{} results grow with
larger jet masses.

\begin{figure}[htb]
\begin{center}
\includegraphics[width=0.49\textwidth]{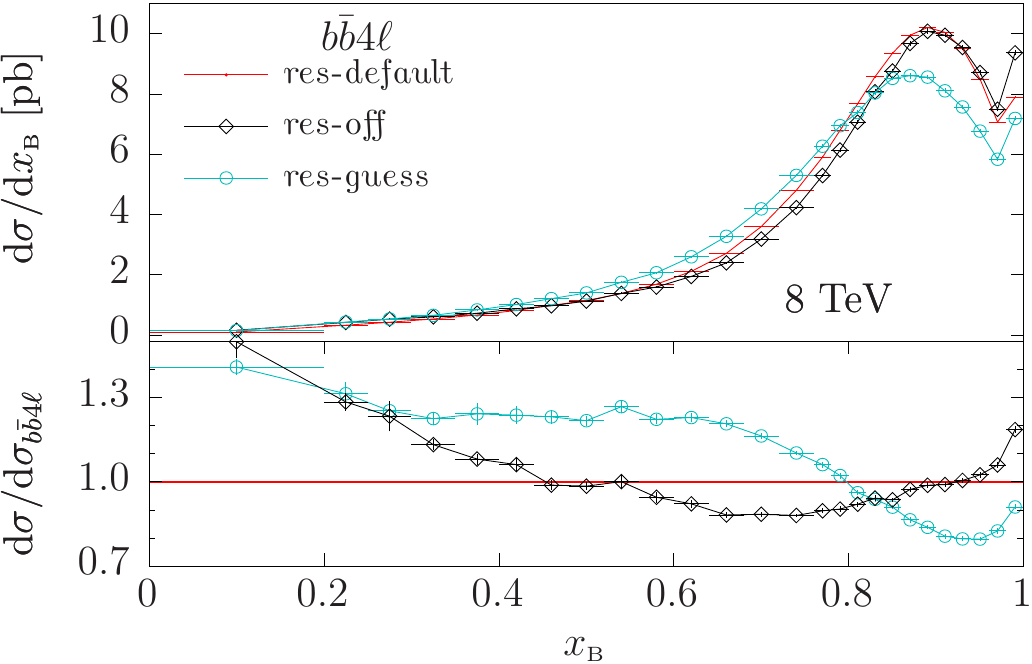}
\includegraphics[width=0.49\textwidth]{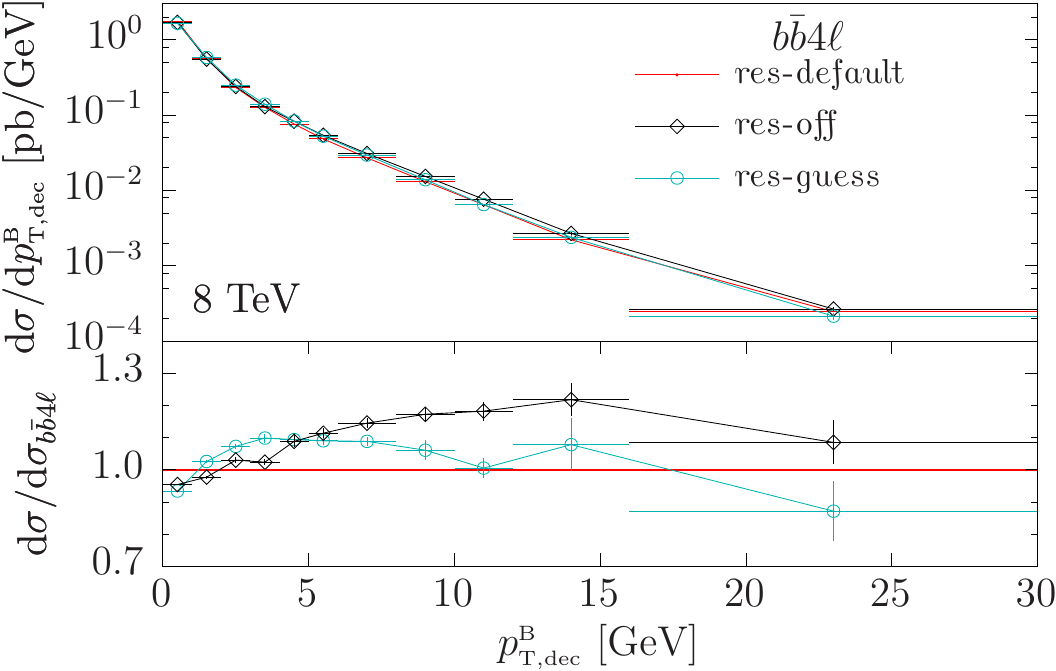}
\end{center}
\caption{$B$ fragmentation function and $B$-hadron transverse momentum in the top
  decay frame. Absolute predictions and ratios as 
in~\reffi{fig:m_w_jbot-res-nores}.}
\label{fig:bfrag-allrad-nores} 
\end{figure}

In~\reffi{fig:bfrag-allrad-nores}
we compare the $B$ fragmentation function and the $B$-hadron transverse momentum
computed in the reconstructed top-decay rest frame. The $\xB$ variable is
defined as the $B$ energy in the reconstructed top rest frame normalized to
the maximum value that it can attain at the given top virtuality, while
\ptdecB{} is the transverse momentum of the $B$ relative to the recoiling $W$
in the same frame.  We find marked differences also for these
distributions. While in the case of the $\ptdecB$ variable we see 
a reasonable consistency between the \noresi{} and \default{} results, 
the agreement deteriorates in the case of the fragmentation function.

We conclude that the consistent treatment of resonances implemented in 
the \bbfourl{} generator yields a narrower peak for the reconstructed top distribution
with respect to a traditional (resonance-blind) NLO+PS matching approach.
Furthermore, a large part of the difference is not related to the
lack of resonance information at the level of the shower generator, and thus
cannot be reduced by using a more sophisticated interface to the shower 
based on a resonance-guessing approach of kinematic nature.

\subsection{Comparison with the {\tt ttb\_NLO\_dec} generator}
\label{sec:ttNLOdec}

In this section we compare the \bbfourl{} generator 
against the {\tt ttb\_NLO\_dec} generator of~\citere{Campbell:2014kua}. 
The standard $\ttbar$ cuts 
of~\refeqs{eq:jet_cuts}{eq:leptonic_cuts} are applied throughout.
We examined a large set of distributions, but here we
only display the most relevant ones, and those that show the largest
discrepancies.

\begin{figure}[htb]
\begin{center}
  \includegraphics[width=0.49\textwidth]{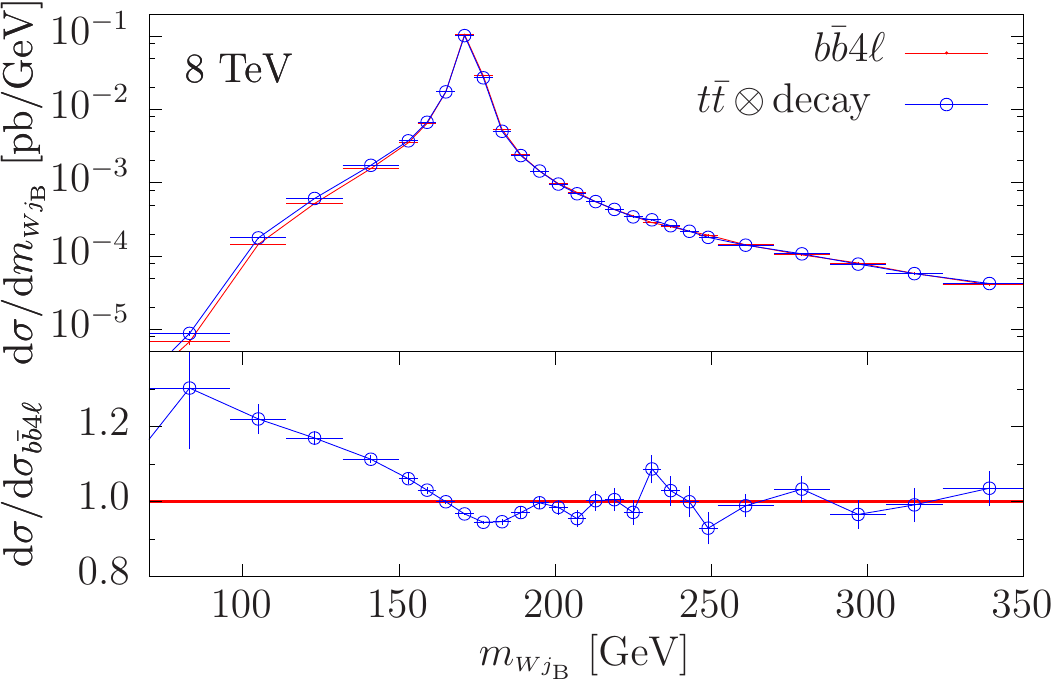}
  \includegraphics[width=0.49\textwidth]{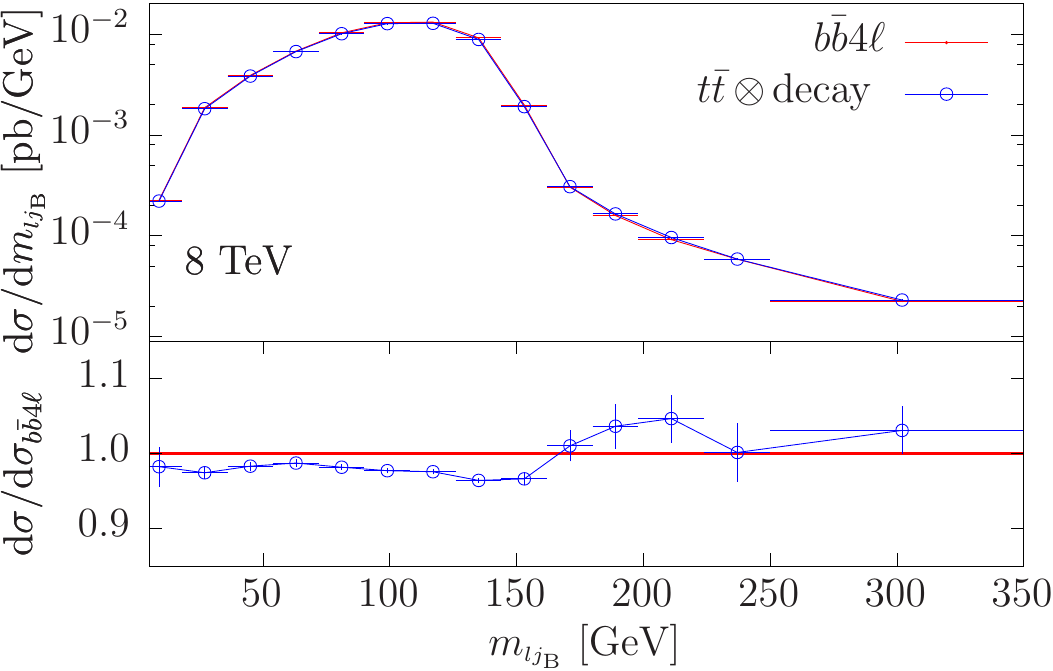}
\end{center}
\caption{Invariant mass of the $W\bj$~(left) and of the $l\bj$~(right)
  systems. Comparison of NLO+PS predictions obtained with the \bbfourl~(\BBFLRES) and the 
  \DEC~(\TTBARDEC) generators. In the ratio plot we illustrate relative deviations with respect to the \BBFLRES{} prediction. }
\label{fig:m_w_jbot-res-dec} 
\end{figure}

We begin by showing in~\reffi{fig:m_w_jbot-res-dec}
the invariant mass distribution of the $W\bj$ and $l\bj$ systems. We 
observe remarkable agreement between the \bbfourl{} and \DEC{} generators,
especially in the description of the reconstructed top peak and of the
shoulder in the lepton-\bj{} invariant mass.  This agreement is quite
reassuring. In fact, in the \DEC{} generator, the separation of radiation in
production and resonance decay is unambiguous, while in \bbfourl{} it is 
based on a probabilistic approach according to a kinematic proximity criterion.
Thus, in the light of \reffi{fig:m_w_jbot-res-dec},
the former generator supports the method of separation of resonance
histories adopted by the latter. On the other hand, off-shell 
and non-resonant effects are implemented in 
the \DEC{} generator in LO approximation, by reweighting the on-shell
result.
Thus the \bbfourl{} results support the validity of this
approximation in the \DEC{} implementation. 
As an indicative estimate of the potential implications 
for precision $m_t$ determination, we have determined that 
in a window of $\pm 30$~GeV around the peak of the $W\bj$ distributions, the 
average $W\bj$ mass computed with the \dec{} generator is roughly 0.1~GeV smaller 
than the one from \bbfourl{}.

The NLO distribution in the mass of the reconstructed top was also examined 
in~\citere{Campbell:2014kua} (sec.~3.2, Fig.~3).  There, the
\DEC{} fixed-order NLO result was compared to the fixed-order NLO result 
of~\citere{Denner:2012yc}, and the former was found to be enhanced by about
10\%{} in a region of roughly 1~GeV around the peak. This comparison was
carried out with massless $b$ quarks, since mass effects were not available
in~\citere{Denner:2012yc}. We computed the same distribution and carried
out the same NLO comparison, using however the \bbfourl{} generator instead
of the result of~\citere{Denner:2012yc} and taking into account $b$-mass effects.
Again, we find the same enhancement in the \DEC{} NLO
result. However, in the fully showered result we see instead a small
suppression of the peak in the \DEC{} relative to the \bbfourl{} generator,
suggesting that the NLO difference tends to be washed out by showering effects.

\begin{figure}[bt]
\begin{center}
  \includegraphics[width=0.49\textwidth]{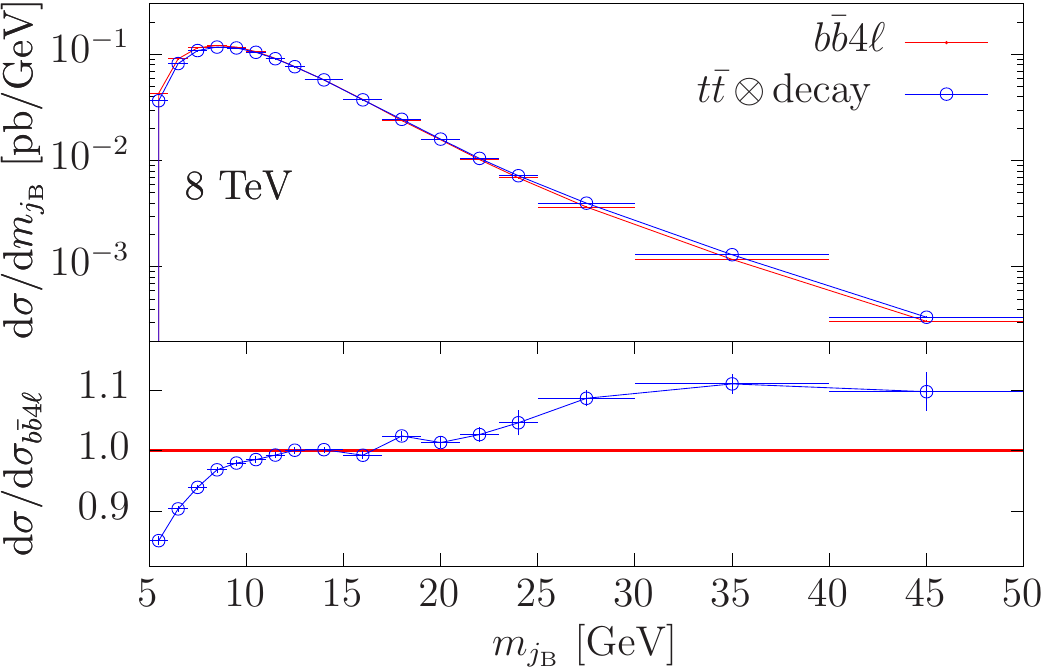}
  \includegraphics[width=0.49\textwidth]{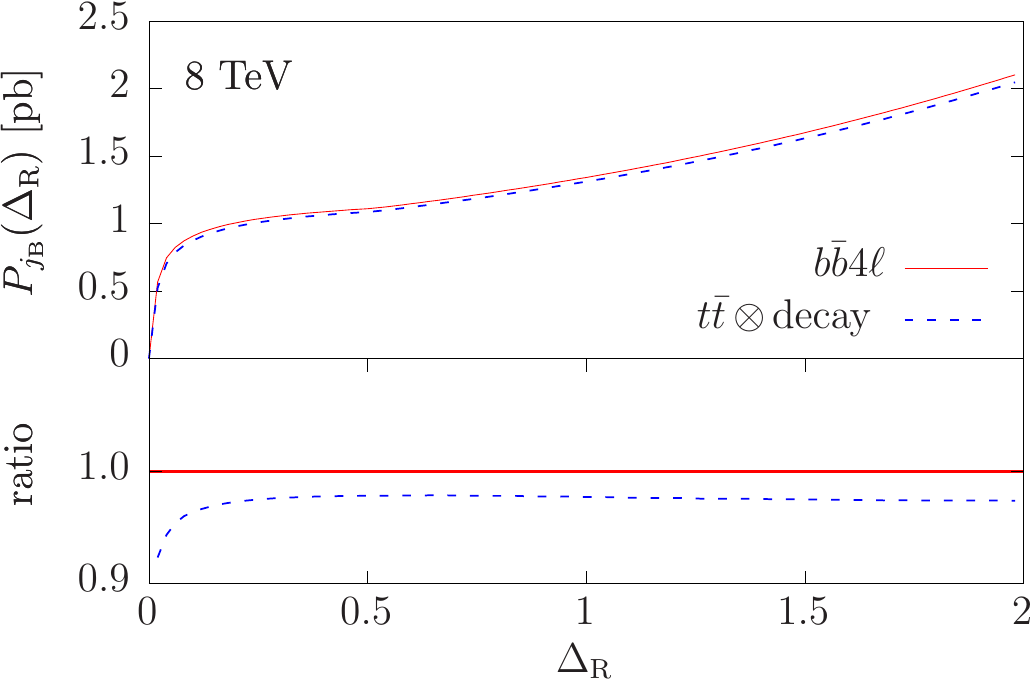}
\end{center}
\caption{Mass~(left) and profile~(right) of the $\bj$. Absolute predictions and ratios as 
in~\reffi{fig:m_w_jbot-res-dec}.}
\label{fig:bjet-allrad-dec} 
\end{figure}

We examined several distributions involving $b$-jets (here again we average over
the $b$- and $\bar{b}$-jet contributions). We found no appreciable difference
for the $b$-jet transverse momentum, while we did find significant differences
in the jet mass and the jet profile, displayed in~\reffi{fig:bjet-allrad-dec}.
Both plots indicate that the \bbfourl{} generator yields slightly wider
$b$-jets as compared to the \DEC{} one.

\begin{figure}[tb]
\begin{center}
  \includegraphics[width=0.49\textwidth]{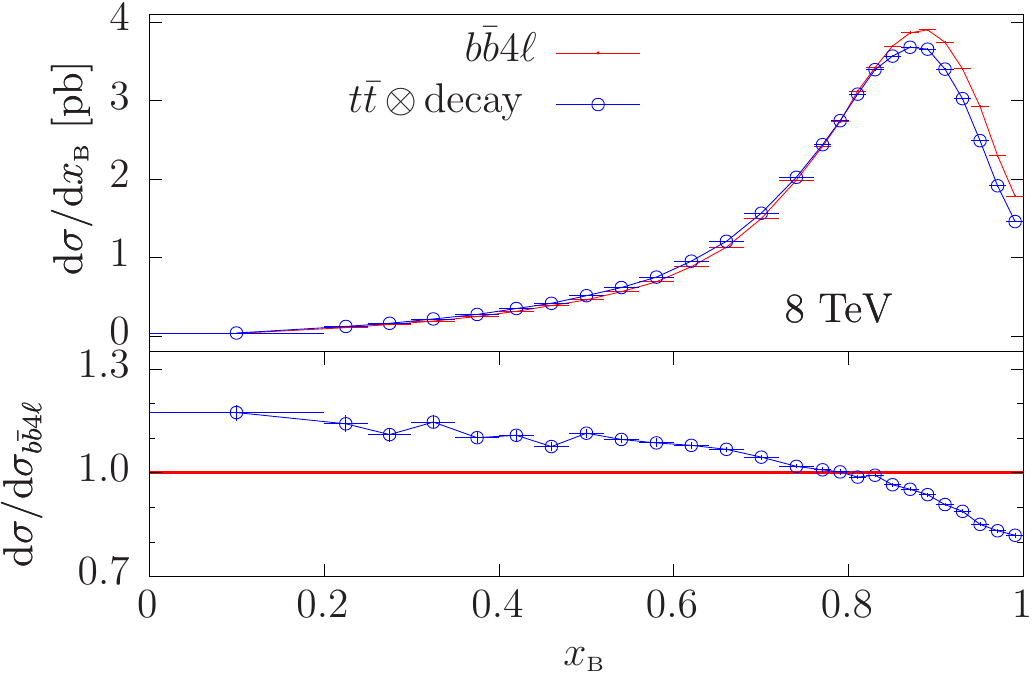}
  \includegraphics[width=0.49\textwidth]{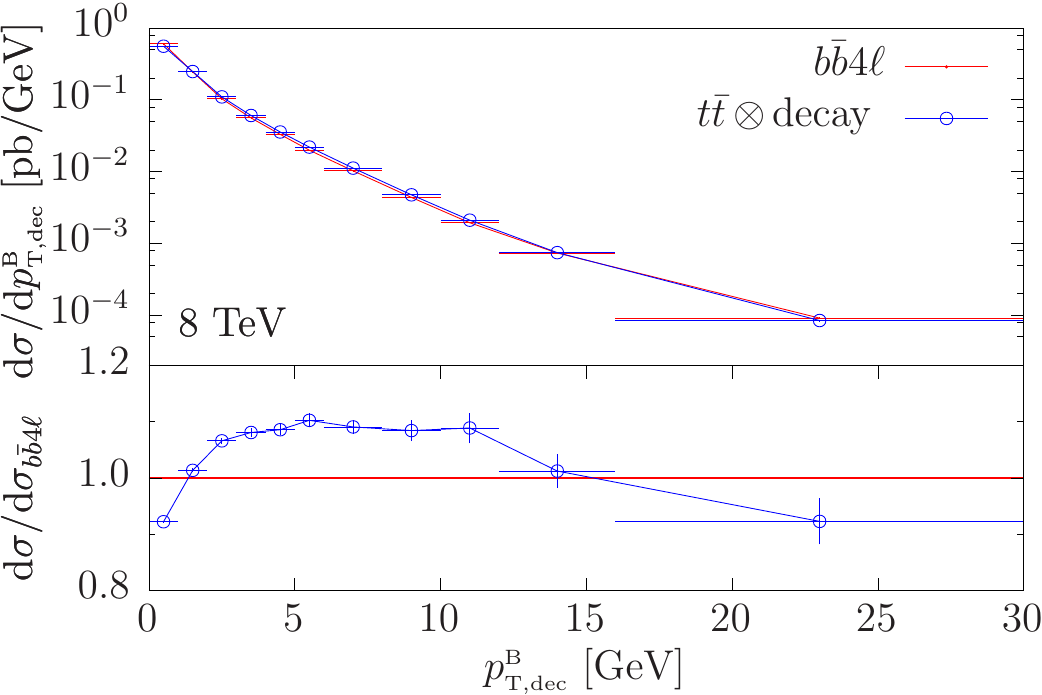}
\end{center}
\caption{The $B$ fragmentation function and transverse-momentum distribution of the
  \ptdecB{} observable. Absolute predictions and ratios as 
in~\reffi{fig:m_w_jbot-res-dec}.}
\label{fig:8TeV_fragb} 
\end{figure}

In~\reffi{fig:8TeV_fragb} we plot
the $B$ fragmentation function and the \ptdecB{} observables.
We find that the fragmentation function is slightly harder, and the \ptdecB{}
distribution is slightly softer in the \bbfourl{} case.
Again, this is consistent with the
observation of slightly reduced radiation from $b$'s in the \bbfourl{} case.
We have verified that this feature persists also when hadronization is
switched off in \PythiaEight{}.

Although the differences in the $b$-jet structure are quite significant,
they are not sufficient to induce an observable shift in the reconstructed mass peak. This
could only happen if the difference in the jet profile caused a consistent
difference in the jet energy, due to energy loss outside the jet-cone. This does
not seem to be the case since the jet profiles become similar in the two
generators already for $\Delta_{\rm R}< 0.5$.

\subsection{Comparison with the {\tt hvq} generator}
\label{sec:RES_HVQ}

\begin{figure}[htb]
\begin{center}
  \includegraphics[width=0.49\textwidth]{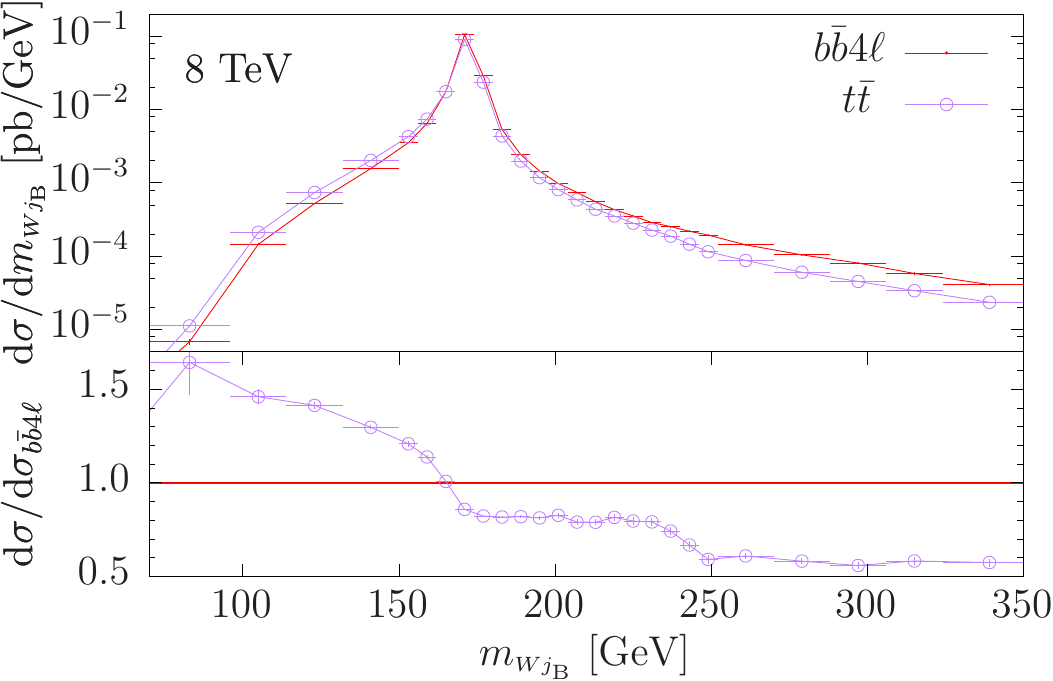}
  \includegraphics[width=0.49\textwidth]{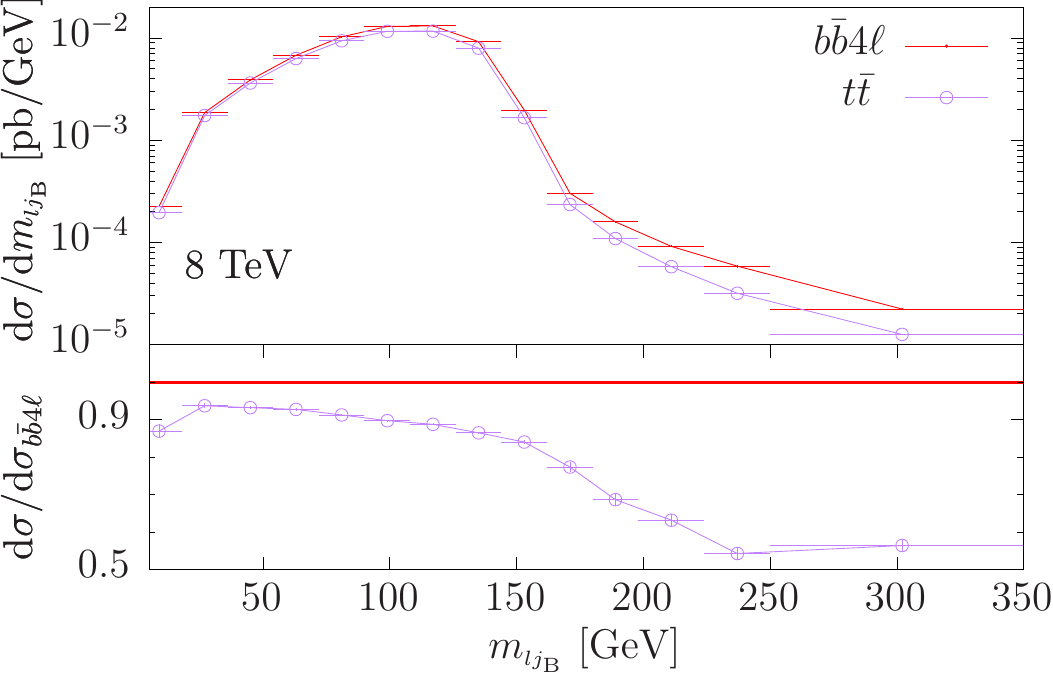}
\end{center}
\caption{Invariant mass of the $W\bj$~(left) and of the $l\bj$~(right)
  systems. Comparison of NLO+PS predictions obtained with the \bbfourl~(\BBFLRES) and the 
  \hvq~(\TTBAR) generators. In the ratio plot we illustrate relative deviations with respect to the \BBFLRES{} prediction.}
\label{fig:m_w_jbot-allrad-hvq} 
\end{figure}

In this section we compare the \bbfourl{} generator against the 
\hvq{} generator of~\citere{Frixione:2007vw}, which is based on on-shell NLO matrix 
elements for \ttbar{} production. 
Again the standard $\ttbar$ cuts of~\refeqs{eq:jet_cuts}{eq:leptonic_cuts} are applied throughout.
The $W\bj$ and $l\bj$~mass distributions, shown in~\reffi{fig:m_w_jbot-allrad-hvq}, 
show reasonably good agreement 
between the two generators as far as the shape of the
$W\bj$ peak and of the $l\bj$ shoulder are concerned.
However, for large top virtualities, i.e. in the tails of both distributions, sizable differences can be appreciated. 
As we will see below, such differences originate from the fact that,
 in this region, the \bbfourl{} generator tends to radiate considerably less,
which results in narrower b-jets as compared to the \hvq{} generator. 
We note that the observed deviations 
with respect to the \hvq{} generator are more drastic than the ones
observed in section~\ref{sec:ttNLOdec} for the \DEC{} generator.
The $m_{W\bj}$ distribution on the left of \reffi{fig:m_w_jbot-allrad-hvq}
additionally suggests a non-negligible shift in the reconstructed top mass between the two generators.
In fact, we determined that in a window of $\pm 30$~GeV
around the peak of the $m_{W\bj}$ distributions, the average $W\bj$ mass computed
with the \hvq{} generator is roughly 0.5~GeV smaller than with the \bbfourl{} one.

\begin{figure}[htb]
\begin{center}
  \includegraphics[width=0.49\textwidth]{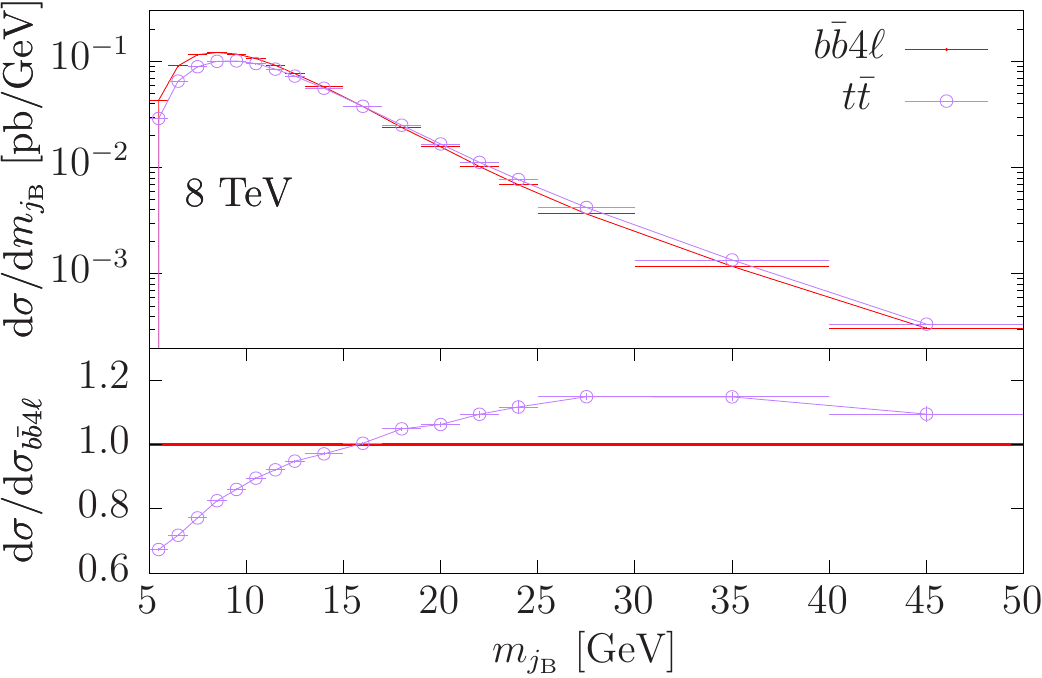}
  \includegraphics[width=0.49\textwidth]{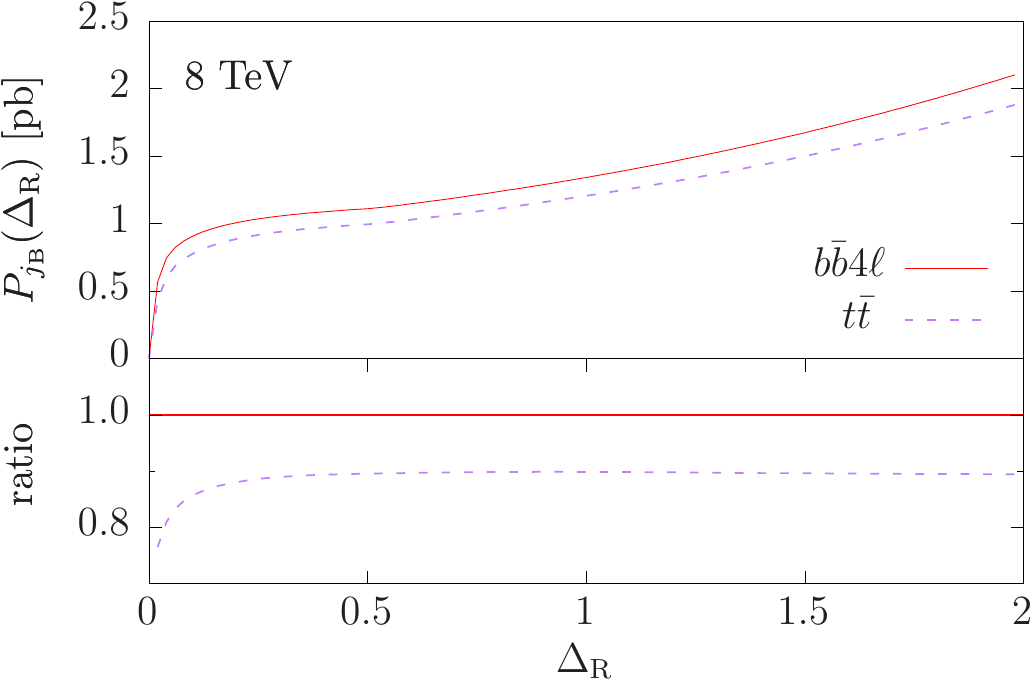}
\end{center}
\caption{Mass~(left) and profile~(right) of $\bj$. Absolute predictions and ratios as 
in~\reffi{fig:m_w_jbot-allrad-hvq}.}
\label{fig:bjet-allrad-hvq} 
\end{figure}

\begin{figure}[htb]
\begin{center}
  \includegraphics[width=0.49\textwidth]{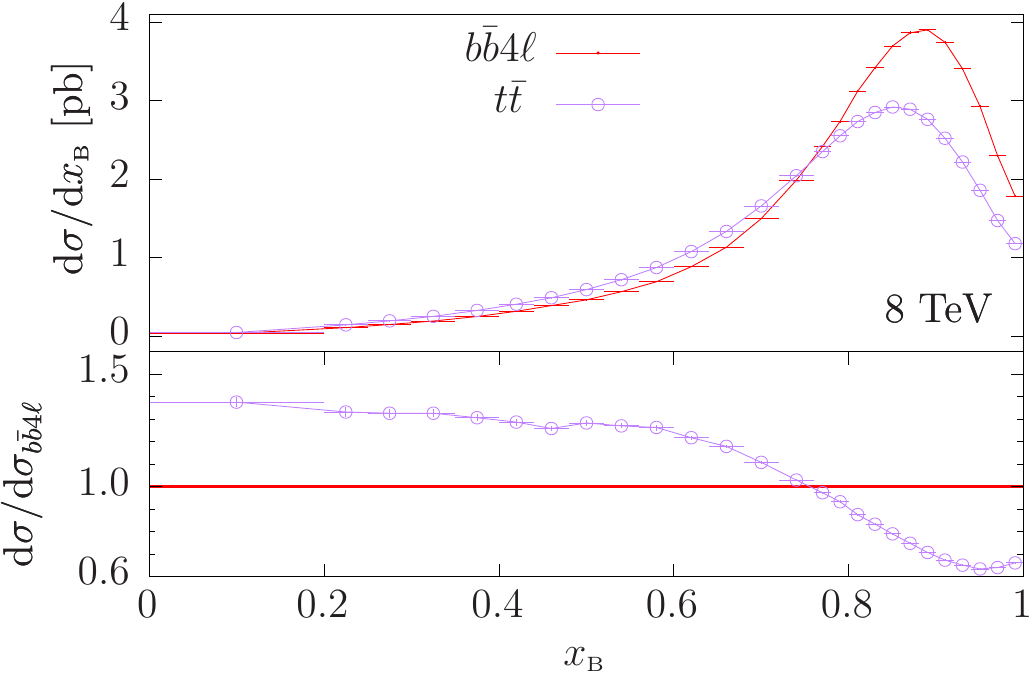}
  \includegraphics[width=0.49\textwidth]{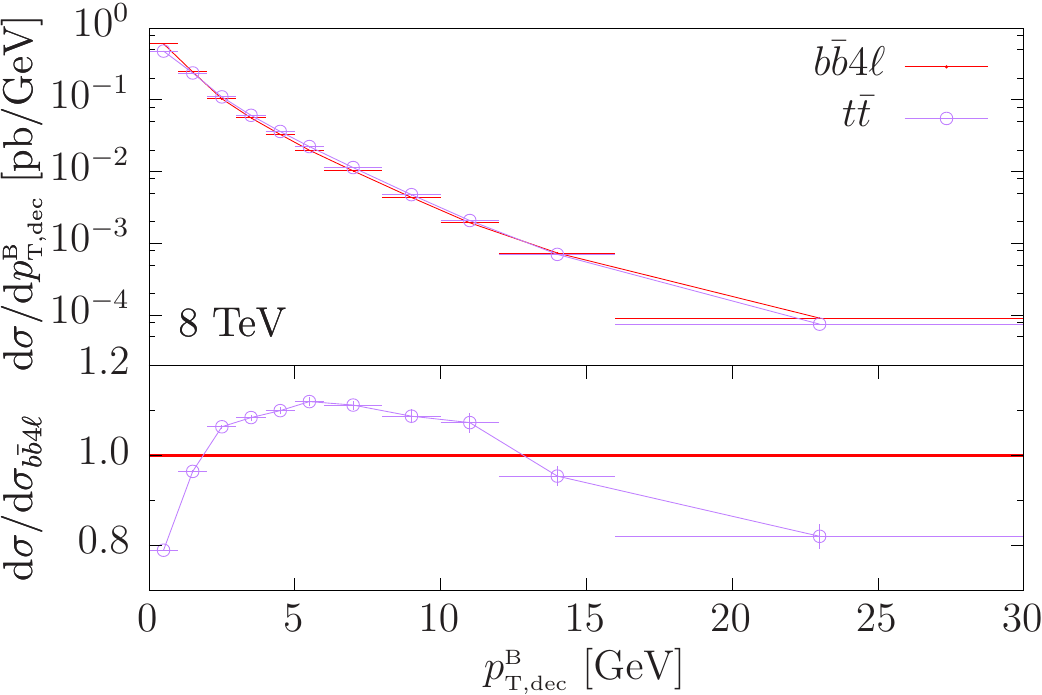}
\end{center}
\caption{The $B$ fragmentation function and \ptdecB{} distribution.  Absolute predictions and ratios as 
in~\reffi{fig:m_w_jbot-allrad-hvq}.}
\label{fig:8TeV_fragb-allrad-hvq} 
\end{figure}

In~\reffi{fig:bjet-allrad-hvq}
we show distributions in the $b$-jet mass and profile, as defined in \refeq{eq:profile}. 
Both plots indicate significantly narrower $b$-jets in the predictions
obtained with the \bbfourl{} generator.
Similarly, as
shown in~\reffi{fig:8TeV_fragb-allrad-hvq},
the \bbfourl{} generator yields a harder $B$ fragmentation function and a softer \ptdecB{}
distribution.
The pattern we observe for the structure of $b$-jets is consistent with the
fact that the \bbfourl{} generator has a reduced radiation in $b$-jets with
respect to \PythiaEight{}. In the \hvq{} generator, radiation from the $b$'s
is handled exclusively by \PythiaEight{}, while, in the \bbfourl{} generator,
the hardest radiation from the $b$ is handled by \POWHEG{}. It should be
stressed, however, that the $B$ fragmentation function has a considerable
sensitivity to the hadronization parameters. It would therefore be desirable
to tune these parameters to $B$ production data in $e^+ e^-$ annihilation,
within the \POWHEG{} framework, in order to perform a meaningful comparison.

\begin{figure}[ht]
\begin{center}
\includegraphics[width=0.6\textwidth]{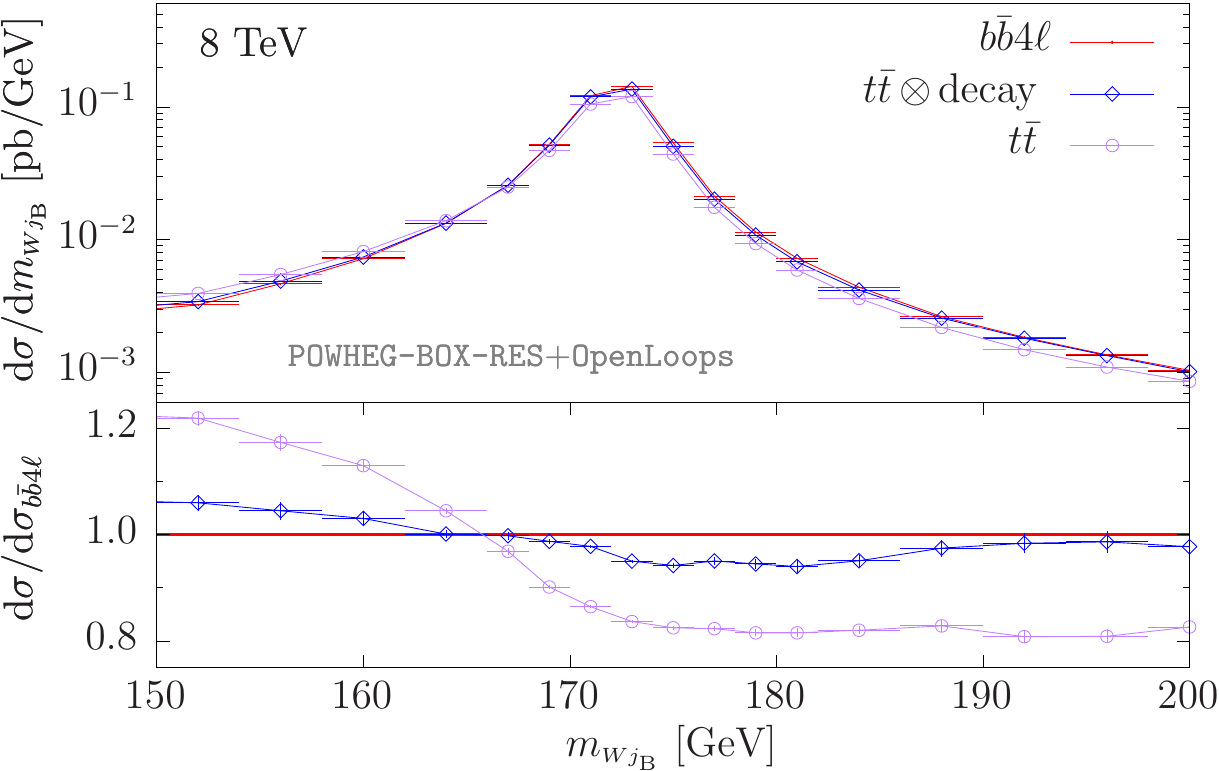}
\end{center}
\caption{The $W\bj$ mass distribution near the top peak for the three
  generators \bbfourl{}~(\BBFLRES), \DEC{}~(\TTBARDEC) and \hvq{}~(\TTBAR). In the ratio plot we illustrate relative deviations with respect to the \BBFLRES{} prediction.}
\label{fig:8TeV_res-allrad_dec_hvq_PY8_m_w_jbot} 
\end{figure}

In~\reffi{fig:8TeV_res-allrad_dec_hvq_PY8_m_w_jbot} we show a summary of
the shape of the reconstructed top peak comparing each of the available \POWHEG{}
generators for $\ttbar$ production: \bbfourl{}, \DEC{} and \hvq{}. We notice a
fair consistency between the \bbfourl{} generator and the \DEC{} one, while larger deviations 
are observed comparing against \hvq{}.

\section{Jet vetoes and single-top enriched observables}
\label{sec:WtPhenomenology}

In this section we investigate the behaviour of the $\bbfourl$ generator in
the presence of $b$-jet and light-jet vetoes.  
Such kinematic restrictions are widely used in order to reduce top
backgrounds in $H\to\WW$ studies and in many other analyses that involve
charged leptons and missing energy.
Also, jet vetoes play an essential role for experimental studies of $Wt$ single-top
production~\cite{Chatrchyan:2014tua,Aad:2015eto}.  
In particular, the separation of \Wt{} and $t\bar{t}$ production typically
relies upon the requirement that one large transverse-momentum $b$-jet is
missing in the first process.

From the theoretical point of view, the separation of \Wt{} and $t\bar{t}$
production is not a clear cut one, since the two processes interfere.  As
pointed out in the introduction, in the $\bbfourl$ generator this problem is
solved by providing a unified description of $\ttbar$ and \Wt{} production
and decay, where also interference effects are included at NLO.  Thus jet
vetoes are expected to enrich the relative single-top content of $\bbfourl$
samples, resulting in significant differences with respect to other
generators that do not include $\Wt$ contributions and interferences at NLO.
The \bbfourl{} generator is particularly well-suited for the study of jet
vetoes also because it includes $b$-mass effects, NLO radiation in
top-production and -decay subprocesses, as well as resummation of multiple
QCD emissions and hadronization effects as implemented in the parton shower.

\begin{figure}[htb]
\begin{center}
\includegraphics[width=0.49\textwidth, trim=0 0 0 0,clip]{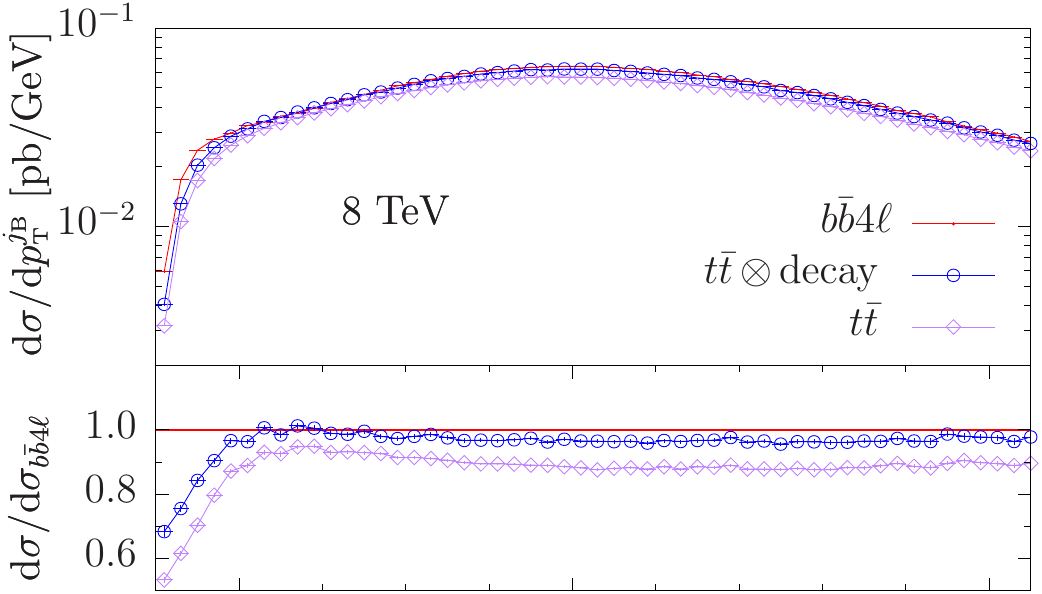} \\
\vspace{-0.099cm}
\includegraphics[width=0.49\textwidth]{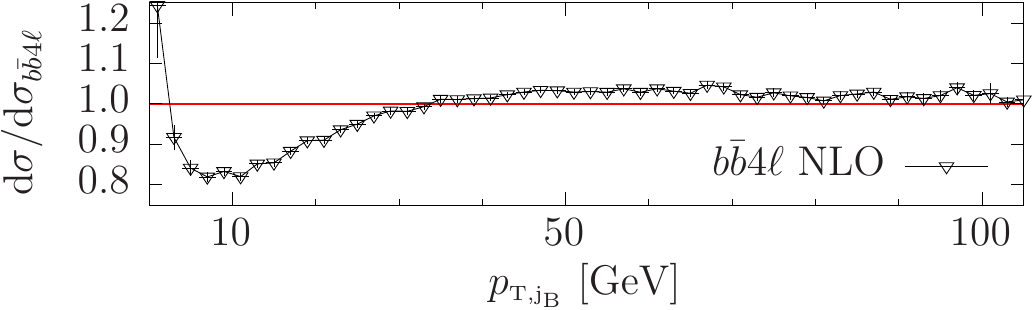}
\end{center}
\caption{Distribution in the $b$-jet transverse momentum: comparison of
  NLO+PS predictions obtained with the three generators \bbfourl~(\BBFLRES),
  \DEC~(\TTBARDEC) and \hvq (\TTBAR).  The middle frame illustrates relative
  NLO+PS deviations with respect to the \BBFLRES{} prediction, while the
  lower frame compares \BBFLRES{} versus corresponding fixed-order NLO results.}
\label{fig:res-dec-hvq-lo-bjpt}
\end{figure}

A first picture of the $b$-jet activity in the three generators, \bbfourl{},
\DEC{} and \hvq{}~(labelled according to~\refta{tab:generators} as
\BBFLRES{}, \ttbardecay{} and \ttbar{} respectively), is provided 
by~\reffi{fig:res-dec-hvq-lo-bjpt}, where we compare NLO+PS distributions in
the transverse momentum of the $b$-jet.
More precisely, the plotted observable 
corresponds to the sum of the $b$- and \mbox{$\bar b$-jet} spectra and was computed
in absence of any acceptance cut. 
Thus it involves potentially enhanced contributions from single-top topologies, which
can lead to significant deviations between the \ttbar{} prediction\footnote{%
  In order to make sure that, apart form the absence of $\Wt$ contributions,
  the \ttbar{} predictions are internally consistent, we have checked that
  off-shell top contributions~(which are modelled through an heuristic
  Breit--Wigner smearing approach in \hvq{}) play only a marginal role for
  the observable at hand.  To this end we have applied cuts to the $t$ and
  $\bar{t}$ virtualities, imposing that they should not differ from the $t$
  pole mass by more than 15~GeV.  The effect of such cuts was found to be
  negligible.}  and the ones that implement off-shell $\fourl\,\bbbar$ matrix
elements.
At large transverse momentum, the various predictions have rather similar
shape, but the \ttbar{} result features a clear deficit of about 10\% with
respect the \BBFL{} and \ttbardecay{} ones.  This can be attributed to the
missing single-top contributions in the \hvq{} generator.
At high $\pt$, thanks to the implementation of $\Wt$ contributions via exact
Born matrix elements for $\ppllllbb$, the \ttbardecay{} prediction is found
to be in good agreement with the \BBFL{} one.
At small transverse momenta, the relative weight of $\Wt$ production becomes
more important, and the deficit of the \ttbar{} prediction grows rather
quickly, reaching up to 50\% for very small transverse momenta.  The
\ttbardecay{} and \BBFL{} predictions remain in good agreement down to
$p_{{\rm T},\bj}\simeq 10$~GeV, but at smaller transverse momenta the
\ttbardecay{} one develops a deficit that grows up to about 25\%.  This can
be attributed, at least in part, to the increased importance of $\Wt$
channels combined with the fact that these channels are not supplemented by
an appropriate NLO correction in the \ttbardecay{} predictions. We also note
that the discrepancy at hand can be interpreted as a kinematic shift of a few
GeV only, while the enhancement of the resulting correction can be attributed
to the pronounced steepness of the absolute $p_{{\rm T},\bj}$ distribution in the
soft region.  Its sign is consistent with the fact that radiation arising
from $\fourl\,\bbbar$ NLO matrix elements is expected to be rather soft in
the presence of single-top contributions with initial-state collinear $g\to
\bbbar$ splittings, while in the \ttbardecay{}{} generator radiation is
always emitted as if all $b$-quarks would arise from top decays, which
results in a harder emission spectrum.
The lower frame of~\reffi{fig:res-dec-hvq-lo-bjpt} illustrates the
relative importance of matching and shower effects in the \bbfourl{}
generator, comparing against corresponding fixed-order NLO predictions.
Again we observe nontrivial shape effects in the soft region.  While they are
not directly related to the differences observed in the middle frame, such
effects highlight the importance of a consistent treatment of radiation and
shower effects at small $b$-jet $\pt$. On the other hand the good agreement
between the \ttbardecay{}{} and \BBFL{} predictions down to 10~GeV suggests
that matching and pure shower effects are reasonably well under control in
the bulk of the phase space.

\begin{figure}[tb]
\begin{center}
\includegraphics[width=0.49\textwidth]{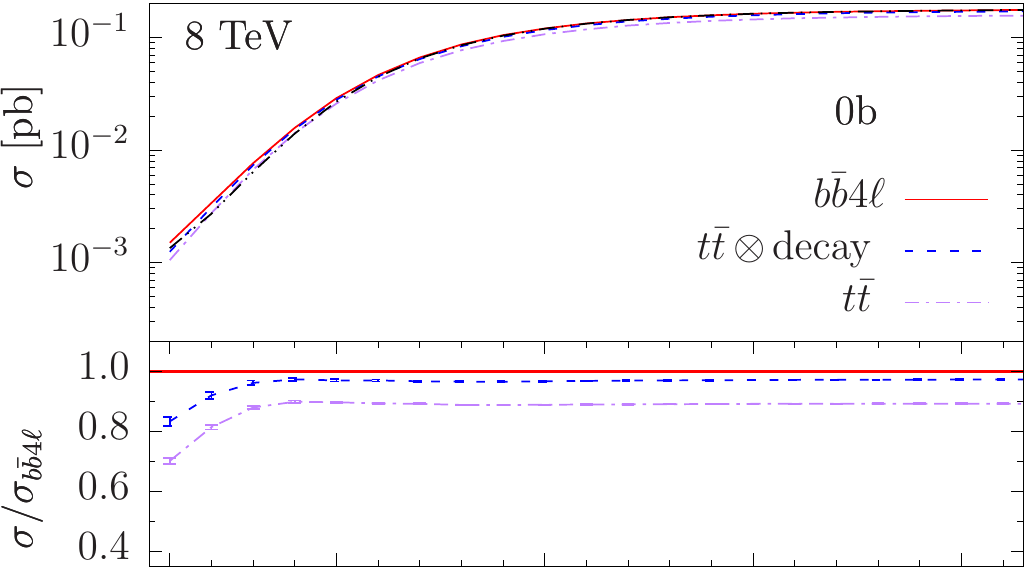}
\includegraphics[width=0.49\textwidth]{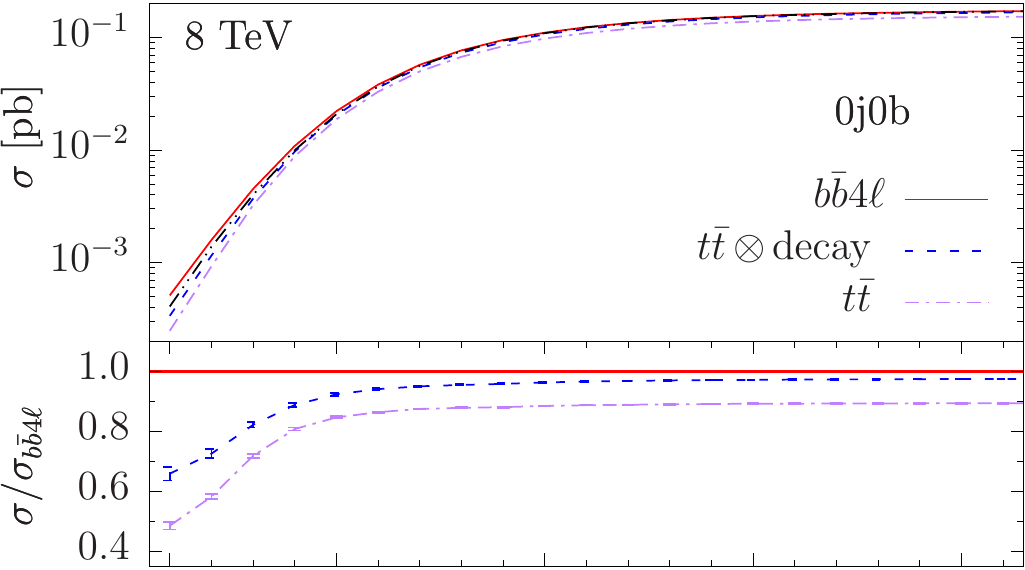}\\
\vspace{-0.068cm}
\includegraphics[width=0.49\textwidth]{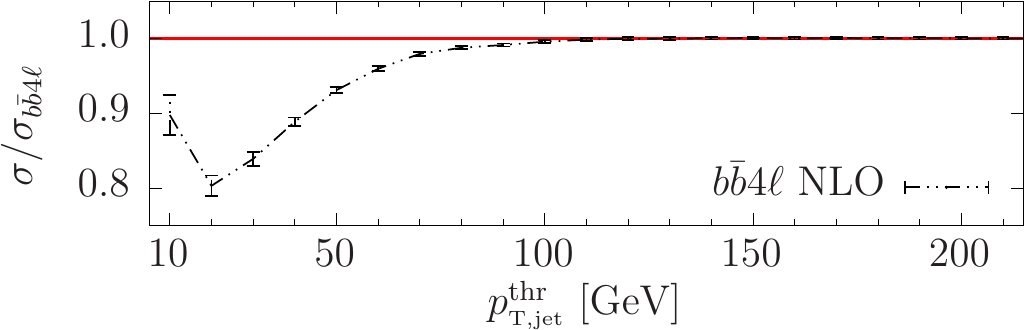}
\includegraphics[width=0.49\textwidth]{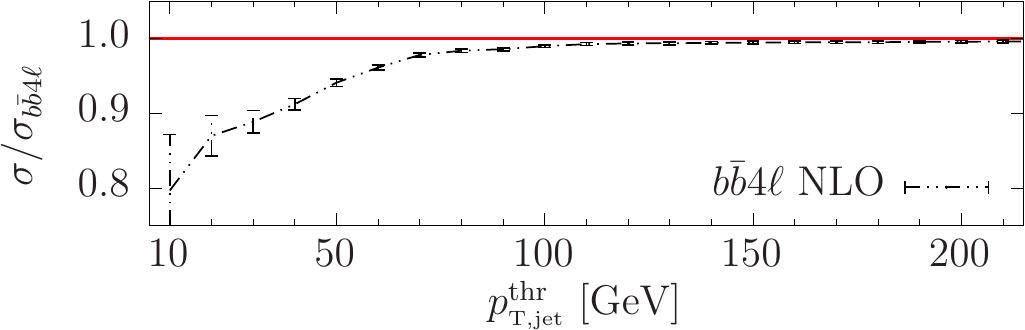}
\caption{Integrated cross sections at 8~TeV in jet bins with zero $b$-jets
  as a function of the jet-$\pT$ threshold.  The left plot is inclusive with
  respect to extra jet radiation~($\nj{}\ge\nb{}=0$), while the right plot is
  exclusive ~($\nj=\nb=0$).  Absolute predictions and ratios as 
in~\reffi{fig:res-dec-hvq-lo-bjpt}.  }
\label{fig:8TeV_zerob}
\end{center}
\end{figure}

Jet-binning and jet-veto effects are studied 
in~\reffis{fig:8TeV_zerob}{fig:8TeV_oneb}. For this analysis we
apply again the lepton selection cuts of~\refeq{eq:leptonic_cuts} and,
at variance with the $b$-jet definition in~\refse{sec:ttbarPhenomenology},
we identify as $b$-jets those
jets containing at least a $b$- or \mbox{$\bar b$-flavoured} hadron, irrespectively of its hardness.%
\footnote{At fixed-order NLO, jet clustering and
  $b$-jet tagging are applied at parton level.}
Events are categorised
according to the number of~(light \textit{or} heavy-flavour) jets, 
\nj{}, and to the number of $b$-jets, \nb{},
in the rapidity range $|\eta|<2.5$,
while we vary the jet transverse-momentum threshold \pTthr{} that defines
jets.

In~\reffi{fig:8TeV_zerob}, to investigate the effect of a $b$-jet veto,
the integrated cross sections is plotted versus the jet-veto threshold,
\pTthr{}.  In the left plot the veto acts only on \mbox{$b$-jets} \mbox{($\nj{}\ge\nb{}=0$),}
while in the right plot a veto against light and $b$-jets is
applied~\mbox{($\nj{}=\nb{}=0$).}
For $\pTthr{} \gtrsim 80$~GeV the vetoed cross section is dominated by $t
\bar t$ production and quickly converges towards the inclusive result.  In
this region we observe few-percent level agreement between the
\ttbardecay{}{} and \BBFL{} predictions, while the on-shell \ttbar{} prediction
features a 10\% deficit due to the missing single-top topologies.
Reducing the jet-veto scale increases this deficit up to $-30\%$ in the case
of the inclusive $\nb{}=0$ cross section.  This finding is well consistent
with the size of finite-width effects reported in~\citere{Cascioli:2013wga}.
In the case of the exclusive zero-jet cross section~($\nj{}=\nb{}=0$, shown
on the right) the deficit of the \ttbar{} prediction
is even more pronounced and reaches up to $-50\%$
at $\pTthr{}=10$~GeV.
Also the \ttbardecay{} results feature a similar, although less pronounced,
deficit as the \ttbar{} ones in the soft region.
This can be attributed to the fact that initial-state radiation in both, the
\hvq{} and \DEC{}, generators is computed with on-shell tops, and thus
overestimates the radiation produced near the single-top kinematic region.

Matching and pure shower effects are illustrated in the lower frames 
of~\reffi{fig:8TeV_zerob}. Both in the inclusive~($\nj{}>\nb{}=0$) and
exclusive~($\nj{}=\nb{}=0$) case we observe that, down to 20~GeV, NLO+PS
predictions feature an increasingly strong enhancement with respect to
fixed-order ones.  This can be attributed to shower-induced losses of $b$-jet
transverse momentum.  In the exclusive case~($\nj{}=\nb{}=0$) this
enhancement is somewhat milder, which we tentatively attribute to the
interplay of parton shower radiation with the additional light-jet veto.

\begin{figure}[tb]
\begin{center}
\includegraphics[width=0.49\textwidth]{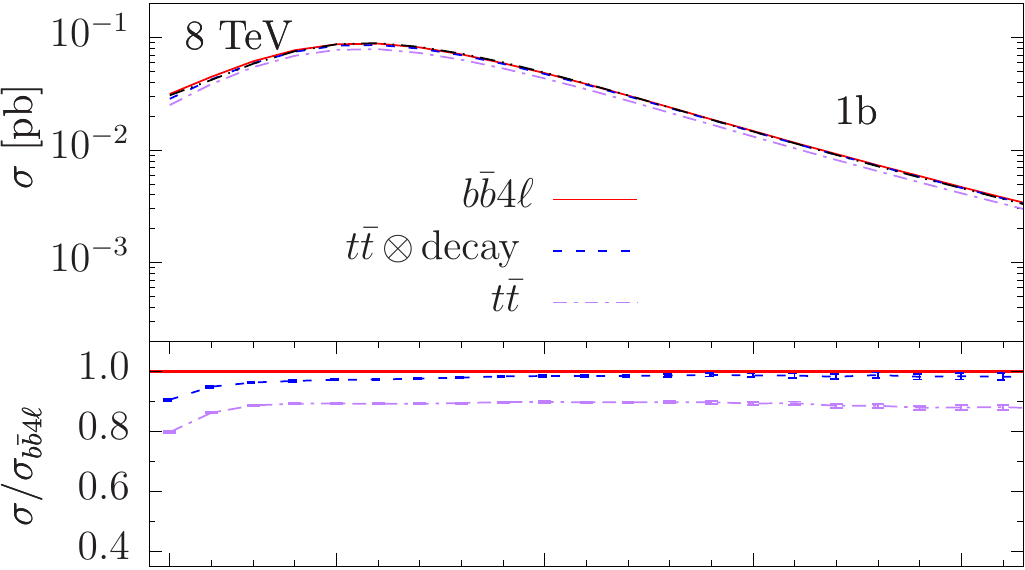}
\includegraphics[width=0.49\textwidth]{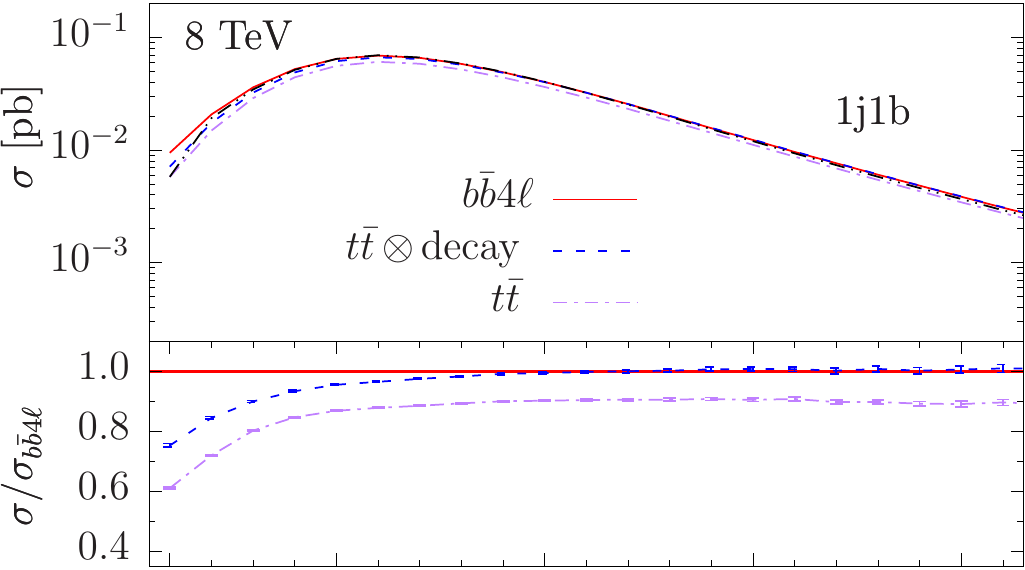}\\
\vspace{-0.065cm}
\includegraphics[width=0.49\textwidth]{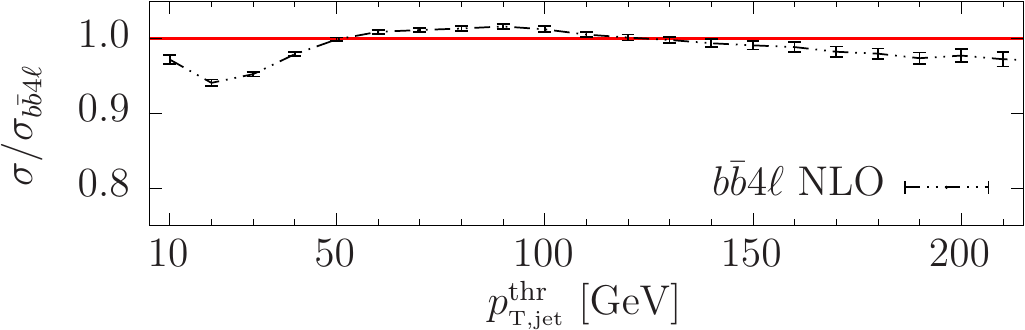}
\includegraphics[width=0.49\textwidth]{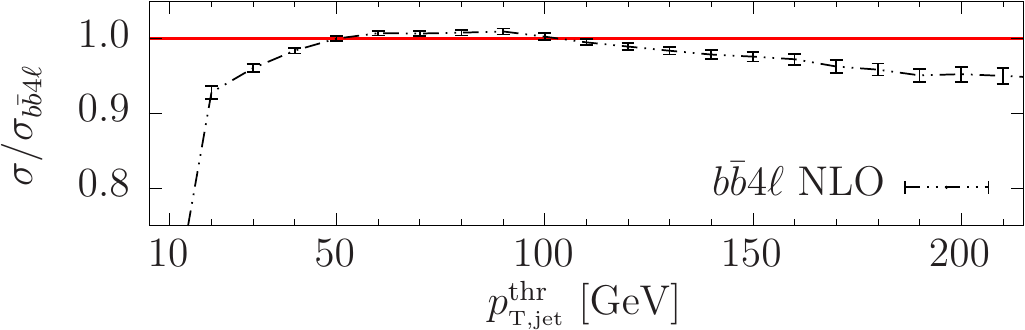}
\caption{Integrated cross sections at 8~TeV in jet bins with one $b$-jet as
  a function of the jet-$\pT$ threshold.  The left plot is inclusive with
  respect to extra jet radiation~($\nj{}\ge\nb{}=1$), while the right plot is
  exclusive~($\nj=\nb=1$).  Absolute predictions and ratios as 
in~\reffi{fig:res-dec-hvq-lo-bjpt}.  }
\label{fig:8TeV_oneb}
\end{center}
\end{figure}

In~\reffi{fig:8TeV_oneb} we plot the cross section with exactly one
$b$-jet above the threshold \pTthr{}, i.e.~we veto additional $b$-jets above
this threshold. Again, inclusive results ($\nj{}\ge \nb{}=1$, shown on the
left) are compared with exclusive ones ($\nj{}=\nb{}=1$, shown on the right).
The one-$b$-jet bin is typically used in $Wt$ single-top analyses. Similarly
as for the zero-$b$-jet case, the difference between the \BBFL{} and \ttbar{}
results points to an increasingly important single-top contribution at small
$\pTthr{}$.  Its quantitative impact is consistent with the fixed-order
results of~\citere{Cascioli:2013wga}, and at $\pTthr{}=30$~GeV it amounts to
about 10\% and $20\%$, respectively, in the inclusive and exclusive cases.
Similarly as for the zero-$b$-jet case, \ttbardecay{} predictions feature a
qualitatively similar but quantitatively less pronounced deficit with respect
to the \BBFL{} predictions.  Matching and shower effects turn out to be
rather mild in the inclusive case, probably due to the fact that the absolute
distribution is not particularly steep in the limit of small transverse
momentum.  In contrast, the exclusive one-jet cross
section~($\nj{}=\nb{}=1$) is much more sensitive to the jet-veto scale,
which leads to sizable matching and shower
effects at small \pTthr{}.

In summary, jet-vetoed cross sections can involve enhanced single-top
contributions that are completely missing in the \ttbar{} predictions
obtained with the \hvq{} generator while they are significantly
underestimated in the \ttbardecay{} predictions 
of the \DEC{}
generator, where single-top contributions are 
implemented via LO reweighting~\cite{Campbell:2014kua}.  In practice such a reweighting approach
ceases to work in phase-space regions far away from the double-resonant
region.

\section{Conclusions}\label{sec:conc}

In this paper we have presented the first Monte Carlo generator that provides
a fully consistent NLO+PS simulation of $\ttbar$ production and decay in the
different-flavour dilepton channel, including all finite-width and
interference effects.
This new generator, dubbed $\bbfourl$, is based on the full NLO matrix
elements for the process $pp\to\enmn\,\bbbar$.  This guarantees NLO accuracy
in $\ttbar$ production and decay, as well as the exact treatment of spin
correlations and off-shell effects in top decay.
Top resonances are dressed with quantum corrections, and also
non-factorisable corrections associated with the interference of radiation in
production and decays are taken into account.
Bottom-quark masses are consistently included, which is quite important for
the accurate modelling of $b$-quark fragmentation.  Moreover, finite
$b$-quark masses permit to avoid collinear singularities from initial- or
final-state $g\to \bbbar$ splittings. This allows for $\WWbb$ simulations in
the full phase space, including regions with unresolved $b$ quarks, which are
indispensable for the simulation of top backgrounds in the presence of jet
vetoes.  It moreover provides a unified NLO description of $\ttbar$ and
single-top $\Wt$ production, including their quantum interference.

The technical problems that arise from infrared subtractions and
NLO+PS matching in the presence of top-quark resonances 
are addressed by means of the
fully general resonance-aware matching method
that was proposed in~\citere{Jezo:2015aia} and implemented in the \RES{}
framework.  This framework, besides allowing for a consistent matching to
shower Monte Carlo generators, also ameliorates the efficiency of infrared
subtraction and phase-space integration in a drastic way, and allows for a
factorised treatment of NLO radiation in off-shell top production and decays.
This represents a significant improvement (especially for what concerns top
decays) with respect to the case where NLO+PS matching is applied to a
single QCD emission.

Technically, the $\bbfourl$ generator was realised by implementing
\OpenLoops{} matrix elements in the \POWHEGBOX{} framework. To this
end we have developed a new and fully flexible interface, which allows
one to set up \POWHEGBOX{}+\OpenLoops{} NLO+PS generators for any
desired process in a rather straightforward way.

We have carried out a thorough study of the impact of the resonance-aware
method.  
To this end, we have compared our results with those
obtained after disabling the
resonance-aware formalism
in such a way that 
the \bbfourl{} generator becomes fully equivalent to a traditional \VTWO{}
implementation.
On the one hand we observed that ignoring resonance structures 
can deteriorate the performance of the generator 
up to the point of rendering it unusable. On the other hand,
we observed considerable distortions in the
reconstructed mass of the top resonances with respect to the full resonance-aware
result.  In essence, the mass distribution becomes wider around the peak and
slightly shifted.  We were able to track the origin of these effects to two
competing causes: the generation of radiation performed by \POWHEG{}, that
is considerably modified in the resonance-aware method, and the generation of
radiation in the shower stage, where the shower Monte Carlo, being unaware of
which groups of particles arise from the same resonance, tends to widen the
resonance peaks. We have also shown that it does not seem to be possible to remedy
this last problem by reconstructing the resonance structure on the basis
of simple kinematic guesses.

Much attention was dedicated to the comparison of the new $\bbfourl$
generator and the \dec{} generator of~\citere{Campbell:2014kua}.  Both
are capable of handling NLO spin correlations and radiation in top
decays. However, off-shell effects are only computed at LO in \dec{} by
reweighting the NLO cross section using the ratio of the full off-shell Born
cross section divided by its zero-width approximation.  
These
two generators are expected to provide similar results in the vicinity of 
top resonances.
In fact, in this
region, we find only modest 
differences between the two.
 In particular, the
top virtuality distribution and distributions involving $b$ jets are in
reasonably good agreement.  Slightly larger differences are found in
distributions involving $B$ hadrons, like for example, the $B$ fragmentation
function, in the top-decay frame.

A section of this work was dedicated to a comparison 
against the \hvq{} generator, which has been heavily used by the LHC experimental
collaborations for the generation of $t\bar{t}$ samples in both Run~I and
Run~II. 
Close to the mass peak, \bbfourl{} and \hvq{} predictions are fairly consistent, 
but the agreement is quickly spoiled
as one moves towards off-shell regions. 
Furthermore, the ratio of the \hvq{} to the
\bbfourl{} results exhibits a negative slope across the resonance peak, and
we found that the average virtuality of the top resonance in a window of
$\pm 30$~GeV around the peak differs by about 0.5~GeV for the two generators.
This calls for dedicated studies of the implications of 
resonance-aware matching in the context of precision $m_t$-measurements.
More sizable differences have been 
observed in the structure of the associated $b$-jets, the \bbfourl{}
generator leading consistently to narrower jets and a harder fragmentation
function for the associated $B$ hadron. 
The above findings should be interpreted 
by keeping in mind that within the \hvq{}
generator radiation in top decays is solely handled by \PythiaEight{}, with
matrix-element corrections turned on by default. 
These matrix-element corrections should improve the 
overall agreement between \hvq{} and \bbfourl{}, and we have verified that
disabling them leads to 
much more pronounced differences between the two generators.

We have included in this work an indicative comparative study of jet-veto
effects when using the \bbfourl{}, \dec{} and \hvq{} generators.  
In the presence of jet vetoes,
the \hvq{} generator alone is clearly not adequate,
since it misses the essential component of associated \Wt
production. Perhaps surprisingly, it turns out 
that also the \dec{} generator does
not perform sufficiently well.
 Since $\Wt$ production effects are
included in this generator only at the level of a leading-order reweighting,
we are led to conclude that the lack of 
NLO accuracy in the simulation of 
$\Wt$ contributions limits the usability of the \dec{} generator 
in single-top enriched regions.
We stress however that the issue of jet-veto effects is
complex, and deserves a dedicated future study.

The theoretical improvements implemented in the $\bbfourl$ generator
are relevant for phenomenological studies and experimental analysis that depends on the
kinematic details of top-decay products. In particular, this new
generator is ideally suited for precision determinations of the
top-quark mass, for measurements of $\Wt$ production, and for analyses
where $\ttbar$ and $\Wt$ production are subject to jet vetoes. The
exact treatment of off-shell and non-resonant effects is also
important for top backgrounds in Higgs and BSM studies based on
kinematic selections with high missing energy or boosted $b\bar{b}$
pairs.

\acknowledgments We thank P.~Maierh\"ofer for valuable help and discussion
and important improvements in \OpenLoops.  We thank P.~Skands for useful
exchanges about the \PythiaEight{} interface. We also wish to thank
A.~Denner, S.~Dittmaier and L.~Hofer for providing us with pre-release
versions of the one-loop tensor-integral library \Collier.  This research was
supported in part by the Swiss National Science Foundation~(SNF) under
contracts BSCGI0-157722 and PP00P2-153027, by the Research Executive Agency
of the European Union under the Grant Agreement PITN--GA--2012--316704~({\it
  HiggsTools}), and by the Kavli Institute for Theoretical Physics through
the National Science Foundation's Grant No. NSF PHY11-25915.  PN and CO would
like to express a special thanks to the Mainz Institute for Theoretical
Physics~(MITP) for its hospitality and support while part of this work was
carried out.

\appendix

\section{Technical details}
\label{app:technicalities}

In this appendix we detail the technical improvements to the \RES{} framework
that have been implemented in order to allow for the implementation of
\ppllllbb.

\subsection{Automatic generation of resonance histories}
\label{app:resonance_histories}

The algorithm for finding the resonance histories is at present at an
experimental level. It has been kept as simple and straightforward as
possible in order to allow for future improvements and modifications.

The algorithm begins with the lists of particle flavours specified in the
user process routine \verb!init_processes!, where the arrays \verb!flst_born!
and \verb!flst_real! are filled. At variance with the \VTWO{} version, one
also has to specify the length of each flavour list in the arrays. The
lengths are stored in the arrays \verb!flst_bornlength!  and
\verb!flst_reallength!. For the process we are considering here~(and in most
cases) the lengths have all the same values~(8 for the Born process and 9 for
the real).  At this stage, no resonance information is provided for the
flavour lists, so the lists of resonance pointers~(\verb!flst_bornres! and
\verb!flst_realres!)  remain initialized to zero, and the user does not need
to modify them.  The powers of the strong and weak coupling constants in the
Born amplitudes~(\verb!res_powst!  and \verb!res_powew!)  must instead be
initialized by the user-process routines.  At the moment we do not consider
the possibility of having multiple Born-level processes with different orders
of the strong and weak coupling constants.  This may be required when
considering mixed strong and electromagnetic radiation being generated with
the \POWHEG{} method, and will require minor modification of the code.

The algorithm proceeds recursively: intermediate particles are added at the
end of the flavour list, and the pointers associated with the particles that
arise from their splitting are appropriately set.

As an example, we consider the production of a $W$ in association with a
quark antiquark pair $ d \bar{u} \to e^- {\bar\nu_e} u \bar{u} $, with two
powers of the strong coupling constant and two powers of the weak one. The
input consists of the following arguments
\begin{verbatim}
  flav    = [1, -2, 11, -12,  2, -2],
  flavres = [0,  0,  0,   0,  0,  0],
  powst   = 2,
  powew   = 2.
\end{verbatim}
The algorithm proceeds as follows:
\begin{itemize}
\item
  The first particle is kept fixed. The second particle is charge reversed,
  so that the process looks like the decay of the first particle into the
  remaining ones. At this stage we then have 
\begin{verbatim}
  flav    = [1, 2, 11, -12,  2, -2],
  flavres = [0, 0,  0,   0,  0,  0],
  powst   = 2,
  powew   = 2.
\end{verbatim}
\item
  We look through all~(ordered) pairs of particles, excluding the first one,
  that have {\tt flavres} equal to zero, and that can be merged into a single
  particle via a strong or weak interaction vertex.  In the example at hand,
  we would find several cases: the second and last entry~(a $u$ and a
  $\bar{u}$) merged into a gluon, a photon or a $Z$; for the third and fourth
  entry~(an electron and its anti-neutrino) merged into a $W^-$; the last two
  entries~(a $u$ and $\bar{u}$) merged into a gluon, a photon or a $Z$.
\item
  For each found possible merging, we prepare a new input for the recursive
  procedure, with a new flavour list including the merged particle and
  updated values of the resonance pointers and of the power of the couplings.
  In our example, after the $e^- {\bar \nu}$ pair is merged into a $W^-$, the
  new input for the recursive procedure looks like this
\begin{verbatim}
  flav    = [1, 2, 11, -12,  2, -2, -24],
  flavres = [0, 0,  7,   7,  0,  0,   0],
  powst   = 2,
  powew   = 1.
\end{verbatim}
Notice that now the {\tt flavres} third and fourth entries~(the $e^-$ and
${\bar \nu}$) contain pointers to their mother resonance~(the $W^-$), added
in the seventh position. The value of {\tt powew} has been updated, since one
electroweak coupling was used for the $W^-$ splitting, and only one is left.
Notice also that there are cases where the same particles can merge into a
different one, as for the $u$ and $\bar{u}$ merging case, and all these new
inputs are passed to the recursive resonance-searching algorithm.
\item
  By proceeding with the recursion, we will reach a point when no further
  merging is possible. Following the example at hand, we may find that the
  $u{\bar u}$ pair is merged into a gluon
\begin{verbatim}
  flav    = [1, 2, 11, -12,  2, -2, -24,  0],
  flavres = [0, 0,  7,   7,  8,  8,   0,  0],
  powst   = 1,
  powew   = 1,
\end{verbatim}
followed by a $g u$ merging into a $u$
\begin{verbatim}
  flav    = [1, 2, 11, -12,  2, -2, -24,  0, 2],
  flavres = [0, 9,  7,   7,  8,  8,   0,  9, 0],
  powst   = 0,
  powew   = 1,
\end{verbatim}
finally followed by a $u W^-$ merging into a $d$
\begin{verbatim}
  flav    = [1, 2, 11, -12,  2, -2, -24, 21,  2, 1],
  flavres = [0, 9,  7,   7,  8,  8,  10,  9, 10, 0],
  powst   = 0,
  powew   = 0.
\end{verbatim}
At this point three conditions are checked: whether no pairs can be further
merged, whether no more powers of the coupling constants are available, and
whether the last added particle coincides with the first one, meaning that
all outgoing particles have been merged into the incoming one.  If any of
these conditions are not met, the configuration is abandoned.
\end{itemize}
The list just found represents a tree diagram for the process at hand. As
such, there is always a unique path in the tree that joins any two external
particles. The path joining the two incoming particle is the $t$-channel
one. It can be found starting from particle 2 and going recursively through
its ancestors, until particle 1 is reached.

The list is processed further, by the subroutine
\verb!clean_resonance_structure!, that performs the following operations. It
first examines the $t$-channel structure of the flavour list. If it finds a
$t$-channel fermion line that emits two electroweak bosons~($W$, $Z$ or
$H$) that can directly couple to each other trough a triple-boson vertex,
with any intermediate emission of photons or gluons, it abandons the
configuration.
This is because another configuration with a richer resonance
structure must exist, i.e.~the structure where the two electroweak bosons
arise from the decay of a single electroweak boson, with splittings
involving the trilinear vector coupling, or the Higgs coupling to a vector
boson. This richer configuration is well-suited to represent the one where
the two electroweak bosons do not originate from another electroweak boson,
and thus the latter configurations need not be considered.  It then carries
out a similar operation on $s$-channel lines. If we find a fermion line that
emits two electroweak bosons, there must be a richer configuration where the
two bosons originate from a single electroweak boson, and we thus abandon
this configuration. Care is taken to handle the special case when the fermion
line becomes a top quark, since the top is treated as a resonance, and the
emission of a Higgs from the fermion line, since it must come from a top, in
this case.

After the elimination procedure is carried out, all $t$-channel resonance
entries, and all $s$-channel resonance entries corresponding to a massless
particle are deleted from the list. The list is then put in a standard form:
the resonances are moved just after the two incoming particles, and the
final-state particles follow. The \verb!clean_resonance_structure! exits. If
the examined flavour structure is to be kept, the program calls a subroutine
that stores it in a temporary array structure, provided there are no other
equivalent configurations already stored. Once all configurations are found
and stored, the subroutine \verb!pwhg_res_histos_born! or
\verb!pwhg_res_histos_real! is called, and the configurations are transferred
from the temporary storage to the global arrays \verb!flst_born*! or
\verb!flst_real*!, that are overwritten with the Born and real flavour
structure including resonance-history information.

The procedure that we have illustrated so far should be appropriate for most
Standard Model processes. We checked that it works also in the case of
single-top production studied in~\citere{Jezo:2015aia}, by replacing the
hand-written resonance histories that we used there with those automatically
generated with the procedure presented here.

\subsection{Colour assignment}
\label{app:colourassignment}

In the \POWHEGBOX{}, colour assignment is mainly performed at the level of
the underlying Born process.
Given a Born flavour configuration and kinematics, one considers the colour
subamplitudes that contribute to the squared Born amplitude, computed in the
large colour limit. A color flow is then chosen with a
probability proportional to the values of the subamplitudes, for that
particular phase-space point.  The \POWHEGBOX{} then generates the QCD
radiation for a particular collinear region, through the splitting
process. The colour configuration for the generated real-emission amplitude
is obtained from the one of the underlying Born by attaching the colour flow
that corresponds to that splitting.

Contributions that do not have singular regions (i.e.~regular contributions)
are instead treated as the Born term itself.  In~\citere{Campbell:2012am}
a corresponding interface was developed such that this colour information
could be extracted from \MadGraphFour{}.
In~\refse{sec:openloops_interface}, we describe an analogous
implementation in \OpenLoops.

\begin{figure}[tbh]
\begin{center}
  \includegraphics[width=0.4\textwidth,angle=-90]{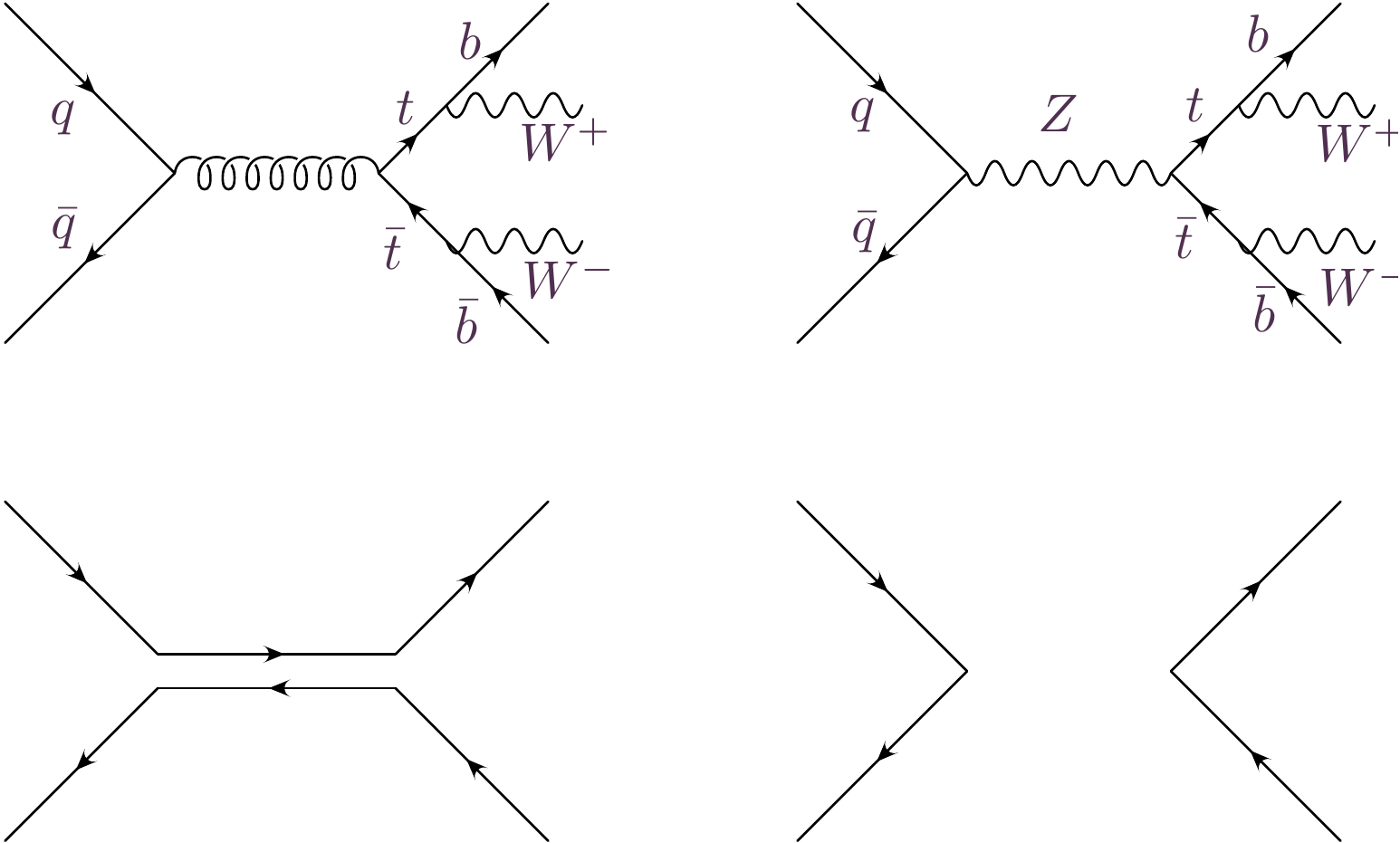}
\end{center}
\caption{Different resonance histories~(top) for identical processes, and
  associated colour flows~(bottom).}
\label{fig:colour-mismatch} 
\end{figure}

When resonance histories are being considered, a modification of this scheme
becomes necessary. In fact, the randomly generated colour assignment may not
be compatible with the resonant structure being considered. Consider for
example the process 
$q\bar q\to q\bar q VV$,
where $V$ is a vector boson, illustrated
in~\reffi{fig:colour-mismatch}.  If the process proceeds via the exchange
of a gluon in the $s$-channel, then the corresponding colour flow is
illustrated in the bottom-left diagram in the figure, and initial- and
final-state quarks are colour connected.  Whereas if the exchanged particle in
the $s$-channel is a colourless vector boson, then the color flow is depicted
in the bottom-right diagram, where it is evident that initial- and
final-state quarks are not colour connected, but there is a colour
connection between the quarks in the initial state, and another colour
connection between the quarks in the final state.

In the \POWHEGBOX{} framework, colour
assignment is independent of the resonance structure, and thus one may end up
assigning the colour flow in the left of the figure to the resonance history
on the right, or vice-versa. In order to remedy to this, we keep generating random
colour configurations, and accept the first one that is compatible with the
resonance history.

\subsection{\OpenLoops interface and settings}
\label{sec:openloops_interface}

The \OpenLoops program is based on a fast numerical recursion for the
generation of tree and one-loop scattering amplitudes~\cite{Cascioli:2011va}.
Combined with the OPP reduction method~\cite{Ossola:2006us} implemented in
\CutTools~\cite{Ossola:2007ax} and the scalar one-loop library
\OneLOop~\cite{vanHameren:2010cp}, or with the tensor integral reduction
methods~\cite{Denner:2002ii,Denner:2005nn,Denner:2010tr} implemented in
\Collier~\cite{Denner:2016kdg}, the employed recursion permits to achieve
very high CPU performance and a high degree of numerical stability. The small
fraction of numerically unstable one-loop matrix elements is automatically
detected and rescued through re-evaluation with \CutTools in quadruple
precision.

The new \POWHEGBOX{}+\OpenLoops interface is implemented via a {\sc
  Fortran90} module called {\tt openloops\_powheg}, which is included in the
\RES{} framework. Internally the \POWHEGBOX{}+\OpenLoops framework automatically compiles,
loads and manages all required \OpenLoops amplitude libraries.
The new interface provides the subroutines {\tt openloops\_born}, 
{\tt openloops\_real}, and {\tt openloops\_virtual} with interfaces identical to
the corresponding \POWHEGBOX{} routines {\tt setborn}, {\tt setreal}, and
{\tt setvirtual}. In particular the {\tt openloops\_born} routine returns,
besides the squared tree-level amplitude $\mathcal{B}$, the corresponding
colour- and spin-correlated tree-level amplitudes $\mathcal{B}_{ij}$ and
$\mathcal{B}_{\mu\nu}$ in the format required by the
\POWHEGBOX{}~\cite{Alioli:2010xd}. Additionally, the interface provides the
routines {\tt openloops\_init}, {\tt openloops\_borncolour} and {\tt
  openloops\_realcolour}.  The former synchronizes all parameters between
\OpenLoops and the \POWHEGBOX{} and should be called at the end of the {\tt
  init\_processes} subroutine of the \POWHEGBOX{}. The latter two provide
the required colour information as outlined in~\refse{app:colourassignment}, 
i.e.~they return a colour-flow of the
squared Born and real matrix elements in the large colour limit, on a
probabilistic basis.  Since the probability priors are determined from the
colour-flow decomposition of the corresponding matrix elements at a given
phase-space point, the colour-trace basis employed internally in
\OpenLoops is converted into a color-flow basis.

Several \OpenLoops internal options and switches can be passed directly
from the \verb!powheg.input! file to the code. In particular,
\OpenLoops offers the possibility to switch between the tensor-integral
reduction methods implemented in \Collier and OPP reduction methods
implemented in \CutTools.  By default \Collier is used, while inserting the
line
\begin{verbatim}
olpreset 1
\end{verbatim}
in the \verb!powheg.input! file, reduction via \CutTools can be selected. 
In a similar way, inserting the line 
\begin{verbatim}
olverbose <OpenLoops verbosity level>
\end{verbatim}
allows to select the verbosity level of \OpenLoops.

While all relevant input parameters are automatically passed by \POWHEGBOX{} to \OpenLoops, further
internal \OpenLoops parameters can be set directly via the routine (member of the {\sc
  Fortran90} module {\tt openloops})
\begin{verbatim}
set_parameter(parameter, value)
  character(*), intent(in) :: parameter
  TYPE, intent(in) :: value
\end{verbatim}
Here, {\tt TYPE} can either be integer, double or character(*) according 
to the parameter to be set, as detailed on \url{http://openloops.hepforge.org/parameters.html}.

\subsubsection*{Implementation of new processes}

In order to set up a new processes within the \POWHEGBOX{}+\OpenLoops framework, 
one should run the script 
{\tt <POWHEG-BOX-RES>/COMMON/OpenLoopsStuff/generate\_process.py} with the
following arguments
\begin{verbatim}
./generate_process.py <library name> -order_ew=<m> -order_qcd=<n> -name=<..>
\end{verbatim}
Here, {\tt <library name>} corresponds to the \OpenLoops amplitude library 
of the desired process and {\tt <m>} and {\tt <n>} denote the 
order of the Born cross section, $\ord(\as^n \aem^m)$, in terms of powers of the 
strong and weak couplings. 
This will setup a rudimentary \POWHEGBOX{} process structure within the directory
{\tt  <POWHEG-BOX-RES>/COMMON/<name>}.  For example the call
\begin{verbatim}
./generate_process.py pplnjjj -order_ew=2 -order_qcd=3 -name=Wjjj
\end{verbatim}
will yield the structure for an NLO+PS generator including all required tree and one-loop amplitudes
for $pp \to W(\to \ell \nu)+3~{\rm jet}$ production.

A user has only to provide the list of contributing flavour structures of the Born and
real subprocesses in the {\tt init\_processes.f} file and the number of intermediate
resonances in the {\tt nlegborn.h} file. An automatic
generation of these structures is currently being validated and will soon be
included.  Currently NLO QCD corrections to any SM process are supported by
this interface, while NLO electroweak corrections will follow in the
future.

\subsection{Optimising the virtual corrections}
\label{app:virtuals}

 \subsubsection*{Fixed-order NLO calculations}
If one is interested in fixed-order NLO results, the most CPU-demanding
contributions come from the computation of the real graphs, that also
implement the cancellation of the collinear and soft singularities. In the
\RES{} code there are options to separate the virtual contribution from the
rest, in such a way that it can be computed with an accuracy that matches the
one of the real contribution, thus saving computer time. More specifically,
the code can be run twice: in the first run, the user can set the flag
\verb!novirtual! to 1 in the \verb!powheg.input! file. In this way, no call
to the calculation of the virtual corrections is done, and the corresponding
distributions do not contain the virtual corrections~(plus other soft
contributions). The code is then rerun by using the same importance sampling
grids used in the first run, with the flag \verb!virtonly! set to 1 and with
lower statistics with respect to the previous run. In this way, the virtual
contributions are called fewer times with respect to the Born and real
contributions of the first run. Finally, the kinematic distributions
obtained in the two steps can then be 
combined.
Details and examples for this
procedure are included in the release of the code.

\subsubsection*{Generation of Monte Carlo events}
If one is interested in generating Monte Carlo events, it is more convenient
to avoid the computation of the virtual corrections for the large number
of events that are vetoed during the generation. This can be done, provided
one renounces to generating events with constant weight. In essence, we
generate events with settings such that the virtual contribution is not
computed, but the cross section and the distributions are sufficiently
similar to the exact result. The events are then reweighted with the full
cross section including the virtual contribution. With this procedure, the
virtual contribution is computed only once for each generated event, instead
of the several tens of event that are computed and then vetoed in a standard
run. In order to do so, one inserts in the \verb!powheg.input! file the lines
\begin{verbatim}
for_reweighting 1
rwl_file '-'
<initrwgt>
<weight id='xx'>  some reweight info </weight>
...
</initrwgt>
\end{verbatim}
and the program generates events with uniform weight with no virtual
corrections. For each \verb!<weight! line, a new weight is generated and
added to the event. These weights are all computed with the inclusion of the
virtual corrections.\footnote{
When running with the {\tt for\_reweighting}  flag set to 1, the \RES{} code
sets the internal flag {\tt flg\_novirtual} to true, and thus the subroutine
that computes the full soft-virtual contributions is forced to return zero.}

We would like to remark an additional technical issue: the subtraction term
in case of a massive fermion emitter~(i.e.~the subtraction term corresponding
to the soft singularity in the $b\to b g$ splitting) is modified in such a
way that it becomes closer to the $P_{qg}(z)$ Altarelli-Parisi splitting
function~(that is to say, we give it a weight $(1+z^2)/(1-z)$ rather than the
original weight $2/(1-z)$).  In fact, we found that if we do not include this
modification, the kinematic distributions before reweighting~(i.e.~those with
no soft-virtual contributions) can develop relatively large, negative values
near the top-mass peak. After reweighting we do get back the correct
results. However, reweighting coefficients can be very large or negative with
the original subtraction term, while, with the modified one, we get sensible
distributions even before reweighting, and no large reweighting factors.

\section{Phenomenological details}
\label{app:phenodetails}

\subsection{Kinematic guess of resonance structures}
\label{app:kinematic_guess}

In this appendix we detail the kinematic procedure for the construction of
resonance information from agnostic events, i.e.~events where no resonance
information is available.

We start at the Les Houches event level and modify each event as
follows:
\begin{itemize}
\item The matching $l\nu$ pairs are assigned to the corresponding $W^\pm$
  bosons, that are added to the event record, with the corresponding
  kinematics: $\pwp = (\plp+\pnu)$ and $\pwm = (\plm+\pnubar)$

\item If no parton is radiated, or if the radiated parton is not a gluon,
  then the top resonances can only be formed by pairing a $W$ and the
  corresponding $b$~(i.e.~no radiation from top quarks is present). In this
  case we compute the resonance enhancement factors
  \begin{eqnarray}
    f_{ t} &=& \frac{m_t^4}{\lq (\pb + \pwp)^2-m_t^2 \rq^2+(\Gamma_t
        m_t)^2}\,, \\ 
    f_{ \bar t} &=& \frac{m_t^4}{\lq (\pbbar + \pwm)^2 -m_t^2\rq^2+(\Gamma_t
      m_t)^2}\,, \\ 
    f_{\sss Z} &=& \frac{m_{\sss z}^4}{\lq (\pwp+\pwm)^2-m_{\sss
        Z}^2\rq^2+(\Gamma_{\sss Z}  m_{\sss Z})^2} \, .
  \end{eqnarray}
We then generate a random number $r$: if $r<f_{\sss Z}/(f_{t}\,f_{ \bar
  t}\,+f_{\sss Z})$ we assume that the event has a resonance history with the
$W$ pair arising from an intermediate $s$-channel $Z$. If not, the $W$'s are
paired with the corresponding $b$ and assigned to a top and an anti-top. In
both cases, the Les Houches event record is adjusted accordingly.

\item If the radiated parton is a gluon, besides the $f$ factors computed in
  the previous item, we compute
    \begin{eqnarray}
    f_{ t}^{\sss (r)} &=&  \frac{m_t^4}{ \lq (\pb +
      \pwp+\pg)^2-m_t^2\rq^2+(\Gamma_t m_t)^2}\,, \\
    f_{ {\bar t}}^{\sss (r)} &=& \frac{m_t^4}{ \lq (\pbbar +
      \pwm+\pg)^2-m_t^2\rq^2+(\Gamma_t m_t)^2} \,.
  \end{eqnarray}
    Furthermore, we also compute:
    \begin{itemize}
     \item $\pt^{\sss g}$: the gluon transverse-momentum relative to the
       beams;
     \item $p_{\sss T,b/{\bar b}}^{\sss g}$: the gluon transverse-momentum 
       relative to the outgoing $b$'s in the partonic CM system;
     \item $p^{\sss g (r)}_{\sss T,b/{\bar b} }$: the gluon
       transverse-momentum relative to the outgoing $b$'s in the $t^{\sss
         (r)}/{\bar t}^{\sss (r)}$ frame built under the assumption that
       radiation arises from top/anti-top decay.
    \end{itemize}
  \item We compute the following weights
    \begin{eqnarray}
      &&w_{\sss Z} = \frac{f_{\sss Z}}{\(\pt^{\sss g}\)^2} 
      + \frac{f_{\sss Z}}{\( p_{\sss T,b}^{\sss g}\)^2} + 
      \frac{f_{\sss Z}}{\( p_{\sss T,\bar b}^{\sss g}\)^2} \,,\\      
     && w_{\sss t{\bar t}} = \frac{f_{\sss t}\,f_{\sss \bar t}}{\(\pt^{\sss
          g}\)^2}\, ,
      \qquad   w_{\sss t}^{\sss  (r)} =
       \frac{ f_{\sss t}^{\sss (r)} }{\( p^{\sss g (r)}_{\sss T,b}\)^2 }\,  , \qquad
      w_{\sss \bar t}^{\sss  (r)} = 
      \frac{f_{\sss \bar t}^{\sss (r)}}{\( p^{\sss g (r)}_{\sss T,\bar b}\)^2 }\, ,
    \end{eqnarray}
    corresponding to the following resonance histories:
    \begin{itemize}
      \item the two $W$'s come from the intermediate $Z$ boson and the
        gluon is associated to initial- or final-state radiation~(from the
        $b$'s);
      \item the $W$'s and $b$'s come from a $ t\bar t$ pair, and the gluon
        from initial-state radiation;
      \item same as before but with the gluon from the top-resonance decay;
      \item same as before but with the gluon from the antitop-resonance decay.
    \end{itemize}
    If the colour of the $b$ is not consistent with the colour assigned to
    the gluon, the corresponding weight is set to zero.  Then, a resonance
    history and gluon assignment are chosen among the four configurations
    considered here with probability proportional to the corresponding $w$
    weight.
\end{itemize}

\begin{figure}[tbh]
\begin{center}
  \includegraphics[width=0.49\textwidth]{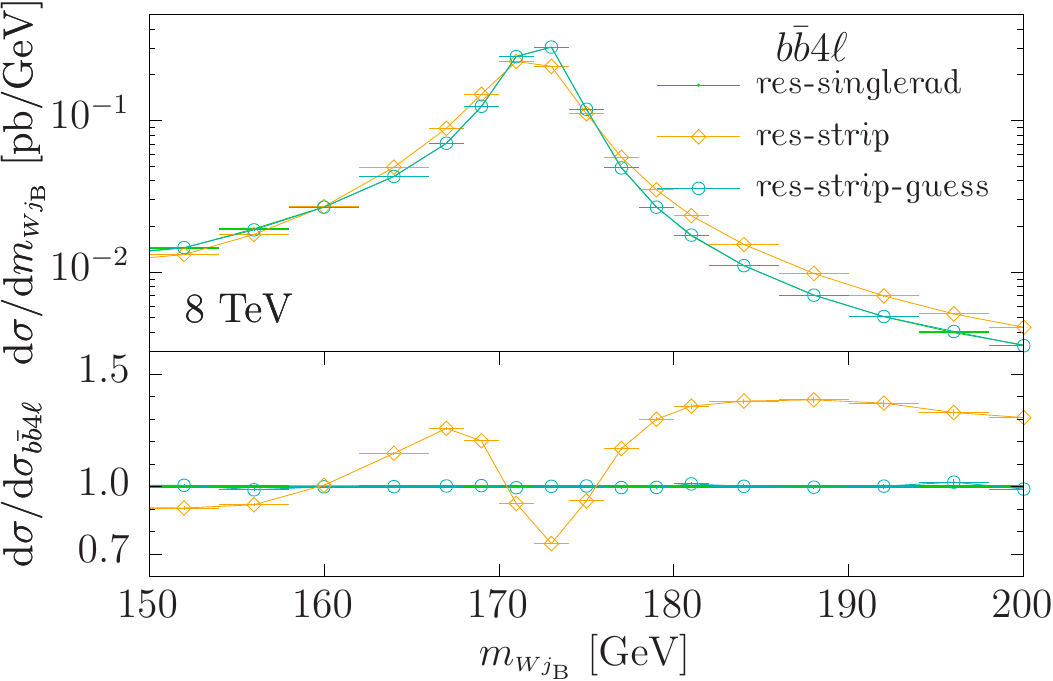}
\end{center}
\caption{Comparison of the effect of removing resonance information and then
  adding it back according to a guess, based upon kinematics, in the invariant
  mass of the $W\bj$ system.}
\label{fig:m_w_jbot-stripres-guessres} 
\end{figure}

In order to validate this procedure, we stripped any resonance information
from the Les Houches events of a resonance-aware \bbfourlNoallrad{} sample,
i.e.~switching off the multiple-radiation scheme. We indicate the
corresponding results with the label \stripres{}. Then, following the
procedure outlined above, we added back guessed resonance information.  The
obtained result, labelled as \stripresguess{}, is displayed in~\reffi{fig:m_w_jbot-stripres-guessres}.
As we can see, the procedure for the kinematic construction of resonance information
reproduces nicely the correct $W\bj$ peak.

\subsection{Comparisons of shower veto schemes}
\label{sec:veto}
When generating radiation from resonance decay, the traditional Les Houches
generic-user process interface is no longer viable. In fact, the
standard~\cite{Boos:2001cv} contemplates only a single scale~(called
\verb!scalup!), and it requires that the shower does not generate any
radiation harder than that scale at the production stage. Radiation in
resonance decays remains, however, unrestricted, while our generator requires
it to be vetoed either by the transverse momentum of the radiation generated
by \POWHEG{} in the decay process~(\verb!allrad 1! case), or by the hardest
radiation scale, irrespective of its origin~(i.e.~either from production or
from resonance decay) in the \verb!allrad 0! case.

The default method for interfacing the \bbfourl{} generator to \PythiaEight{}
was taken from~\citere{Campbell:2014kua}, and it is described in
appendix~A of that paper. In essence, the procedure was to examine the
showered event, compute the transverse momentum of \PythiaEight{} radiation
in top decays, and veto it if higher than the corresponding \POWHEG{} one.

\PythiaEight{} provides with its own mechanism for vetoing radiation from
resonance decay. One should implement a virtual function
\verb!canSetResonanceScale!  that returns a true value if \PythiaEight{} is
to use this mechanism.  Furthermore, one should also implement a function
\verb!scaleResonance!  that \PythiaEight{} invokes in each event for each
resonance, returning the scale for vetoing radiation in decay.  We also
implement this mechanism in our generator. It is activated by setting the
flag \verb!pythiaveto 1! in the \verb!powheg.input! file.  

\begin{figure}[tbh]
\begin{center}
  \includegraphics[width=0.49\textwidth]{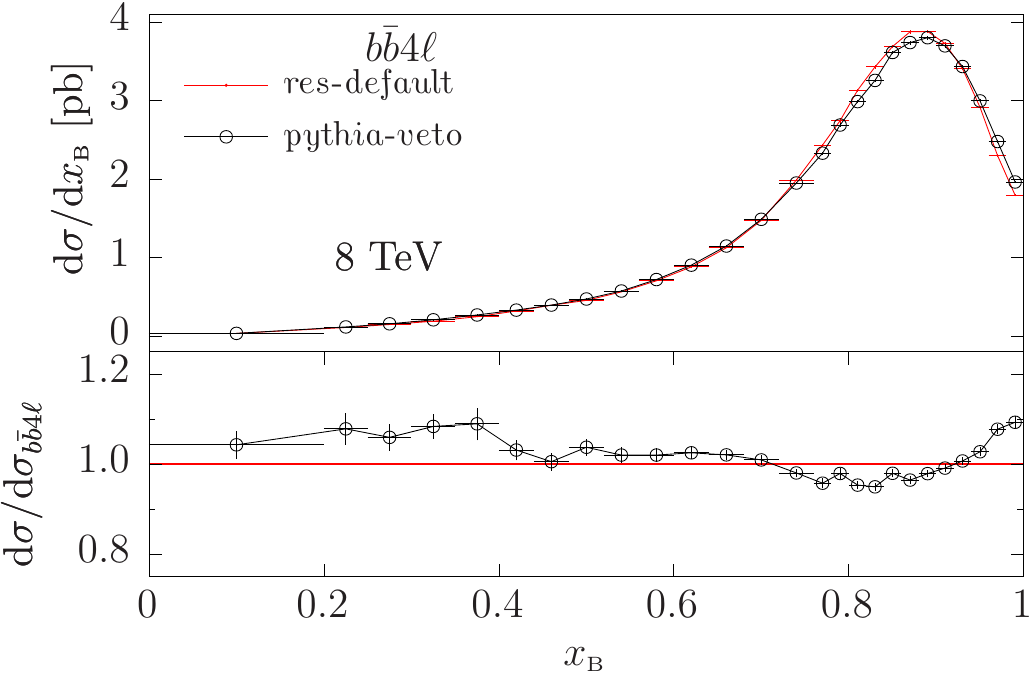}
  \includegraphics[width=0.49\textwidth]{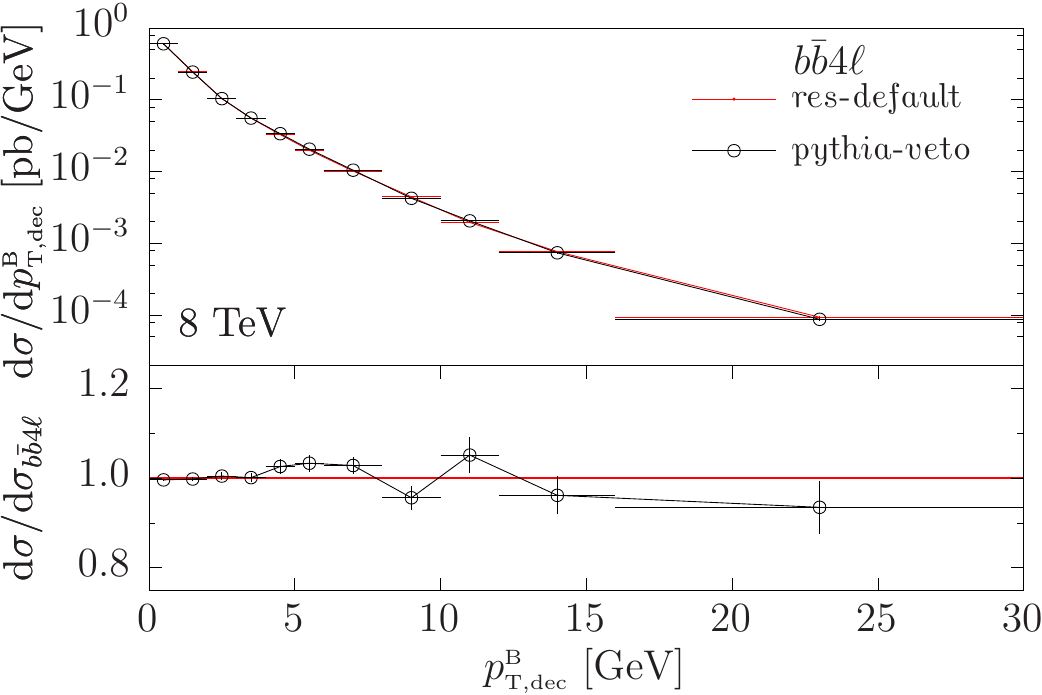}
  \includegraphics[width=0.49\textwidth]{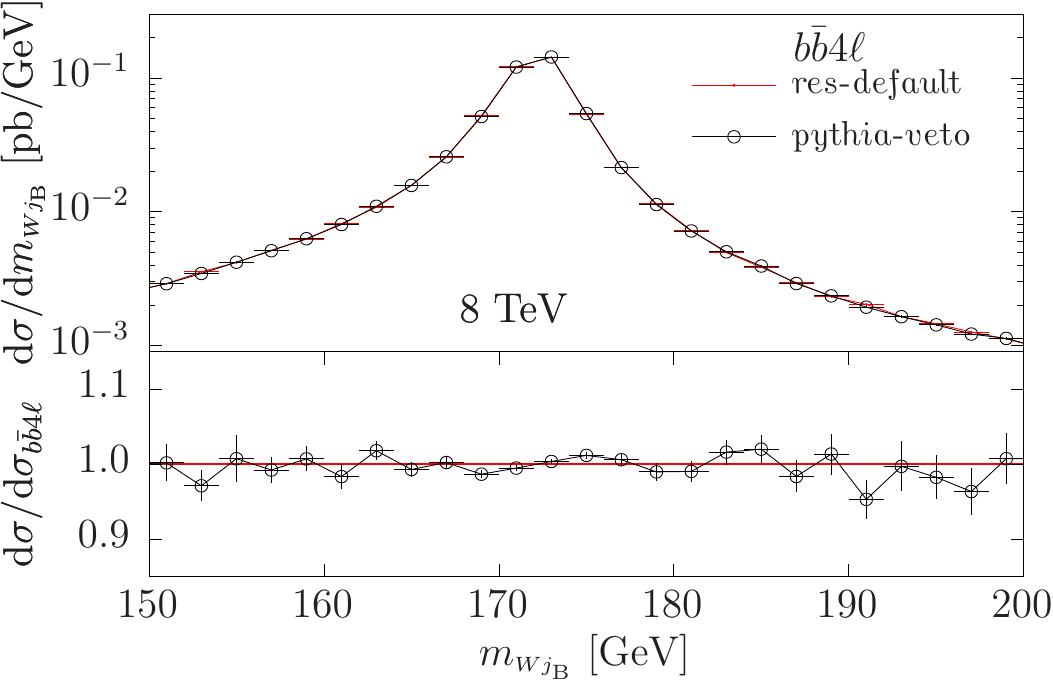}
  \includegraphics[width=0.49\textwidth]{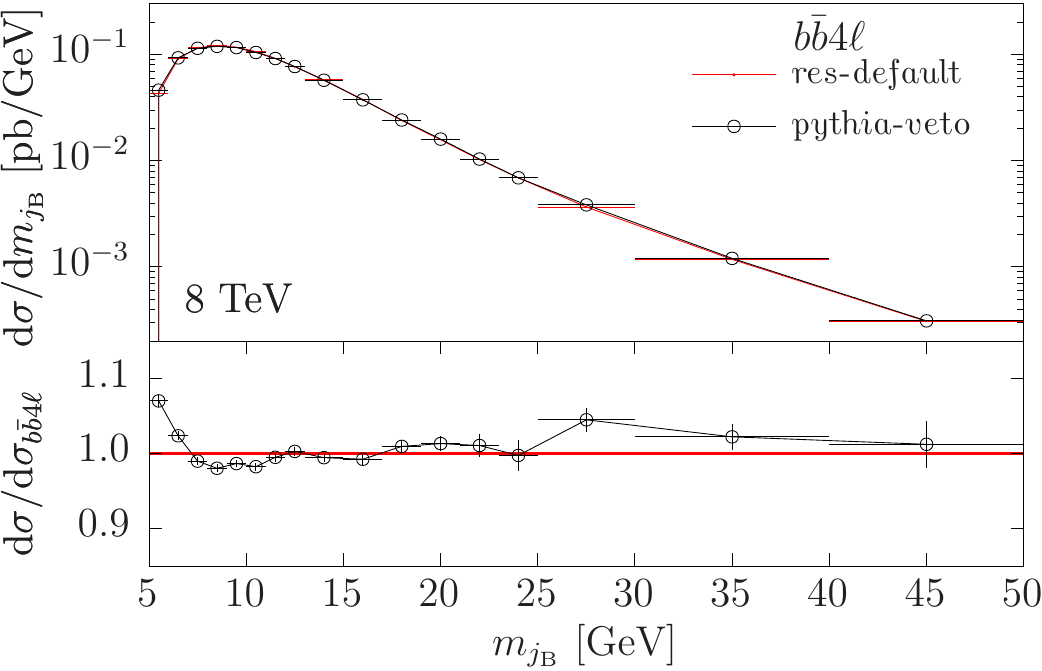}
\end{center}
\caption{Comparison of two veto schemes on the $B$ fragmentation function, on 
  \ptdecB{}, on the mass of the $W\bj$ system and on the mass of $\bj$
  distributions.}
\label{fig:fig-weveto-pyveto-bfrag} 
\end{figure}

We show in~\reffi{fig:fig-weveto-pyveto-bfrag}
the comparison of results obtained with the two veto mechanisms.  In these
plots, as well as in all the others that we have examined, we have found very
good agreement between the two veto schemes.  We notice that the difference
in the ratio of the $m_\bj$ distribution at small masses~(one of the few
distributions where we found mild discrepancies) is taking place in a region
where the cross section is getting small, and is thus of little relevance.

We conclude that the internal \PythiaEight{} method for vetoing
resonance radiation in decay is suitable for use with the \bbfourl{}
generator, and we can thus recommend its use.

\subsection{Impact of the multiple radiation scheme}
\label{sec:ALLRAD_NOALLRAD}
\begin{figure}[htb]
\begin{center}
  \includegraphics[width=0.49\textwidth]{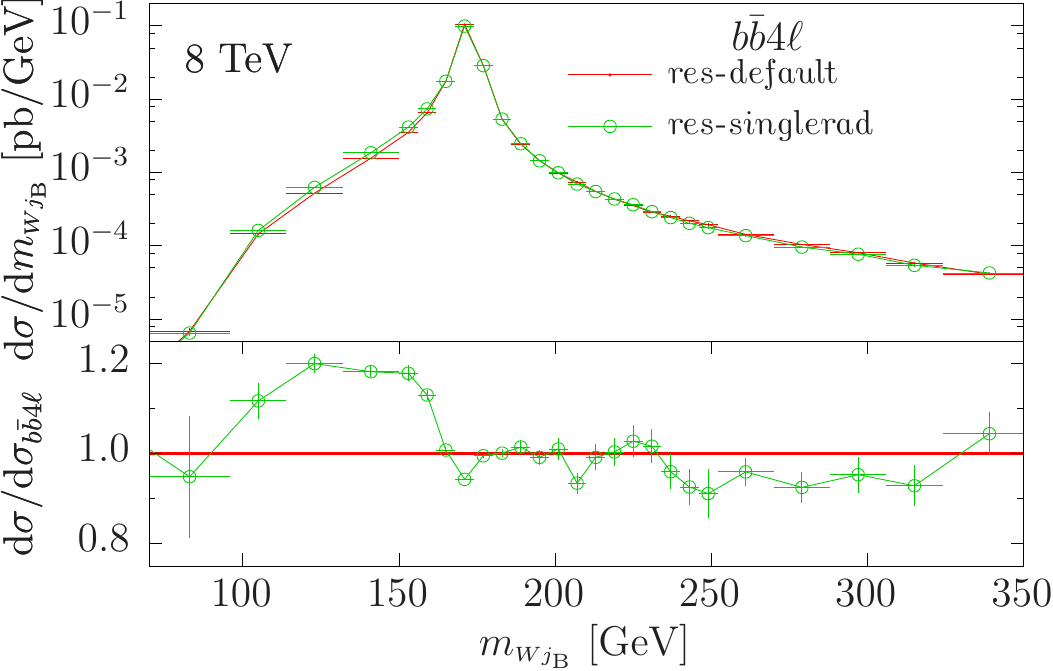}
  \includegraphics[width=0.49\textwidth]{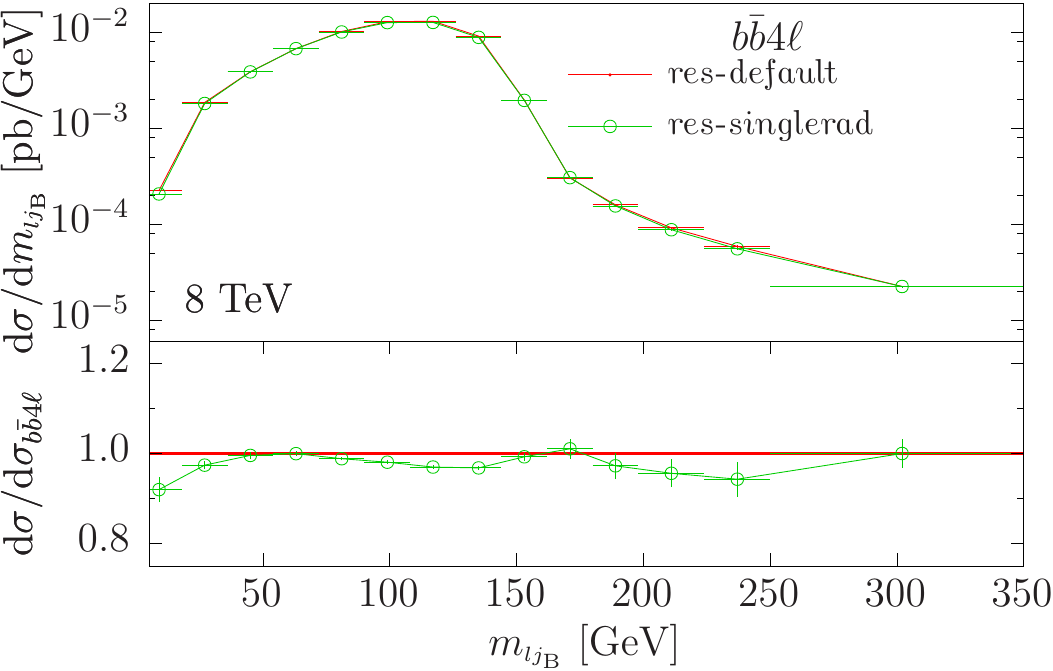}
\end{center}
\caption{Invariant mass of the $W\bj$~(left) and of the $l\bj$~(right)
  systems. We compare NLO+PS resonance-aware predictions with (\resdefault) and without (\bbfourlNoallrad{}) employing the multiple radiation scheme. In the ratio plot we illustrate relative deviations with respect to \resdefault.}
\label{fig:m_w_jbot-allrad-noallrad} 
\end{figure}

\begin{figure}[htb]
\begin{center}
  \includegraphics[width=0.49\textwidth]{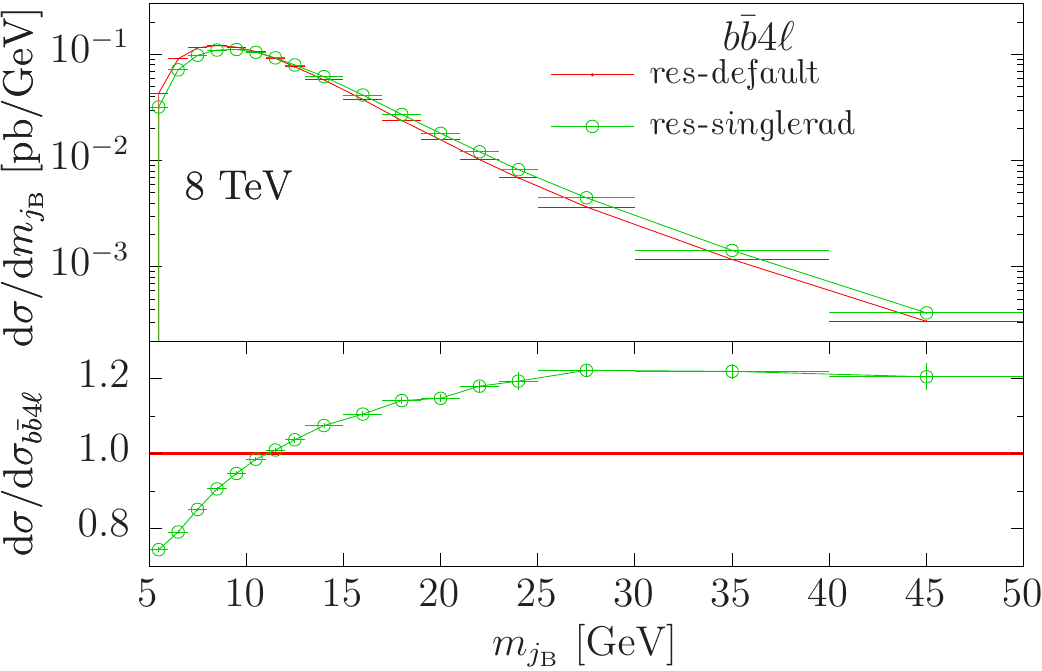}
  \includegraphics[width=0.49\textwidth]{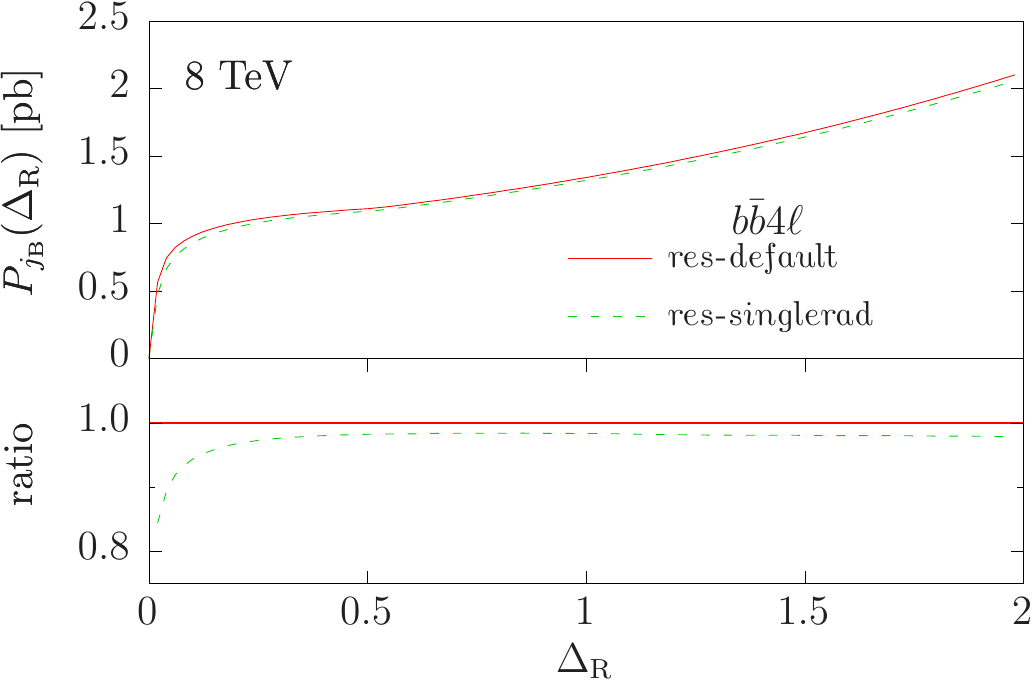}
  \includegraphics[width=0.49\textwidth]{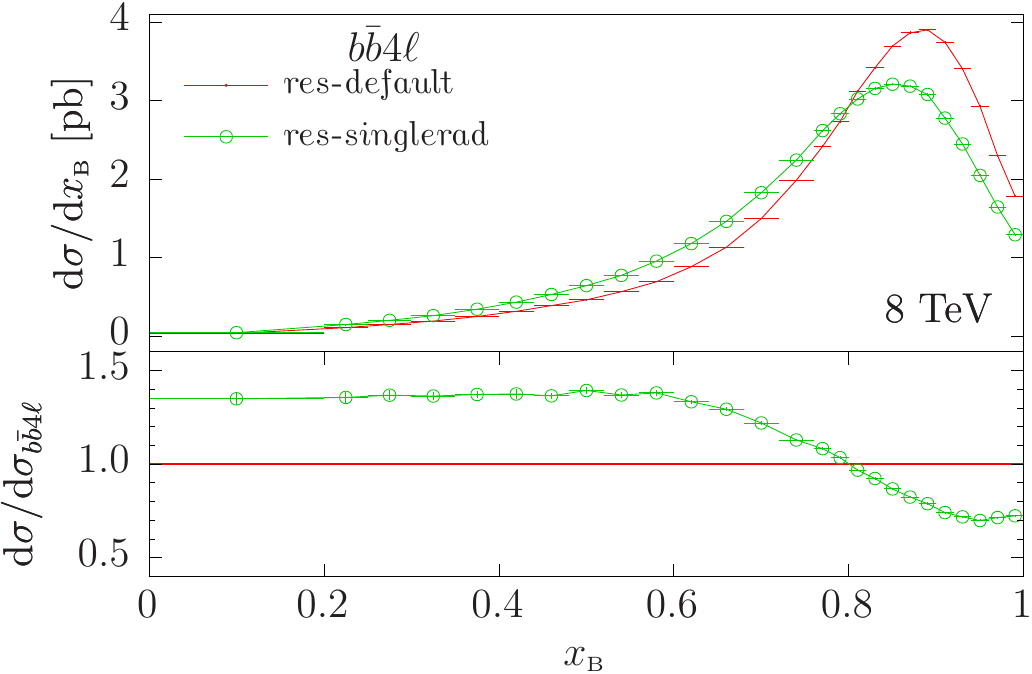}
  \includegraphics[width=0.49\textwidth]{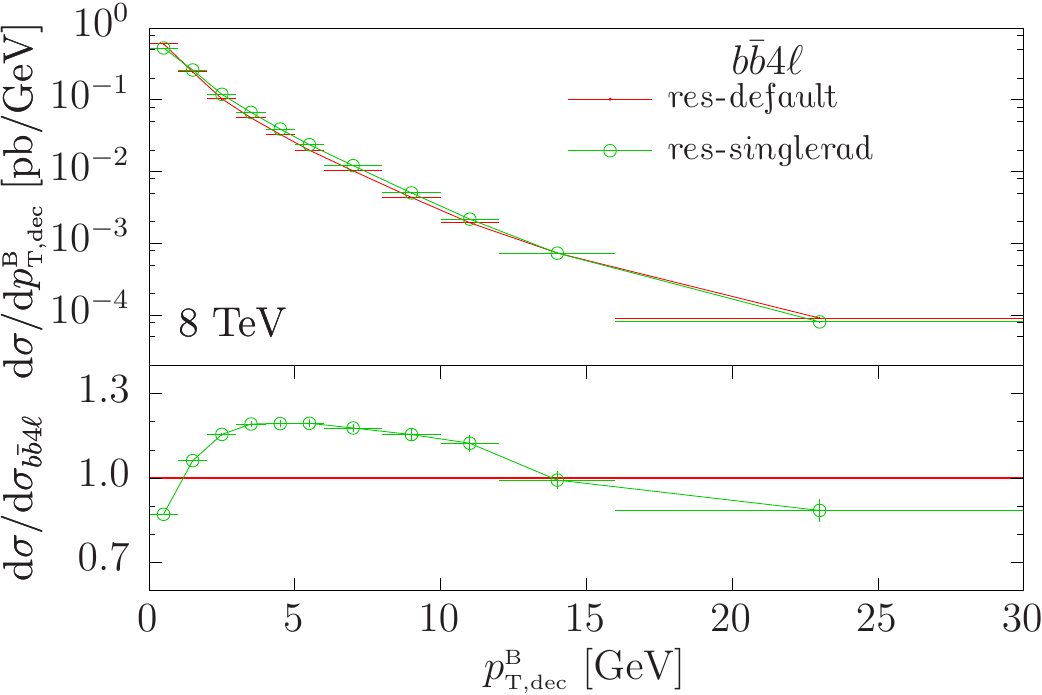}
\end{center}
\caption{Mass~(left top) and profile~(right top) of the $b$-jet, $\bj$, and for
  the $B$ fragmentation function~(left bottom) and transverse momentum
  distribution of the $B$ hadron in the top decay frame, \ptdecB{}~(right bottom).  Absolute predictions and ratios as 
in~\reffi{fig:m_w_jbot-allrad-noallrad}. }
\label{fig:bjet-allrad-noallrad} 
\end{figure}

\begin{figure}[htb]
\begin{center}
  \includegraphics[width=0.49\textwidth]{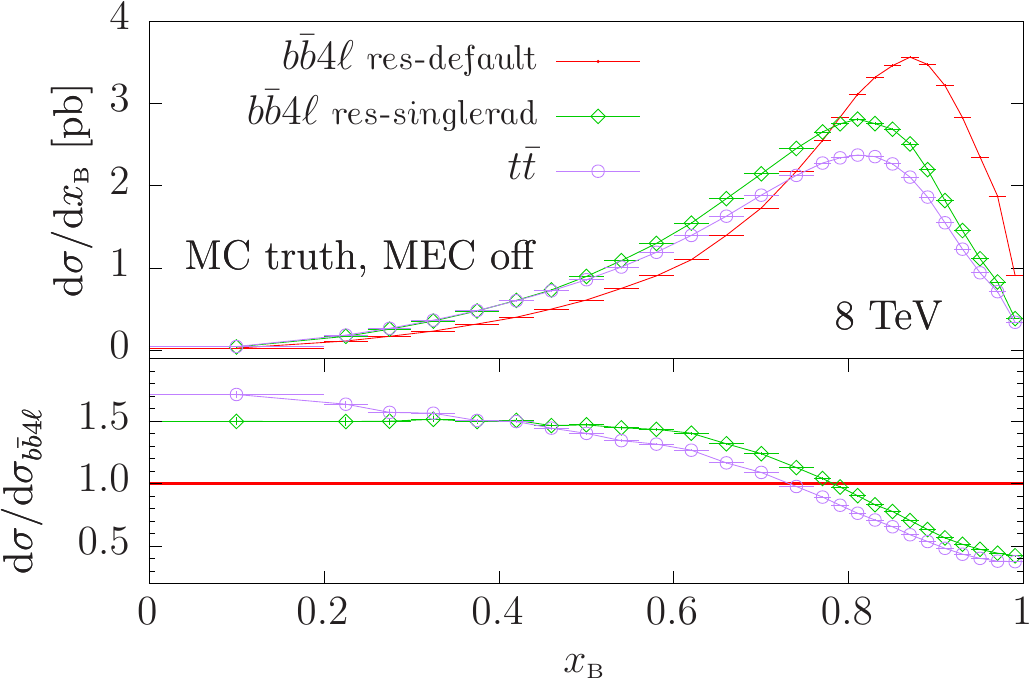}
  \includegraphics[width=0.49\textwidth]{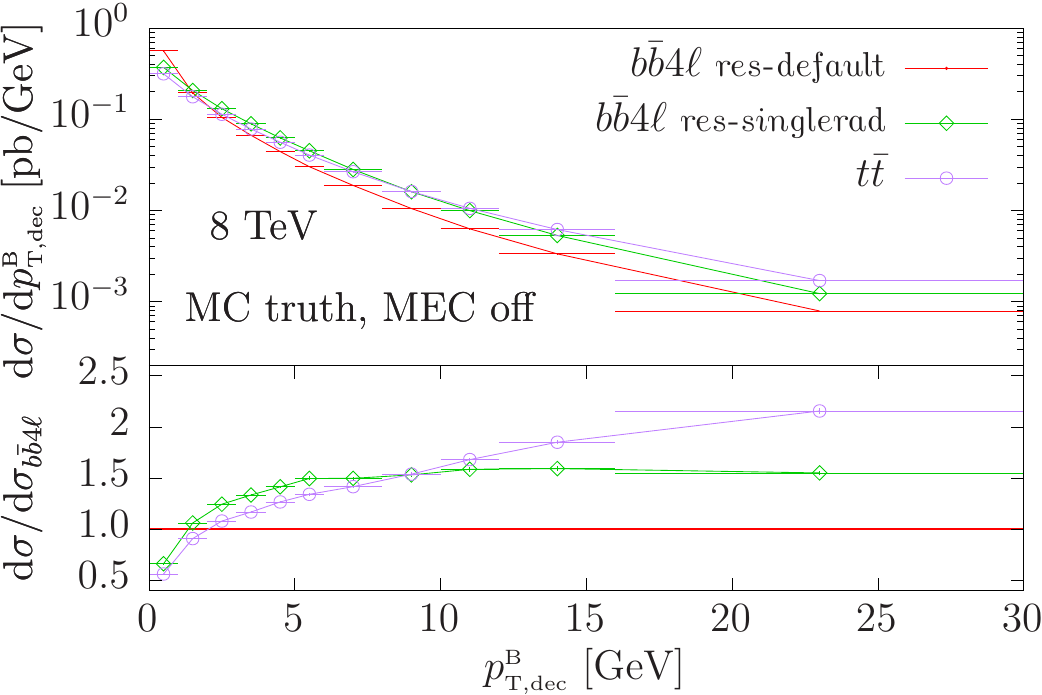}
\end{center}
\caption{Predictions for the $B$ fragmentation function and transverse momentum distribution of the $B$ hadron in the top decay frame, \ptdecB{} obtained with the \bbfourl{} generator in its default mode employing the multiple radiation scheme (\resdefault), without employing this scheme (\bbfourlNoallrad) and corresponding predictions obtained with the 
\hvq{} generator (\TTBAR). In these predictions the top reference frame is determined according to the Monte Carlo truth~(MC truth) and the \TTBAR predictions are obtained switching off matrix-elements corrections in \PythiaEight{}.
}
\label{fig:8TeV_fragb-allrad-noallrad-MCtruth} 
\end{figure}

Difference in $\ttbar$ observables induced by the multiple-radiation scheme
of \refeq{eq:allrad}
were already discussed at length in~\citere{Campbell:2014kua} for the
\DEC{} generator.  It was found there that, by switching off the
multiple-radiation scheme~(\verb!allrad! 0), radiation from top decays are
mostly handled by the shower generator. In fact, the absolute hardest
radiation is more often produced by the initial state, in part because of the
larger colour charge, and in part due to the wider phase space available.
Here we present some comparisons as a brief reminder of the relevant issues.
In this section we apply our default cuts 
defined in~\refeqs{eq:jet_cuts}{eq:leptonic_cuts}.
We begin by
showing in~\reffi{fig:m_w_jbot-allrad-noallrad}
the invariant mass distribution of the $W\bj$ and of the $l\bj$
systems. There is a good agreement between the two distributions, except for
the region of low top virtuality in the left plot.  On the other hand,
observables that are sensitive to the $B$ and \bj{} properties display larger
differences, as can be seen in~\reffi{fig:bjet-allrad-noallrad}.
In view of the large differences in the fragmentation function and \ptdecB{} distribution, we
compare in~\reffi{fig:8TeV_fragb-allrad-noallrad-MCtruth}
the same quantities computed using as reference frame the top quark at the
level of Monte Carlo truth~(``MC truth'', usually identified with the last
top quark appearing in the shower output list) rather than the reconstructed
top.  We also add to this comparison the output of the \hvq{} generator. In
this last case, we switch off \PythiaEight{} matrix element
corrections~(MEC), for the purpose of determining whether the use of our
generator, even if the \verb!allrad! feature is switched off, brings out some
improvement with respect to a generic shower treatment of top decays.  We see
from the figures that by using the MC truth for the top reference frame
brings the \bbfourl{} and \bbfourlNoallrad{} results in better agreement, at
least as far as the \ptdecB{} distribution is concerned.

The comparison of the \bbfourl{}, \bbfourlNoallrad{} and \hvq{} results for
the \ptdecB{} distribution is particularly enlightening.  If we focus upon
radiation in the top decay, in the \bbfourl{} case the hardest radiation is
always generated by \POWHEG{}. In the \bbfourlNoallrad{} case, \POWHEG{} is
mostly responsible for radiation with a large value of the \ptdecB{}
observable, since it must be harder than the radiation generated in
production. The region of small \ptdecB{} is thus more often determined by
the shower.  In the \hvq{} case, radiation in the top decay is handled only
by the shower, that has only leading logarithmic accuracy, and thus fails at
large values of the \ptdecB{} observable. This is why we see a large
discrepancy between the \hvq{} and the \bbfourlNoallrad{} at large values of
the \ptdecB{} observable.

\bibliography{paper}

\end{document}

%% file: texmacros.tex
\newcommand\POWHEGBOX{{\tt POWHEG-BOX}}
\newcommand\POWHEG{{\tt POWHEG}}
\newcommand\MCatNLO{{\tt MC@NLO}}
\newcommand\PythiaEight{{\tt Pythia8}}
\newcommand\Herwigseven{{\tt Herwig7}}
\newcommand\PythiaEightPone{{\tt Pythia8.1}}
\newcommand\PythiaEightPtwo{{\tt Pythia8.2}}
\newcommand\MadGraphFour{{\tt MadGraph4}}
\newcommand\RES{{\tt POWHEG-BOX-RES}}
\newcommand\VTWO{{\tt POWHEG-BOX-V2}}
\newcommand\ttNLOdec{{\tt ttb\_NLO\_dec}}
\newcommand\DEC{\ttNLOdec}

\newcommand\TTBAR{\ttbar}
\newcommand\TTBARDEC{\ttbardecay}
\newcommand\BBFL{$b \bar b 4\ell$}
\newcommand\BBFLRES{\BBFL{}}

\font\dunhxi=cmu10   

\newcommand{\fntres}{\dunhxi}
\newcommand{\prfx}{res}
\newcommand\resdefault{{\fntres \prfx-default}}
\newcommand\bbfourlPYdefault{{\fntres \prfx-default}}

\newcommand\default{\bbfourlPYdefault}

\newcommand\bbfourl{{\tt bb4l}\xspace}
\newcommand\bbfourlNoallrad{{\fntres \prfx-singlerad}\xspace}
\newcommand\bbfourlPYnoallrad{{\fntres \prfx-singlerad}}

\newcommand\dec{{\tt ttb\_NLO\_dec}}

\newcommand\noresi{{\fntres \prfx-guess}}
\newcommand\noresiPY{{\fntres \prfx-guess}}
\newcommand\nores{{\fntres \prfx-off}}
\newcommand\noresPY{{\fntres \prfx-off}}
\newcommand\stripres{{\fntres \prfx-strip}}
\newcommand\stripresPY{{\fntres \prfx-strip}}

\newcommand\stripresguess{{\fntres \prfx-strip-guess}}
\newcommand\hvq{{\tt hvq}}

\newcommand{\MCFM}{{\tt MCFM}\xspace}
\newcommand{\OpenLoops}{{\tt OpenLoops}\xspace}
\newcommand{\CutTools}{{\tt CutTools}\xspace}
\newcommand{\OneLOop}{{\tt OneLOop}\xspace}
\newcommand\Collier{{\tt COLLIER}\xspace}

\newcommand{\mathd}{\mathrm{d}}

\newcommand{\tmop}[1]{\ensuremath{\operatorname{#1}}}

\newcommand{\GeV}{\text{GeV}\xspace}
\newcommand{\GF}{{G_{\sss\mu}}}
\newcommand{\ri}{\mathrm{i}}

\newcommand{\kt}{\ensuremath{k_{\sss T}}\xspace}
\newcommand\sss{\mathchoice%
{\displaystyle}%
{\scriptstyle}%
{\scriptscriptstyle}%
{\scriptscriptstyle}%
}
\newcommand{\pt}{\ensuremath{p_{\sss T}}\xspace}

\newcommand{\ord}{\mathcal{O}}
\newcommand\enmn{\fourl}
\newcommand{\ttbar}{\ensuremath{t \bar t}\xspace}
\newcommand{\bbbar}{\ensuremath{b \,\bar b}\xspace}
\newcommand{\ttbardecay}{\ensuremath{t \bar t\,\otimes\,}decay\xspace}
\newcommand{\Wt}{\ensuremath{W t}\xspace}
\newcommand{\WW}{\ensuremath{W^+W^-}\xspace}
\newcommand{\WWbb}{\ensuremath{W^+W^- b \bar b}\xspace}
\newcommand{\WWb}{\ensuremath{W^+W^- b}\xspace}

\newcommand{\ppllllbbX}{\ensuremath{\ppllllbb + X}\xspace}
\newcommand{\ppllllbb}{\ensuremath{pp \,\to \,\enmn\, \bbbar}\xspace}

\newcommand{\Pt}{\ensuremath{t}\xspace}

\newcommand{\PW}{\ensuremath{W}\xspace}
\newcommand{\PZ}{\ensuremath{Z}\xspace}
\newcommand{\PH}{\ensuremath{H}\xspace}
\newcommand{\nj}{\ensuremath{n_{j}}\xspace}
\newcommand{\nb}{\ensuremath{n_{b}}\xspace}

\newcommand{\bj}{\ensuremath{{j_{\rm \sss B}}\xspace}}
\newcommand{\bbj}{\ensuremath{j_{\rm \sss {\bar B}}}\xspace}

\newcommand{\ptdecB}{\ensuremath{p^{\rm\sss B}_{\sss \rm T,dec}}\xspace}

\newcommand\pwp{p_{\sss W^+}}
\newcommand\pwm{p_{\sss W^-}}
\newcommand\plp{p_{\sss \ell^+}}
\newcommand\plm{p_{\sss l^-}}
\newcommand\pnu{p_{\sss \nu}}
\newcommand\pnubar{p_{\sss \bar\nu}}
\newcommand\pb{p_{\sss b}}
\newcommand\pbbar{p_{\sss \bar{b}}}
\newcommand\pg{p_{\sss g}}

\def\beq{\begin{equation}}
\def\beqn{\begin{eqnarray}}
\def\eeq{\end{equation}}
\def\eeqn{\end{eqnarray}}

\def\lq{\left[} 
\def\rq{\right]}

\def\({\left(} 
\def\){\right)} 
 
\newcommand     \MSB            {\ifmmode {\overline{\rm MS}} \else
                                 $\overline{\rm MS}$\fi}

\newcommand\nlf{n_{\rm lf}}

\newcommand\as{\alpha_{\sss\rm S}}

\newcommand\aem{\alpha_{\sss\rm EM}}

\newcommand\TF{T_{\sss\rm F}}

\newcommand\muf{\mu_{\sss\rm F}}
\newcommand\mur{\mu_{\sss\rm R}}
\newcommand\mb{m_{b}}
\newcommand\fourl{\ell^+\nu_{\sss\ell}\, l^-\bar{\nu}_{\sss l}}

\newcommand\pT{\ensuremath{p_{\rm T}}}
\newcommand\pTthr{\ensuremath{p_{\rm \sss T,jet}^{\rm\sss thr}}}

\newcommand\xB{\ensuremath{x_{\sss \rm B}}}
\newcommand\Etbj{\ensuremath{P_{\bj}}}
\newcount\minutes 
\newcount\scratch 
\def\timestamp{%
\scratch=\time 
\divide\scratch by 60 
\edef\hours{\the\scratch} 
\multiply\scratch by 60 
\minutes=\time 
\advance\minutes by -\scratch 
---$\,$\hours:\null 
\ifnum\minutes< 10 0\fi 
\the\minutes}

\def\refeq#1{\mbox{Eq.~(\ref{#1})}}
\def\refeqs#1#2{\mbox{Eqs.~(\ref{#1})--(\ref{#2})}}
\def\reffi#1{\mbox{Fig.~\ref{#1}}}
\def\reffis#1#2{\mbox{Figures~\ref{#1}--\ref{#2}}}
\def\refta#1{\mbox{Table~\ref{#1}}}

\def\refse#1{\mbox{Section~\ref{#1}}}
\def\refses#1#2{\mbox{Sections~\ref{#1}--\ref{#2}}}
\def\refapp#1{\mbox{Appendix~\ref{#1}}}
\def\citere#1{\mbox{Ref.~\cite{#1}}}

\definecolor{mygray}{gray}{0.5}

